\documentclass[aps,pra,twocolumn,superscriptaddress,floatfix,showkeys]{revtex4}

\usepackage[utf8]{inputenc}
\usepackage{graphicx}
\usepackage{physics}
\usepackage{xcolor, soul}
\usepackage{float}
\usepackage{xfrac} % for \sfrac
\usepackage{gensymb} %for the \degree sign
\usepackage{units} %nicefrac
\usepackage{amssymb}
\usepackage[colorlinks=true,linkcolor=blue,urlcolor=black,citecolor=black]{hyperref}
\newcommand{\drm}{\mathrm{d}}
\newcommand{\me}{\mathrm{e}}

\usepackage{bbold}
\usepackage[normalem]{ulem} %makes underlining possible: \uline \uwave \sout
\usepackage{color}%makes \textcolor{grey}{text} available

\usepackage[normalem]{ulem}

\graphicspath{ {./Figures/} }

\begin{document}

\author{Georgia M. Nixon}
\affiliation{Cavendish Laboratory, University of Cambridge, J.J. Thomson Avenue, Cambridge CB3 0HE, United Kingdom\looseness=-1}
\email{gmon2@cam.ac.uk}

\author{F. Nur \"{U}nal}
\affiliation{Cavendish Laboratory, University of Cambridge, J.J. Thomson Avenue, Cambridge CB3 0HE, United Kingdom\looseness=-1}

\author{Ulrich Schneider}
\affiliation{Cavendish Laboratory, University of Cambridge, J.J. Thomson Avenue, Cambridge CB3 0HE, United Kingdom\looseness=-1}

\keywords{Floquet engineering, quantum simulation, ultracold atoms, local control, optical lattices}

% \title{Local periodic driving in optical lattices}
\title{Individually tunable tunnelling coefficients in optical lattices using local periodic driving}
%\title{\gn{Generating locally tunable tunnelling coefficients in optical lattices using periodic driving}}
% Floquet engineering locally tunable tunnelling coefficients in optical lattices
% \title{Local Floquet Hamiltonian Engineering in Optical Lattices}

\begin{abstract}
Ultracold atoms in optical lattices have emerged as powerful quantum simulators of translationally invariant systems with many applications in e.g.\ strongly-correlated and topological systems. 
However, the ability to locally tune all Hamiltonian parameters remains an outstanding goal that would enable the simulation of a wider range of quantum phenomena.
Motivated by recent advances in quantum gas microscopes and optical tweezers, we here show theoretically how local control over individual tunnelling links in an optical lattice can be achieved by incorporating local time-periodic potentials. 
We propose to periodically modulate the on-site energy of individual lattice sites and employ Floquet theory to demonstrate how this provides full individual control over the tunnelling amplitudes in one dimension. 
We provide various example configurations realising interesting topological models such as extended Su-Schrieffer-Heeger models that would be challenging to realise by other means. 
Extending to two dimensions, we demonstrate that local periodic driving in a Lieb lattice engineers a 2D network with fully controllable tunnelling magnitudes. In a three-site plaquette, we show full simultaneous control over the relative tunnelling amplitudes and the gauge-invariant flux piercing the plaquette, providing a clear stepping stone to building a fully programmable 2D tight-binding model. We also explicitly demonstrate how utilise our technique to generate a magnetic field gradient in 2D.
This local modulation scheme is applicable to many different lattice geometries.
\end{abstract}

\maketitle

%%%%%%%%%%%%%%%%%%%%%%%%%%%%%%%%%%%%%%%%%%%%%%%%%%%%%%%%%%%%%%%%%%%%%%%%%%%%%%%%%%%%%%%%%%%%%%
\section{Introduction}
The realisation of fully-programmable, coherent quantum systems is an important but challenging task in modern science and technology. Analogue quantum simulation
% ~\cite{Manin-ComputableNonComputable-1980, feynman_simulating_1981} %, which involves studying quantum many-body problems using experimental quantum systems , 
is a vibrant area of research yielding insights into a variety of problems that are computationally and analytically intractable \cite{Cirac-Zoller-GoalsOpportunitiesQuantum-2012, Georgescu14_RMP, Daley-Zoller-PracticalQuantumAdvantage-2022}.
As a prime example, ultracold atoms in optical lattices provide a pristine environment to simulate translationally invariant quantum systems given their high degree of control over model parameters, refined measurement techniques, long coherence times and scalability to large system sizes \cite{Bloch-UltracoldQuantumGases-2005, Bloch-Zwerger-ManybodyPhysicsUltracold-2008, Schafer-Takahashi-ToolsQuantumSimulation-2020, Gross-Bloch-QuantumSimulationsUltracold-2017}. As such, their facility to simulate elusive physical phenomena encompasses a vast range of applications including condensed matter models \cite{greiner_quantum_2002, struck_quantum_2011, simon_quantum_2011,Jotzu14_Nat, glaetzle_designing_2015, Mazurenko-Greiner-ColdatomFermiHubbard-2017, Asteria19_NatPhys, Wintersperger20_NatPhys,  Unal20_PRL,  Zhao22_CommPhys_EuQuench}, statistical physics~\cite{Deissler10_NatPhys,Schreiber-Bloch-ObservationManybodyLocalization-2015,Kondov-DeMarco-DisorderInducedLocalizationStrongly-2015, Choi-Gross-ExploringManybodyLocalization-2016,Yu23_arXiv_BoseGlass}, cosmology~\cite{Viermann-Oberthaler-QuantumFieldSimulator-2022, Song-Schneider-RealizingDiscontinuousQuantum-2022}, high-energy physics~\cite{Zohar-Reznik-QuantumSimulationsLattice-2015} and quantum chemistry \cite{Arguello-Luengo-Cirac-AnalogueQuantumChemistry-2019}.
In particular, Floquet driving in optical lattices has been established as a powerful tool for Hamiltonian engineering~\cite{eckardt_colloquium_2017,Cooper19_RMP} and has allowed for the realisation of novel phenomena inaccessible to static systems~\cite{Weitenberg-Simonet-TailoringQuantumGases-2021}. Rapid experimental developments have delivered the realisation of dynamical localisation~\cite{lignier_dynamical_2007, eckardt_exploring_2009, creffield_expansion_2010} and the simulation of artificial electric and magnetic fields, giving access to paradigmatic topological models~\cite{Aidelsburger-Bloch-ExperimentalRealizationStrong-2011, Aidelsburger-Bloch-RealizationHofstadterHamiltonian-2013, Miyake-Ketterle-RealizingHarperHamiltonian-2013,Jotzu14_Nat, Aidelsburger-Goldman-MeasuringChernNumber-2015, Tarnowski19_NatCom}, frustrated systems \cite{eckardt_frustrated_2010, struck_engineering_2013} and dynamical gauge fields \cite{clark_observation_2018,barbiero_coupling_2019, gorg_realization_2019,schweizer_floquet_2019}.

Traditionally, optical-lattice-based quantum simulators have concentrated on simulations of translationally-invariant systems such as those inspired by condensed matter~\cite{Bloch-UltracoldQuantumGases-2005} due to their lack of local control capabilities. However, rapid experimental progress is being made towards single-site control and manipulation of optical lattices. Quantum gas microscopes~\cite{bakr_quantum_2009, Sherson-Kuhr-SingleatomresolvedFluorescenceImaging-2010, GrossBakr21_NatPhys} provide access and control to the on-site energies of individual lattice sites, which has, for instance, enabled the preparation of low entropy states~\cite{Mazurenko-Greiner-ColdatomFermiHubbard-2017, GrossBakr21_NatPhys} and has allowed edge states in topological systems to be populated and probed~\cite{Martinez23_arXivWP,Braun23_arXivWP}. Despite this, progress in controlling individual on-site potentials and the ability to independently tune individual tunnelling links has remained a challenge.
At the same time, optical tweezers have been implemented to trap and control individual atoms and to arrange them in programmable patterns~\cite{Thompson-Lukin-CoherenceRamanSideband-2013, Lester-Regal-RapidProductionUniformly-2015, Bernien-Lukin-ProbingManybodyDynamics-2017, Barredo-Browaeys-SyntheticThreedimensionalAtomic-2018, Andersen-OpticalTweezersBottomup-2022, Trisnadi-Chin-DesignConstructionQuantum-2022}.

In this manuscript, we demonstrate how to achieve full local control over tunnelling elements by combining the renowned power of periodic driving with the local accessibility afforded by quantum gas microscopes and optical tweezers.
%that can be employed to independently create optical potentials on each lattice site. 
%coupling the Floquet physics of periodic modulation generated by an acousto-optic deflector (AOD), with the local accessibility afforded by a quantum gas microscope.  The quantum gas microscope can be used to create an optical-tweezer-like potential on an isolated lattice minima. 
In particular, we consider periodically modulating the on-site energy of individual lattice sites and employ Floquet-Bloch theory to study the effect of this drive within the tight-binding description.
%This translates into adjusting the on-site energy of a single lattice site in the tight-binding description of the system. We can periodically modulate the intensity of the potential to vary the on-site energy of a lattice site in time. 
% he local driving of on-site energy will renormalise local tight-binding model parameters in a controllable way. In one-dimension, shaking a single site renormalises the two neighbouring tunnelling elements \cite{eckardt_colloquium_2017}. 
In this way, any nearest-neighbour tunnelling element can be addressed by controlling the relative modulation between the two involved lattice sites~\cite{eckardt_colloquium_2017, Hubner-Sheikhan-MomentumresolvedFloquetengineeredPair-2023}.   Building on this, we demonstrate how driving individual lattice sites with different amplitudes can generate any desired sequence of tunnelling amplitudes, and show for example how to engineer the Su-Schrieffer-Heeger (SSH) and the extended SSH models. This full local control in 1D can be used to simulate physics beyond condensed matter such as the quantum properties of a black hole \cite{benhemou23_arxiv_blackhole}.

We then move towards two-dimensional (2D) systems and first show that we can engineer a fully programmable square lattice with real and positive tunnelling coefficients by driving specific sub-lattice sites of a Lieb lattice. Recognising the impact of complex tunnelling phases in 2D, we show that both the relative tunnelling amplitudes and the gauge-invariant plaquette flux can be fully and independently tuned in a three-site plaquette. We demonstrate how to engineer locally tunable flux in 2D to simulate e.g. a gradiented magnetic field. Our results illustrate the versatility of local periodic driving in controlling system parameters and investigating diverse quantum phenomena.

The manuscript is organised as follows. In Section \ref{Sec:FloquetEngineering}, we introduce the framework for analysing Floquet driving in tight-binding Hamiltonian models. In Section \ref{Sec:CapabilitiesInOneDimension}, we show how to independently control individual tunnelling links in one dimension using local driving and provide examples of realising topological one-dimensional (1D) systems. 
In Section \ref{Sec:ControlOver2DPlaquette}, we explore 2D settings, first showing that localised modulations can be used to engineer an effective, programmable 2D square lattice by considering a driven Lieb lattice. We investigate control over gauge-invariant fluxes in 2D, demonstrating full programmability over relative tunnelling elements and flux values piercing a triangular plaquette. We also implement local driving on a square lattice to achieve locally tunable flux values.   We discuss details for physical implementation of local driving in experiments and demonstrate its feasibility in Section \ref{Sec:PhysImplement}. We conclude with an outlook in Section \ref{Sec:Conclusion}.

%%%%%%%%%%%%%%%%%%%%%%%%%%%%%%%%%%%%%%%%%%%%%%%%%%%%%%%%%%%%%%%%%%%%%%%%%%%%%%%%%%%%%%%%%%%%%%
\section{Floquet Engineering}
\label{Sec:FloquetEngineering}
%\subsection{General Framework}
%\label{Sec:GeneralFramework}
We first introduce the notation and summarise the general framework for periodically driving optical lattices, which will be employed in subsequent sections to investigate local Floquet engineering in various scenarios. 
Our starting point is a tight-binding description for non-interacting particles with the Hamiltonian 
\begin{equation}
    H(t) = H_0 + V(t),
     \label{Eq:GeneralHt}
\end{equation}
where $H_0 = -J \sum_{\langle i, j \rangle } \hat{a}_{i}^{\dagger} \hat{a}_j $
and $V(t) =  \sum_j W_j (t) \hat{n}_j$. 
Here, $J$ is the bare tunnelling amplitude between nearest-neighbour sites $\langle i, j \rangle$ and  $\hat{a}_j^{\dagger}$ ($\hat{a}_j$) and $\hat{n}_j$ are the creation (annihilation) and number operators at site $j$. 
The time-independent term $H_0$ describes the tunnelling within a static optical lattice in the tight-binding regime, as routinely realised in experiments using quantum gas microscopes~\cite{GrossBakr21_NatPhys}.
To this, we add time-periodic, on-site potential terms $W_j (t)=W_j(t+T)$ with frequency $\omega=2\pi/T$, that can be created by focusing additional off-resonant laser beams (such as those used for optical tweezers) onto individual sites and modulating their intensity. The periodic nature of the modulation ensures $H(t)=H(t+T)$ and allows one to apply Floquet-Bloch theory~\cite{goldman_periodically_2014,eckardt_colloquium_2017} to analyse this driven system.

%There is a system period $T$ such that $w_j(t) = w_j(t+T)$ for all sites $j$. Equation (\ref{Eq:GeneralHt}) describes a general case where onsite-energy functions $w_j(t)$ may be unique to each site $j$. The sum $\sum_{ \langle i,j \rangle }$ is over nearest neighbours. 
In general, it is useful to work with a gauge-transformed Hamiltonian by going to the ``lattice" frame where onsite potential modulations are removed and the effect of driving is recast onto time-dependent Peierls phases~\cite{eckardt_colloquium_2017}. This restores the discrete translational symmetry of the lattice and separates the different energy scales allowing for high-frequency expansions~\cite{bukov_universal_2015,eckardt_high-frequency_2015,eckardt_colloquium_2017}. The Hamiltonian in the lattice frame takes the form,
\begin{equation}
\tilde{H}(t) = -J \sum_{\langle i,j \rangle } e^{i \phi_{ij}(t) } \hat{a}_{i}^{\dagger} \hat{a}_j,
\label{Eq:HTildeGeneral}
\end{equation}
with time-dependent Peierls phases
\begin{equation}
\phi_{ij}(t) =  \int_{t_0}^{t} \left[ W_{j} ( \tau ) - W_i ( \tau ) \right]  \drm  \tau +    \phi^0_{ij},
\label{Eq:TimeDependentPeierlsPhases}
\end{equation}
which describe effective vector potentials under minimal coupling (we set $\hbar=1$ in Eq.~(\ref{Eq:TimeDependentPeierlsPhases}) and hereafter). The constant $\phi_{ij}^0$ is a gauge choice that we choose to ensure the Peierls phases have a vanishing time average $\int_0^T\phi_{ij}(t) \mathrm{d} t  = 0 $~\cite{eckardt_colloquium_2017}. We emphasise that the engineered Peierls phase of a given tunnelling link between two neighbouring sites depends solely on the difference in the modulation between the two sites. Namely, if $W_i(t) = W_{j}(t)$, the Peierls phase vanishes $\phi_{ij}(t) = 0$ and we recover the bare tunnelling coefficient. 
% This has been employed in theoretical proposals to realise optical solenoids and charge pumps of integer or fractional values~\cite{wang_floquet_2018, raciunas_creating_2018}. 

Under periodic driving, the relevant quantity defining the spectrum is the time evolution operator,  
\begin{equation}
    \tilde{U}(t, t_0) = \mathcal{T} \mathrm{e}^{-i \int_{t_0}^{t}  \tilde{H} (\tau) \drm \tau },
    \label{Eq:TimeEvolutionOperatorGeneralTilde}
\end{equation}
where $\mathcal{T}$ indicates time ordering. While energy is not conserved for time-dependent $\tilde{H}(t)$, a {\it quasienergy} $\epsilon$ can be defined through the phases of the eigenvalues of the time evolution operator over one period, $\tilde{U}(t_0+T, t_0)|\psi(t_0)\rangle=e^{-i\epsilon T}|\psi(t_0)\rangle$. This gives rise to a stroboscopic description of the periodic system by
\begin{equation}
    \tilde{U}(t_0 + nT, t_0) = e^{-i nT H_S^{t_0}},
    \label{Eq:StroboscobicTimeEvolution}
\end{equation}
where $H_S^{t_0}$ is the time-independent ``stroboscopic Hamiltonian" and $n$ is an integer~\cite{holthaus_floquet_2016, eckardt_colloquium_2017}.
%The stroboscopic Hamiltonian is an important concept for the coherent control of time-periodically driven quantum systems.  The stroboscopic Hamiltonian is defined to reproduce the time evolution generated by the Hamiltonian over one driving cycle from $t_0$ to $t_0 + T$.
The gauge choice $t_0$ is dubbed the ``Floquet gauge" and there is a family of stroboscopic Hamiltonians $H_S^{t_0}$ parameterised by  $t_0 \in [0, T]$. These are all related by unitary transformations,  $H_S^{t } = P(t, t_0) H_S^{t_0} P^{\dagger} (t, t_0)$, where $P(t,t_0)$ defines the time-periodic ``micromotion'' operator describing time evolution in-between stroboscopic instances, $P(t, t_0) = \tilde{U}(t, t_0) e^{iH_S^{t_0}(t-t_0)}$. Hence, for a full period (stroboscopically), the effect of the micromotion reduces to the identity $P(t_0+T, t_0) = \mathbb{1}$. Although the stroboscopic Hamiltonian is generally difficult to characterise analytically, it can be calculated numerically (see Appendix~\ref{App:NumericalCalculationStrobHam}).
% Each exactly describes one-cycle time evolution from it's initial time as in Eq. (\ref{Eq:StroboscobicTimeEvolution}).  
% Although the stroboscopic Hamiltonian $H^{\mathrm{F}}_{t_0}$ depends on the initial time $t_0$ within a period due to the micromotion, its eigenvalues  $\epsilon$ are independent of this Floquet gauge. 

It is possible and convenient to work with a representation of the stroboscopic Hamiltonian that is expressly $t_0$-independent \cite{bukov_universal_2015}. This Floquet-gauge invariant Hamiltonian is labelled the ``effective Hamiltonian", $H_{\mathrm{eff}}$ and offers an efficient description for an experimental system where the express initial time of modulation may vary. 
The effective Hamiltonian is related to the stroboscopic Hamiltonian by a unitary transformation, $H_S^{t_0} = Q(t_0) H_{\mathrm{eff}} Q^{\dagger}(t_0)$, which relates to the micromotion operator through $P(t, t_0) = Q(t) Q^{\dagger}(t_0)$~\cite{eckardt_colloquium_2017}.
%The unitary operator $P(t, t_0) = Q(t) Q^{\dagger}(t_0) $ is the ``micromotion" operator describing time-evolution in-between stroboscopic instances. 
Accordingly, the generic time evolution including the micromotion for any given time interval can be expressed by using either the stroboscopic Hamiltonian or the effective Hamiltonian, $\tilde{U}(t_2, t_1) = P(t_2, t_0) e^{-iH_S^{t_0} (t_2 - t_1)} P^{\dagger} (t_1, t_0) = Q(t_2) e^{-iH_{\mathrm{eff}} (t_2 - t_1)} Q^{\dagger}(t_1)$. Both descriptions have played important roles in the characterisation of Floquet systems~\cite{goldman_periodically_2014,bukov_universal_2015,eckardt_colloquium_2017}.

One powerful technique in studying Floquet systems involves an infinite series expansion in the orders of $1/\omega$ to describe the effective Hamiltonian, $H_{\mathrm{eff}} = \sum_{n=0}^{\infty}  \frac{1}{\omega^n} H_{\mathrm{eff}}^{(n)}$~\cite{bukov_universal_2015}. 
For sufficiently large driving frequencies, the effective Hamiltonian can be well approximated by the first two terms,
\begin{align}
    H_{\mathrm{eff}}^{(0)} &= \tilde{H}_0 \label{Eq:HighFrequencyExpansionFirstTerm} ,\\
     H_{\mathrm{eff}}^{(1)} &= \sum_{m=1}^{\infty} \frac{1}{m} [\tilde{H}_m, \tilde{H}_{-m}],
     \label{Eq:HighFreqencyExpansionSecondTerm}
\end{align}
where $\tilde{H}_m = \frac{1}{T} \int_0^T e^{-i m \omega t} \tilde{H}(t) \, \dd t$ are the Fourier harmonics of the periodic time-dependent Hamiltonian. In the following, we consider both the low-order series expansion of the effective Hamiltonian to analytically assess Floquet-gauge independent behaviour as well as the numerically calculated stroboscopic Hamiltonian to characterise the full system including higher-order effects in our investigations of local periodic driving.
%Similarly, a high-frequency expansion can be employed for the stroboscopic Hamiltonian, see e.g.~Ref.~\cite{bukov_universal_2015}.

%%%%%%%%%%%%%%%%%%%%%%%%%%%%%%%%%%%%%%%%%%%%%%%%%%%%%%%%%%%%%%%%%%%%%%%%%%%%%%%%%%%%%%%%%%%%%%
\section{Local modulations in a 1D chain}
\label{Sec:CapabilitiesInOneDimension}

\subsection{Driving a single site}
\label{SSec:DrivingASingleSite}
\begin{figure}
	\centering\hspace{-3.5mm}
 \includegraphics{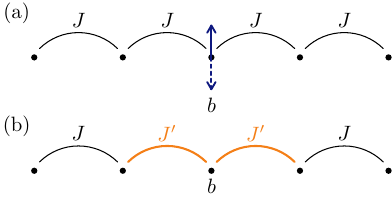}
	\caption{ (a) Sinusoidal modulation of the onsite energy of a single lattice site $b$, with a form $A \cos (\omega t)$. (b) In the effective Hamiltonian given by Eqs.~(\ref{Eq:EffectiveGFramework})-(\ref{Eq:SingleSiteShakeRenormalisedTunnellings}), the tunnelling coefficients on either side of the driven site are renormalised to $J'=J\mathcal{J}_0(A/\omega)$.}
	\label{Fig:SingleShakeDiagram}
\end{figure}

As a key building block of the proposed scheme, we now describe how sinusoidally modulating the onsite energy of a single site in a one-dimensional lattice affects the effective Hamiltonian by employing the Floquet-Bloch theory introduced in Sec.~\ref{Sec:FloquetEngineering}. When driving a single site labelled $b$ as illustrated in Fig.~\ref{Fig:SingleShakeDiagram}, the corresponding time-dependent Hamiltonian is given by Eq.~(\ref{Eq:GeneralHt}) with $V(t) = A \cos(\omega t) \, \hat{n}_b$ where $A$ is the driving amplitude and $\omega$ is the frequency of modulation. 

Using the first term of the high-frequency expansion given by Eq.~(\ref{Eq:HighFrequencyExpansionFirstTerm}), the effective  Hamiltonian becomes
\begin{equation}
    H_{\mathrm{eff}}^{(0)} = -J \sum_{\langle i,j \rangle} \epsilon_{ij} \hat{a}_i^{\dagger} \hat{a}_j,
    \label{Eq:EffectiveGFramework}
\end{equation}
where the renormalisation of  the tunnelling amplitudes is given by
\begin{equation}
\epsilon_{ij} = 
\begin{cases}
 \mathcal{J}_0 \left( \frac{A}{ \omega} \right),   & \text{if $b \in \{i,j\} $},\\
1, & \text{otherwise}. \\
\end{cases} 
\label{Eq:SingleSiteShakeRenormalisedTunnellings}
\end{equation}
Here, $\mathcal{J}_0$ is the Bessel function of first kind \cite{eckardt_colloquium_2017}. Crucially, only the tunnelling to and from the modulated site $b$ is renormalised.  We note that the first-order term given by Eq.~(\ref{Eq:HighFreqencyExpansionSecondTerm}) vanishes in all of the one-dimensional models we study in Sec.~\ref{Sec:CapabilitiesInOneDimension} (see Appendix~\ref{App:VanishingSecondOrderHFExpansionTerm} for details). Therefore, leading-order corrections to the effective Hamiltonian in Eq.~(\ref{Eq:EffectiveGFramework}) scale as $1/\omega^2$.

\begin{figure}
	\centering\hspace{-3.5mm}
 \includegraphics{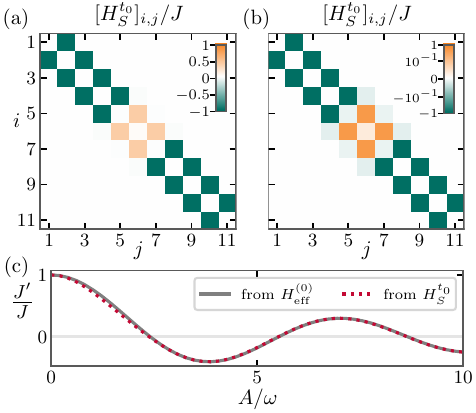}
	\caption{(a) Numerically simulated matrix elements of the stroboscopic Hamiltonian $H_S^{t_0}$ for the sinusoidal modulation of a single site ($b=5$) as in Fig.~\ref{Fig:SingleShakeDiagram}, here shown for $t_0=0$, $A=90$ and $\omega=25$ expressed in units of $J$. The nearest-neighbour tunnelling amplitudes around the driven site are renormalised (orange). (b) Same as (a) but using a log scale colourbar, showing that the stroboscopic Hamiltonian includes non-zero but very small longer range tunnelling coefficients. (c) The renormalised nearest-neighbour tunnelling amplitude $J'$ calculated numerically from $[H_S^{t_0}]_{b,b+1}$ (red dots) as a function of $A/\omega$ for $\omega=10$. This is well captured by a Bessel function $\mathcal{J}_0(A/\omega)$ (grey line) as predicted by the first term of the effective Hamiltonian, see Eqs.~(\ref{Eq:EffectiveGFramework})-(\ref{Eq:SingleSiteShakeRenormalisedTunnellings}). } 
	\label{Fig:SingleSiteShakeStroboscopicHamiltonian}
\end{figure}

Figure~\ref{Fig:SingleSiteShakeStroboscopicHamiltonian}(a) shows the matrix elements of the numerically calculated stroboscopic Hamiltonian $H_S^{t_0}$ (see Appendix~\ref{App:NumericalCalculationStrobHam}) when site $b=5$ is driven with frequency $\omega=25$, amplitude $A=90$ and when considering $t_0=0$. We present all energies in units of the bare tunnelling amplitude $J$ in this paper. We see that the tunnelling elements on either side of the driven site have been renormalised. The effects of higher-order terms are captured by the stroboscopic Hamiltonian and are strongly suppressed by the scaling with $1/\omega^2$~\cite{eckardt_high-frequency_2015,bukov_universal_2015}. They can only be seen when $H_{S}^{t_0}$ is displayed on a log scale as in Fig.~\ref{Fig:SingleSiteShakeStroboscopicHamiltonian}(b) and controlled by tuning the driving parameters. In Fig.~\ref{Fig:SingleSiteShakeStroboscopicHamiltonian}(c), we display the renormalised nearest-neighbour tunnelling amplitude given by $H_S^{t_0}$ as a function of driving amplitude (red dots), which agrees  perfectly with the zeroth-order Bessel function (grey line) predicted by the effective Hamiltonian in Eqs.~(\ref{Eq:EffectiveGFramework})-(\ref{Eq:SingleSiteShakeRenormalisedTunnellings}) to a high level of accuracy. %Figure \ref{Fig:SingleSiteShakeStroboscopicHamiltonian}(c) confirms that the zeroth-order Bessel function gives a good description of the renormalised nearest-neighbour tunnelling. 

In this Floquet gauge, all tunnelling coefficients are real. Although the Bessel function in Eq.~(\ref{Eq:SingleSiteShakeRenormalisedTunnellings}) can become negative, we point out that negative (or complex) tunnelling amplitudes in a one-dimensional tight-binding model with only nearest-neighbour tunnelling are physically indistinguishable from positive ones of the same magnitude, as the phase is gauge dependent and can be absorbed into the definition of the Wannier functions~\cite{Vanderbilt01_PRB_wannier}. %In general, complex phases of tunnelling amplitudes can be gauged away in 1D. 
We note that in cold atom experiments, time-of-flight measurements take the system outside of the tight-binding description and enable one to reveal the associated difference between canonical and mechanical quasi-momentum even in 1D~\cite{struck_tunable_2012}.
Crucially, the sign or complex phase of a tunnelling amplitude cannot be eliminated by an appropriate gauge choice whenever closed loops are possible e.g.~due to long-range tunnelling, new effective dimensions, or 2D lattices~\cite{Wang21_NJP}. We explore controlling the resulting physically significant flux values arising from complex tunnelling phases in 2D in Sec.~\ref{Sec:ControlOver2DPlaquette}.

Recent experimental advances allowing for independent, single-site accessibility in optical lattices \cite{Trisnadi-Chin-DesignConstructionQuantum-2022, Young-Kaufman-Tweezerprogrammable2DQuantum-2022} open up a plethora of possibilities for local periodic modulation. In the following, we present various configurations for such local sinusoidal driving using different amplitudes and relative phases of modulation to simulate and explore various lattice Hamiltonians.

%%%%%%%%%%%%%%%%%%%%%%%%%%%%%%%%%%%%%%%%%%%%%%%%%%%%%%%%%%%%%%%%%%%%%%%%%%%%%%%%%%%%%%%%%%%%%%

\subsection{Modifying a single tunnelling link}
\label{Sec:ModifyingSingleLink}
\begin{figure}
   \includegraphics{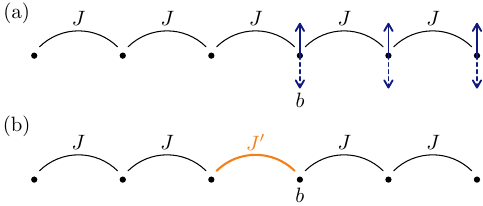}
 \caption{ Controlling a single tunnelling link. (a) All sites $j \geq b$ marked with purple arrows have their onsite energies periodically modulated via the same function $A \cos (\omega t)$ in Eq.~(\ref{Eq:PotentialTermModifySingleTunnellingElement}). (b) The relative modulation between neighbouring sites is only non-zero between sites $b-1$ and $b$ and the resulting effective Hamiltonian has one tunnelling amplitude renormalised to $J'= J \mathcal{J}_0 \left( \frac{A}{\omega} \right)$. }
\label{Fig:SingleTunnellingModification}
\end{figure}

While the modulation in Sec.~\ref{SSec:DrivingASingleSite} leads to the renormalisation of a pair of tunnelling amplitudes, here we elucidate how to modify a single link only. This can be achieved by driving all sites to one side of the target link as proposed in Ref.~\cite{wang_floquet_2018} and illustrated in Fig.~\ref{Fig:SingleTunnellingModification}(a). One can employ a driving potential acting on all lattice sites $j>b$ with the same form,
\begin{equation}
    V(t) = A \cos(\omega t ) \sum_{j > b} \hat{n}_j,
    \label{Eq:PotentialTermModifySingleTunnellingElement}
\end{equation} 
in the Hamiltonian given by Eq.~(\ref{Eq:GeneralHt}).
%Central to this approach is the idea that it is only relative modulation between neighbouring sites that engineers non-zero, time-dependent Peierls phases in Eq. (\ref{Eq:TimeDependentPeierlsPhases}). Subsequently, only non-trivial Peierls phases generate renormalised tunnelling elements in the effective Hamiltonian that deviate from $H_0$, as shown in Eq. (\ref{Eq:CycleAverage}).  By shaking all sites to one side of a chosen tunnelling link, the relative modulation between neighbouring sites is only non-zero across this chosen link. This ensures the subsequent effective Floquet Hamiltonian in Eq. (\ref{Eq:CycleAverage}) has only one renormalised tunnelling link. 
Accordingly, the first term of the high-frequency expansion is given by Eq.~(\ref{Eq:EffectiveGFramework}) with the renormalisation factor
\begin{equation}
\epsilon_{ij} =  
\begin{cases}
 \mathcal{J}_0 \left( \frac{A}{ \omega} \right),   & \text{if } \{i,j\}=\{b, b+1\},\\
1, & \text{otherwise}. \\
\end{cases} 
\label{Eq:SingleTunnellingModEffectiveHamiltonian}
\end{equation}
Namely, only a single tunnelling link between site $b$ and $b+1$ is modified by the Bessel function as shown in Fig.~\ref{Fig:SingleTunnellingModification}(b). %The the second order term of the high-frequency expansion vanishes as is shown in Appendix~\ref{App:VanishingSecondOrderHFExpansionTerm}.  

%%%%%%%%%%%%%%%%%%%%%%%%%%%%%%%%%%%%%%%%%%%%%%%%%%%%%%%%%%%%%%%%%%%%%%%%%%%%%%%%%%%%%%%%%%%%%%
\subsection{Driving with different amplitudes to achieve full control}
\label{SSec:DrivingWithDifferentAmplitudes}
We now consider sinusoidally driving all sites with the same frequency but allow for different driving amplitudes on each site. This allows for a fully programmable one-dimensional chain of tunnelling elements and we use this description to substantiate applications in subsequent sections. The time-dependent, on-site potential term is given by
%We now present a formalism for treating a sinusoidally driven lattice in the general case \note{with one global frequency.} We will use this description to substantiate examples and applications in Sec. \ref{Sec:SuSchriefferHeegerModel} and \ref{Sec:SSH3}. \note{When all sites may be driven,} the time-dependent potential term in Eq. (\ref{Eq:GeneralHt}) can be written as
\begin{equation}
V(t) = \cos (\omega t) \sum_j A_j \hat{n}_j,
\label{Eq:PotentialTermSinusoidalDrivingOneDimension}
\end{equation}
where now each site $j$  can have a different driving amplitude $A_j$.
%The values $A_j$ give driving amplitudes at each site $j$ and $\omega$ is the ubiquitous driving frequency.  
The Hamiltonian in the lattice frame can be written as
\begin{equation}
    \tilde{H}(t) = -J \sum_{\langle i,j \rangle} e^{i\frac{A_i - A_j}{ \omega} \sin (\omega t)} \hat{a}_i^{\dagger} \hat{a}_j.
\end{equation}
 The Fourier harmonics of this system are $
\tilde{H}_m = -J \sum_{\langle i,j \rangle} \hat{a}_i^{\dagger} \hat{a}_j \mathcal{J}_m \left( \frac{A_i - A_j}{\omega} \right)
$, where $\mathcal{J}_m$ is the Bessel function of the $m^{\mathrm{th}}$ kind. Since the pairs of Fourier components ($H_{m}, H_{-m}$) in Eq.~(\ref{Eq:HighFreqencyExpansionSecondTerm}) commute with each other, the second order term of the high-frequency expansion again vanishes (see Appendix~\ref{App:VanishingSecondOrderHFExpansionTerm}).
The effective Hamiltonian is well approximated by the first term of the high-frequency expansion in Eq.~(\ref{Eq:HighFrequencyExpansionFirstTerm}),
%of the effective Hamiltonian given by Eq. (\ref{Eq:HighFreqencyExpansionSecondTerm}) is zero as inverse Fourier components commute. The effective Hamiltonian is therefore well approximated by the first term of the high-frequency expansion as in Eq. (\ref{Eq:HighFrequencyExpansionFirstTerm}) giving 
\begin{equation}
\tilde{H}_{\mathrm{eff}}^{(0)} = -J \sum_{\langle ij \rangle} \hat{a}_i^{\dagger} \hat{a}_j
 \mathcal{J}_0 \left( \frac{|A_i - A_j|}{ \omega} \right).
 \label{Eq:FirstOrderHFTermOneDSinusoidalDrive}
 \end{equation}
Crucially, Eq.~(\ref{Eq:FirstOrderHFTermOneDSinusoidalDrive}) highlights how only the relative modulation between neighbouring sites determines the renormalisation of tunnelling elements.
%Crucially, Eq. (\ref{Eq:FirstOrderHFTermOneDSinusoidalDrive}) demonstrates that non-zero relative motion between neighbouring sites is essential to renormalising the effective Hamiltonian tunnelling elements. When $A_j - A_{j+1} = 0$, the Bessel function is unity and we regain bare tunnelling $J$ between sites $j$ and $j+1$. 
By controlling the individual driving amplitudes for each lattice site, it is possible to tune each tunnelling link individually, providing a fully programmable one-dimensional chain.
In Sections~\ref{Sec:SuSchriefferHeegerModel} and~\ref{Sec:SSH3}, we use this feature to, as an example, reverse-engineer a sequence of driving amplitudes that realise several important models featuring topologically nontrivial phenomena.

\begin{figure} 
	\centering\hspace{-3.5mm}
 \includegraphics{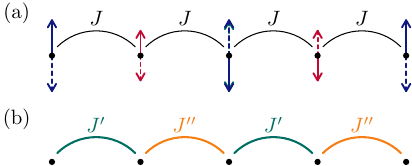}
	\caption{ (a) The SSH model can be generated by locally driving the on-site energies of a one-dimensional chain sinusoidally with the same frequency $\omega$ and amplitudes alternating with a four-site sequence given by Eq.~(\ref{Eq:SSHModelShakeFunctionProtocol}). Purple (red) arrows illustrate the site is driven with the amplitude $C_1 (C_2)$, however the sign of the drive is inverted every second site moving down the chain, corresponding to a $\pi$ phase shift. (b) This driving pattern results in an effective Hamiltonian with alternating tunnelling coefficients $J' = J \mathcal{J}_0 \left(\nicefrac{|C_1 - C_2|}{ \omega}\right)$ and $J'' = J \mathcal{J}_0 \left(\nicefrac{|C_1 + C_2|}{ \omega}\right)$.}  
	\label{Fig:SSHDiagram}
\end{figure}

%%%%%%%%%%%%%%%%%%%%%%%%%%%%%%%%%%%%%%%%%%%%%%%%%%%%%%%%%%%%%%%%%%%%%%%%%%%%%%%%%%%%%%%%%%%%%%
\subsection{Su-Schrieffer-Heeger Model} \label{Sec:SuSchriefferHeegerModel}
The Su-Schrieffer-Heeger (SSH) model~\cite{su_solitons_1979} describes non-interacting particles in a 1D lattice with alternating tunnelling coefficients. As a paradigmatic example of a topologically non-trivial system in one dimension~\cite{ryu_topological_2010}, it has attracted attention both from theory and experiment. The SSH model and its generalisation to the Rice-Mele model have also been realised in optical lattices~\cite{atala13_NatPhys, nakajima_topological_2016, lohse_thouless_2016, deLeseleuc-Browaeys-ObservationSymmetryprotectedTopological-2019}.  The SSH model can be engineered in a one-dimensional optical lattice using a local periodic driving scheme by employing 
two distinct driving amplitudes ($C_1, C_2$) combined with two different phases of the modulation. In particular, we consider on-site potential energy terms given by Eq.~(\ref{Eq:PotentialTermSinusoidalDrivingOneDimension}) with the form
\begin{equation}
    A_j = 
\begin{cases}
C_1,  & \text{if j = 0 (mod 4)},\\
C_2,  & \text{if j = 1 (mod 4)}, \\
-C_1,  & \text{if j = 2 (mod 4)},\\
-C_2, & \text{if j = 3 (mod 4)}, 
\end{cases} 
\label{Eq:SSHModelShakeFunctionProtocol}
\end{equation}
as illustrated in Fig.~\ref{Fig:SSHDiagram}(a). This results in 
\begin{equation}
|A_j(t) - A_{j+1}| = 
\begin{cases}
 | C_1 - C_2  |, & \text{if j = 0 (mod 2)},\\
| C_1 + C_2 |, & \text{if j = 1 (mod 2)}, \\
\end{cases} 
\label{Eq:SSHModelRelativeShakeCases}
\end{equation}
where the relative driving strength between neighbouring sites alternates between two different values. The effective Hamiltonian given by Eq.~(\ref{Eq:FirstOrderHFTermOneDSinusoidalDrive}) then becomes
\begin{equation}
    H_{\mathrm{eff}}^{(0)} = -J \sum_{j} \epsilon_{j} \left( \hat{a}_j^{\dagger} \hat{a}_{j+1} + \hat{a}_{j+1}^{\dagger} \hat{a}_{j} \right)  ,
    \label{Eq:SSHEffectiveHamiltonian}
\end{equation}
where the effective tunnelling coefficients are renormalised in the desired alternating pattern (see Fig.~\ref{Fig:SSHDiagram}(b)),
\begin{equation}
\epsilon_{j} =  
\begin{cases}
 \mathcal{J}_0 \left( \frac{|C_1 - C_2|}{ \omega} \right),   & \text{if j = 0 (mod 2)},\\
 \mathcal{J}_0 \left( \frac{|C_1 + C_2|}{ \omega} \right),  & \text{if j = 1 (mod 2)}. \\
\end{cases}  \label{Eq:SSHEffectivetunnellings}
\end{equation}

\begin{figure}
        \centering\hspace{-3.5mm}
        \includegraphics{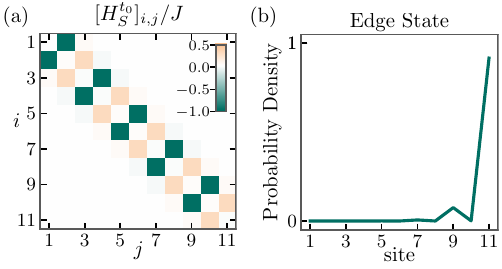}
        \label{Fig:SSHModel}
    \caption{Engineering the SSH model using local periodic driving. (a) The matrix elements of the numerically generated stroboscopic Hamiltonian $H_S^{t_0}$ for $t_0=0$ and where sites are driven as in Eq.~(\ref{Eq:SSHModelShakeFunctionProtocol}) with $C_1 = 29$, $C_2 = 13$ and $\omega=25$. The off-diagonal matrix elements correspond to renormalised tunnelling coefficients and alternate between two values. (b) The probability distribution of the zero-energy eigenstate is exponentially localised at one edge of the chain.}
    \label{Fig:SSHModelStroboscopic}
\end{figure}

To verify the effective Hamiltonian with alternating tunnelling amplitudes, we numerically calculate the stroboscopic Hamiltonian for a small chain which captures the Floquet evolution exactly.
Figure~\ref{Fig:SSHModelStroboscopic}(a) depicts the matrix elements of the stroboscopic Hamiltonian $H_S^{t_0}$ with alternating signs of the engineered tunnelling coefficients. 
% We demonstrate the matrix elements of the stroboscopic Hamiltonian $H_S^{t_0}$ in Fig.~\ref{Fig:SSHModelStroboscopic}(a), which captures the Floquet evolution exactly and illustrates the alternating sign of the engineered tunnelling coefficients clearly
Depending on the edge termination, the SSH model supports zero-energy modes localised at the boundary that are protected by a non-trivial Zak phase~\cite{su_solitons_1979,Cooper19_RMP}. Indeed, this can be seen in the Floquet-engineered Hamiltonian. We show the probability density of the zero-energy eigenstate of the stroboscopic Hamiltonian in Fig.\ref{Fig:SSHModelStroboscopic}(b), which is localised at the edge of the chain and decays exponentially into the bulk, as expected. 

The SSH model can also be engineered using the period-4 sequence of driving amplitudes $\{ C_1, C_1 + C_2, C_2, 0\}$ which ensures all sites are driven in phase. This may simplify experimental implementation (see Section~\ref{Sec:PhysImplement}), however would require larger absolute driving amplitudes to achieve the same SSH model compared to Eq.~(\ref{Eq:SSHModelShakeFunctionProtocol}).

%%%%%%%%%%%%%%%%%%%%%%%%%%%%%%%%%%%%%%%%%%%%%%%%%%%%%%%%%%%%%%%%%
\subsection{Extended Su-Schrieffer-Heeger models}
\label{Sec:SSH3}
Our local driving scheme to realise the SSH model can be easily extended to models with larger unit cells, where the extended Hilbert space is higher-dimensional and can give access to larger winding numbers \cite{Maffei-Massignan-TopologicalCharacterizationChiral-2018}. We label the extended SSH models where tunnelling coefficients cyclically alternate within a set of three or four different values, respectively, the SSH3 and SSH4 models~\cite{Xie-Yan-TopologicalCharacterizationsExtended-2019}. A conventional approach to realise such extended SSH models would require additional lasers with hard-to-achieve wavelengths to induce the extended spatial periodicity. These extensions, however, are straightforward with our local driving scheme.

\begin{figure}
    \centering\hspace{-3.5mm}
    \includegraphics{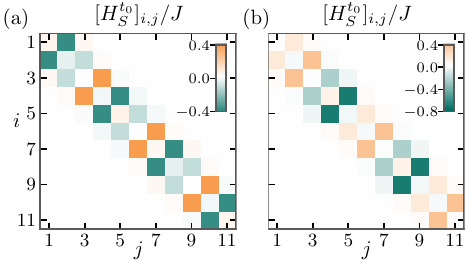} 
    \caption{ Numerically calculated stroboscopic Hamiltonians $H_S^{t_0}$ for $\omega=25$ and $t_0=0$ for extended SSH models. (a) The SSH3 model is generated by driving lattice sites with amplitudes as in Eq.~(\ref{Eq:SSH3ModelShakeFunctionProtocol}) with $C_1=-40$, $C_2=2.5$, $C_3=-53$. The resulting stroboscopic Hamiltonian contains renormalised tunnelling elements that follow a period-three sequence, well-captured by the effective description in Eq.~(\ref{Eq:SSH3EffectiveHamTunnellings}). (b) The SSH4 model is realised by employing driving amplitudes $C_1=17$, $C_2=-50$, $C_3=38$, $C_4=-10$. }
    \label{Fig:SSH34ModelStroboscopic}
\end{figure}

To generate the SSH3 model, expanding on Eq.~(\ref{Eq:SSHModelShakeFunctionProtocol}),
we employ three distinct driving amplitudes. Namely, we modulate the on-site energies given by Eq.~(\ref{Eq:PotentialTermSinusoidalDrivingOneDimension}) with the local driving amplitudes chosen as,
\begin{equation}
    A_j = 
\begin{cases}
C_1,  & \text{if j = 0 (mod 6)},\\
C_2,  & \text{if j = 1 (mod 6)}, \\
C_3,  & \text{if j = 2 (mod 6)},\\
-C_1,& \text{if j = 3 (mod 6)},\\
-C_2,& \text{if j = 4 (mod 6)}, \\
-C_3,& \text{if j = 5 (mod 6)}. 
\end{cases} 
\label{Eq:SSH3ModelShakeFunctionProtocol}
\end{equation}

The relative modulation between neighbouring sites is hence,
\begin{equation}
|A_j(t) - A_{j+1}| = 
\begin{cases}
 | C_1 - C_2  |, & \text{if j = 0 (mod 3)},\\
| C_2 - C_3 |, & \text{if j = 1 (mod 3)}, \\
| C_1 + C_3 |, & \text{if j = 2 (mod 3)}, \\
\end{cases} 
\label{Eq:SSH3ModelRelativeShakeCases}
\end{equation}
which results in the effective Hamiltonian given by Eq.~(\ref{Eq:SSHEffectiveHamiltonian}) with the tunnelling coefficients renormalised as
\begin{equation}
\epsilon_{j} = \begin{cases}
 \mathcal{J}_0 \left( \frac{|C_1 - C_2|}{ \omega} \right),   & \text{if j= 0 (mod 3)},\\
 \mathcal{J}_0 \left( \frac{|C_1 - C_2|}{ \omega} \right),  & \text{if j= 1 (mod 3)}, \\
  \mathcal{J}_0 \left( \frac{|C_1 + C_3|}{ \omega} \right),  & \text{if j= 2 (mod 3)}. \\
\end{cases}
  \label{Eq:SSH3EffectiveHamTunnellings}
\end{equation}
Fig.~\ref{Fig:SSH34ModelStroboscopic}(a) shows the numerically calculated stroboscopic Hamiltonian for the SSH3 model, verifying that tunnelling coefficients alternate between three different values.

In general, our scheme consists of a sequence of local driving amplitudes with a spatial period that is double the desired number of effective tunnelling coefficients. %i.e.~to generate three independently-tunable, renormalised tunnelling coefficients as in the SSH3 model,site period.
% a driving-amplitude set with a period of six for the SSH3 model ensures to generate three independently-tunable renormalised tunnelling coefficients.
For instance, the SSH4 model can be realised by a period-eight sequence of local driving amplitudes as $\{C_1, C_2, C_3, C_4, -C_1, -C_2, -C_3, -C_4\}$, which results in the stroboscopic Hamiltonian displayed in Fig.~\ref{Fig:SSH34ModelStroboscopic}(b).  

We emphasise that the above extended SSH models, which give access to higher winding numbers, are merely examples of the possibilities. The full individual control of local tunnelling terms for instance enables one to study strained 1D materials where the effect of strain can be tuned spatially, or to encode curved space-times to investigate the quantum dynamics around black holes~\cite{benhemou23_arxiv_blackhole}.

%%%%%%%%%%%%%%%%%%%%%%%%%%%%%%%%%%%%%%%%%%%%%%%%%%%%%%%%%%%%%%%%%
\section{Control over a two-dimensional system}
\label{Sec:ControlOver2DPlaquette}

We now extend our discussion of local periodic driving to two dimensions, where this technique opens up many possibilities for engineering locally defined Hamiltonians.
As an exposition, we study examples of 2D lattices generating novel Hamiltonians unobtainable by other means. In subsection \ref{SubSec:LiebLattice}, we demonstrate how locally driving a Lieb lattice allows us to simulate a fully programmable 2D  network with controllable connectivity. %, which can pose interesting connections beyond optical lattice systems. 
In subsection~\ref{Sec:TriangleFullControl}, we consider a triangular plaquette as a building block of a 2D lattice and expose full control on not only relative tunneling amplitudes but also on the plaquette flux.
In subsection \ref{SubSec:LocalFlux}, we extend our consideration to a larger square lattice geometry, utilising laser-assisted tunnelling to  engineer locally tunable fluxes in 2D. Outside of condensed matter, local driving can be utilised to simulate Black hole physics in 2D \cite{benhemou23_arxiv_blackhole}.

\subsection{Lieb lattice}
\label{SubSec:LiebLattice}
\begin{figure}
\centering\hspace{-3.5mm}
    \includegraphics{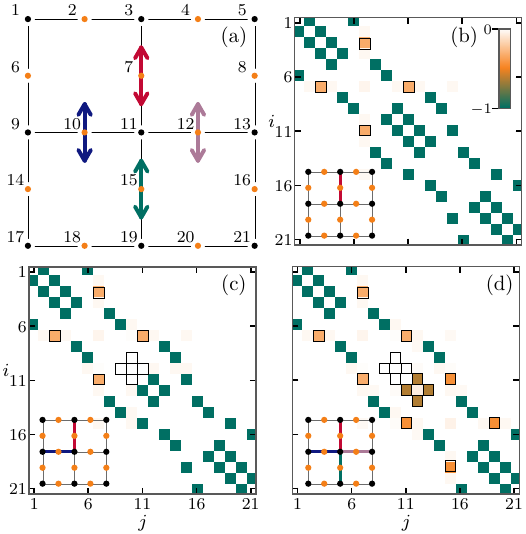}
    \caption{(a) The Lieb lattice consists of three sublattices: The $A$ sites (black dots) form a square lattice with the $B$ and $C$ sites (orange dots) sitting on the edges of the squares. By driving the $(B,C)$ sites with local amplitudes depicted by the coloured arrows, we can control the effective $2^{\mathrm{nd}}$ order couplings between the $A$ sites to create an effective square lattice with local tunability. (b) The matrix elements of the numerically calculated stroboscopic Hamiltonian normalised by the bare tunnelling $\nicefrac{H_{S}^{t_0}}{J}$ when only site $7$ of the Lieb lattice in (a) is driven with $A_7 = 72$ and $\omega=40$. Nearest neighbour tunnelling matrix elements to site $7$ (accentuated by black squares) have been renormalised by $\epsilon_7 = 0.34$ in Eq.~\ref{Eq:LiebEffectiveHam} while all other tunnelling elements remain fixed at $-J$. The inset shows the effective model where only two tunnelling links have been renormalised. (c) Additionally driving site $10$. We utilise zeros of the Bessel function to control the connectivity in sublattice $A$; by choosing $A_{10} = 96$, given $\mathcal{J}_0 \left( \nicefrac{86}{40} \right) = 0$, we turn off links between sites $9$ and $11$. (d) Additionally, sites $12$ and $15$ are driven to independently tune their respective tunnelling amplitudes.} 
    \label{Fig:LiebLattice}
\end{figure}

The Lieb lattice is a decorated square lattice with three sites per unit cell: one of them forms a standard square lattice which we label $A$ (black dots in Fig.~\ref{Fig:LiebLattice}(a)), and the other two lie on each side of the square which we label sublattices $B$ and $C$ (orange dots).
By driving each of the $(B,C)$ sites locally, we can individually control all effective links between $A$ sites.
%The system can then be thought of as a 2D square lattice on sublattice $A$ with locally tunable links.

The time-dependent system can be described by Eq.~\eqref{Eq:GeneralHt} with $V(t) = \sum_{j \in (B,C)} A_j \cos(\omega t) \hat{n}_j$ where $A_j$ gives the driving amplitude at site $j \in (B,C)$. Similarly to section \ref{SSec:DrivingASingleSite}, the resulting Hamiltonian can be expressed as 
\begin{equation}
    H_{\mathrm{eff}}^{(0)} = -J  \sum_{\substack{j \in (B,C)  \\ \langle i,j\rangle }} \epsilon_j \left[ \hat{a}_i^{\dagger} \hat{a}_j+ \hat{a}_j^{\dagger} \hat{a}_i \right],
    \label{Eq:LiebEffectiveHam}
\end{equation}
with tunnelling renormalisation factors $\epsilon_j = \mathcal{J}_0 \left( \frac{A_j}{\omega} \right)$ individually tunable by the local driving amplitudes $A_j$.  Figure \ref{Fig:LiebLattice}(a) depicts an example of this implementation where four sites surrounding a given $A$ site are driven independently and Fig.~\ref{Fig:LiebLattice}(b)-(d) show the resulting stroboscopic Hamiltonian.  
When considering only the square lattice on sublattice $A$, the renormalised system represents a new 2D square network with tunnellings that are independently tunable. For instance, by utilising the zero crossings of the Bessel function, we can control the connectivity in the square lattice by tuning tunnelling elements to zero as shown in Fig.~\ref{Fig:LiebLattice}(c). This capability can be used to create an effective square 2D lattice with bespoke connectivity that enables novel applications in quantum simulations, such as interesting dimensional crossovers between 1D and 2D, and in e.g.\ quantum annealing. We note that second-order tunnellings between neighbouring $A$ sites will not acquire any complex phase given the two first-order tunnelling components are complex conjugates. 
% By utilising the optical lattice in the Mott-regime to simulate spin-Hamiltonians (see Sec.~\ref{Sec:PhysImplement}), this set-up can be used to simulate spatially anisotropic systems such as spin-glasses \cite{Edwards-Anderson-TheorySpinGlasses-1975}.

\subsection{ Full control of a triangular plaquette}
\label{Sec:TriangleFullControl}
In a 1D nearest-neighbour tight-binding model, complex phases of tunnelling amplitudes can always be gauged away and therefore do not contribute to the dynamics of the system. In 2D, in contrast, complex tunnelling amplitudes around a plaquette give rise to a finite, gauge-invariant~\cite{Wang21_NJP} Aharonov-Bohn flux piercing the plaquette through the Peierls substitution~\cite{Peierls33}. This lies at the heart of the celebrated Harper-Hofstadter model describing the quantum Hall effect~\cite{Miyake-Ketterle-RealizingHarperHamiltonian-2013,Aidelsburger-Bloch-RealizationHofstadterHamiltonian-2013}.
While global periodic driving is a well-established technique in optical lattices to create homogeneous artificial gauge fields and various topologically nontrivial phenomena~\cite{OkaAoki09_PRB,Miyake-Ketterle-RealizingHarperHamiltonian-2013,Aidelsburger-Goldman-MeasuringChernNumber-2015,Jotzu14_Nat, Tarnowski19_NatCom,Asteria19_NatPhys, Wintersperger20_NatPhys},  full local control has remained an open question. 
As a starting point in this direction, we show that local periodic driving of individual lattice sites enables full control over both (i) the relative renormalised tunnelling amplitudes and (ii) the total flux piercing the plaquette.

%The full control over individual complex tunnelling coefficients across the whole range in a 2D system, however, remains an open question. 
%and tunability of both flux and tunneling coefficients in a 2D system has been so far unachievable in optical lattice experiments. 

\begin{figure}
\centering
    \includegraphics{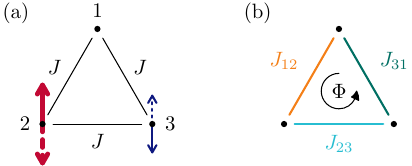}
 \caption{ (a) We modulate two of the three sites in a triangular plaquette with independent sinusoidal functions (arrows) as in Eq.~(\ref{Eq:TriangleTimeDependentHamiltonian}) that can feature different amplitudes, frequencies %(depicted by the different arrow lengths), frequencies  %(different arrow thicknesses) 
 and a relative phase $\varphi$ at $t=0$. The driving frequencies on the two sites are related by a rational number $\alpha/\beta$ ensuring the presence of a global period. (b) The time-independent effective Hamiltonian harbours renormalised tunnelling coefficients $J_{12}$, $J_{23}$, $J_{31}$. In the chosen gauge, $J_{12},J_{31}$ are real, and $J_{23}$ can be complex whenever $\varphi \notin{0,\pi}$, giving rise to an effective flux $\Phi$ piercing the plaquette.}
 \label{Fig:Triangle}
\end{figure}

To demonstrate this, we independently modulate the on-site energies of two sites in a triangular plaquette (see Fig.~\ref{Fig:Triangle}), with the time-dependent Hamiltonian given by
\begin{equation}
\hat{H} (t) = -J \sum_{\langle i,j \rangle } \hat{a}_i^{\dagger} \hat{a}_j + A_2 \cos(\alpha \omega_0 t) \hat{n}_2 + A_3 \cos (\beta \omega_0 t + \varphi) \hat{n}_3.
\label{Eq:TriangleTimeDependentHamiltonian}
\end{equation}
Here, $\omega_0$ is the base driving frequency and ($\alpha, \beta$) are integers allowing the two driven sites to be modulated at distinct yet commensurate frequencies and $T = 2 \pi / \omega_0$ is the global period. 
%We can take into account different modulation frequencies on neighbouring sites which allows us to explore a wider range of parameters in the Floquet-engineered system.
We consider independent driving amplitudes $A_2$ and $A_3$ on sites two and three respectively, and a relative phase $\varphi$ at $t=0$ between the two modulations.

By using the high-frequency expansion discussed in Sec.~(\ref{Sec:FloquetEngineering}), the time-independent, effective Hamiltonian can be approximated in the lowest order as
\begin{equation}
\hat{H}^{(0)}_{\mathrm{eff}} = -J_{12}^0 \hat{a}_1^{\dagger} \hat{a}_2 -J_{23}^0 \hat{a}_2^{\dagger} \hat{a}_3 -J_{31}^0 \hat{a}_3^{\dagger} \hat{a}_1 + h.c.
\label{Eq:2DEffectiveHamitonian}
\end{equation}
where $J_{12}^0$, $J_{23}^0$, $J_{31}^0$ are the potentially complex renormalised tunnelling coefficients given by
%Onsite energy terms may also be introduced but we neglect them in Eq. (\ref{Eq:2DEffectiveHamitonian}) as these can be offset if required by a static potential. 
\begin{gather}
J_{12}^{0} = J \mathcal{J}_0 \left(\frac{A_2}{ \alpha \omega_0}\right) ,\label{Eq:TriangleJ12Renormalised}\\ 
J_{23}^{0} = J \mathcal{J}_0^{\alpha, \beta} \left( \frac{A_2}{\alpha \omega_0}, \frac{A_3}{\beta \omega_0}, \mathrm{e}^{i\varphi} \right) \quad \quad \quad \quad \quad \quad \>  \> \nonumber \\
\quad \quad \quad := J \frac{1}{T}\int_{-T/2}^{T/2} \me^{i \frac{A_3}{ \beta \omega_0} \sin (\beta \omega_0 t + \varphi) - i \frac{A_2}{ \alpha \omega_0} \sin (\alpha \omega_0 t)} \mathrm{d} t, \label{Eq:TriangleJ23Renormalised} \\
J_{31}^{0}  = J  \mathcal{J}_0 \left(\frac{A_3}{ \beta \omega_0}\right) ,
\label{Eq:TriangleJ31Renormalised}
\end{gather}
 and $h.c.$ denotes the hermitian conjugate.
In this chosen gauge, $J^0_{12}$ and $J^0_{31}$ are real and $\mathcal{J}_0^{\alpha, \beta}$ is the generalised two variable, one parameter Bessel function \cite{Korsch-Witthaut-TwodimensionalBesselFunctions-2006}. Crucially, a finite phase shift $\varphi\notin\{0,\pi\}$ and different modulation frequencies $\alpha \neq \beta$  allows for breaking time-reversal symmetry and the tunnelling amplitude $J_{23}^{0}$ connecting the two driven sites can become complex (see Appendix~\ref{Sec:2DZeroRelativePhaseRemovesPlaquetteFlux}), 
%In a 1D nearest-neighbour tight-binding model, complex phases of tunnelling amplitudes can always be gauged away and are therefore physically undetectable. In 2D, in contrast, complex tunnelling amplitudes give rise to a finite Aharonov-Bohn flux piercing a plaquette through the Peierls substitution that cannot be gauged away~\cite{Peierls33}. This lies at the heart of e.g.~the celebrated Harper-Hofstadter model describing the quantum Hall effect in tight-binding models~\cite{Miyake-Ketterle-RealizingHarperHamiltonian-2013,Aidelsburger-Bloch-RealizationHofstadterHamiltonian-2013}.  The required breaking of time-reversal symmetry can be achieved via Floquet engineering~\cite{OkaAoki09_PRB,eckardt_colloquium_2017}. In our model, time-reversal symmetry is broken when $\varphi\notin\{0,\pi\}$ and
giving rise to a finite gauge-independent flux 
\begin{equation}
 \Phi=\mathrm{arg} (J_{12}^0 \times J_{23}^0 \times J_{31}^0 )   
 \label{eq:flux}
\end{equation}
around the plaquette as illustrated in Fig.~\ref{Fig:Triangle}(b).

This setup demonstrates that multi-frequency relative modulations strongly enhance the possibilities of this scheme. Unfortunately, the more complex expressions in Eq.~\eqref{Eq:TriangleJ23Renormalised} make analytic solutions harder, but the systems remain amenable to straightforward numerical optimisation.

%%%%%%%%%%%%%%%%%%%%%%%%%%%%%%%%%%%%%%%%%%%%%%%%%%%%%%%%%%%%%%%%%
\subsubsection{ Full control over tunnelling strengths}
\label{Sec:TriangleControlOverRelativeTunnelling}
\begin{figure}
    \centering
    \includegraphics[width=1\linewidth]{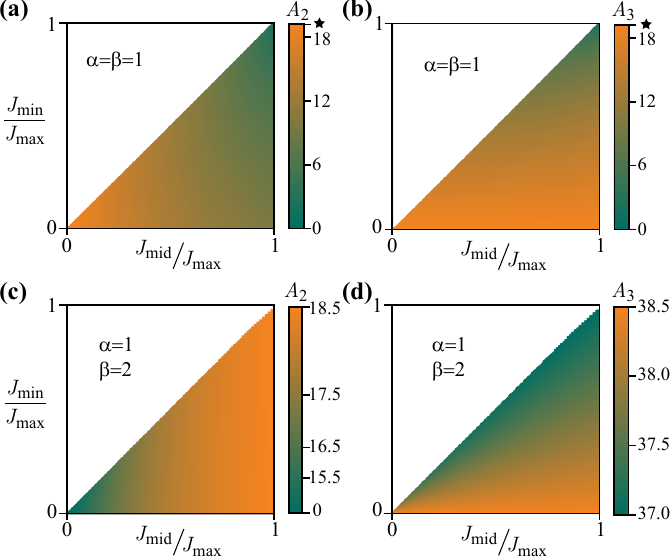}
    \caption{%Any relative strength between the three renormalised tunnelling coefficients $|J_{\mathrm{min}}|$, $|J_{\mathrm{mid}}|$, $|J_{\mathrm{max}}|$ can be realised in our local driving model.  
(a,b) The values of $A_2$ (a) and $A_3$ (b) resulting in given ratios of effective tunnelling coefficients according to Eqs.~(\ref{Eq:TriangleJ12Renormalised})-(\ref{Eq:TriangleJ31Renormalised}).
Full coverage over the lower-right unit triangle demonstrates that all ratios are accessible. This uses $\omega_0=8$, $\alpha=\beta=1$, $\varphi=0$ and the maximum driving amplitudes required correspond to the first zero of the Bessel function ($\mathcal{J}_0\left(19.2/8\right) \approx 0$) and is marked by a star ($\star$).  (c,d) Same as above but for $\alpha=1,\beta=2$, again showing full coverage and demonstrating the flexibility of the scheme: There are many choices of driving parameters resulting in the same relative tunnelling ratio. Note the different ranges of amplitudes and that the colorbar in (c) uses a non-linear (power-law) scale for visual clarity.}
    \label{Fig:Control_over_relativeTunnelling}
\end{figure}

In our triangular plaquette, we first show how local, onsite driving in the form of Eq.~(\ref{Eq:TriangleTimeDependentHamiltonian}) provides full control over the relative strength of the renormalised tunnelling amplitudes by numerically calculating the effective Hamiltonian given by Eqs.~(\ref{Eq:2DEffectiveHamitonian})--(\ref{Eq:TriangleJ31Renormalised}). 
Rather than the absolute strength of the tunnelling coefficients, we note that the physics of the plaquette depends only on the ratio between the tunnelling amplitudes and therefore we focus our discussion accordingly. 
We consider in-phase local modulations ($\varphi=0$) ensuring all renormalised tunneling coefficients are real (see Appendix~\ref{Sec:2DZeroRelativePhaseRemovesPlaquetteFlux}) and label them from smallest to largest in magnitude as $(J_{\mathrm{min}}, J_{\mathrm{mid}},J_{\mathrm{max}})$. Figure~\ref{Fig:Control_over_relativeTunnelling} depicts how tuning the driving amplitudes $A_2$ and $A_3$ results in tuning the relative tunnelling strengths 
\begin{equation}
(x,y) = (J_{\mathrm{min}}/J_{\mathrm{max}},J_{\mathrm{mid}}/J_{\mathrm{max}}),
\label{eq:xy}
\end{equation}
for a fixed frequency $\omega_0=8$. %[As an aside, we note that the two relevant, independent parameters in this analysis are $A_2/\omega_0$ and $A_3/\omega_0$ however we fix $\omega_0=8$ here and show in Appendix.~\ref{App:FluctuationsFromInitialTime} that this constitutes a feasible frequency with which to utilise the first-order approximation of the effective Hamiltonian, while keeping $A_2$, $A_3$ in experimentally feasible regions throughout our analysis.]

We observe that the range of accessible relative tunnelling coefficients covers all possible values such that the lower-right unit triangle is fully filled.
%Since only the relative hopping strengths are physically relevant, this demonstrates that any desired ratio between the three renormalised tunnelling amplitudes can be achieved.  
Even for a fixed base frequency $\omega_0$, there are in fact multiple driving parameters that result in the same ratios of tunnelling strength, highlighting the flexibility of this scheme~\cite{Wang21_NJP}. 
When the two sites are driven with the same frequency ($\alpha=\beta=1$), see (a,b),  Eq.~(\ref{Eq:TriangleJ23Renormalised}) simplifies to $J_{23}^0 = J \mathcal{J}_0 \left(\nicefrac{|A_3 - A_2|}{ \omega_0} \right)$, and the  maximum driving amplitudes required correspond to the first zero of the Bessel function ($\mathcal{J}_0\left(19.2/8\right) \approx 0$).
%For each value of $(A_2,A_3)$, we depict the relative tunnelling amplitudes $J_{23}/J_{12}$ and $J_{31}/J_{12}$, which are all real since $\varphi=0$.
In contrast, in Figure~\ref{Fig:Control_over_relativeTunnelling}(c,d) the two sites are driven at different frequencies as $\alpha=1$ and $\beta=2$. This affects the range of the required driving amplitudes, but we nonetheless obtain  full coverage of all relative tunnelling strengths. 

The possibility  to  employ different frequencies, even on the same lattice site, significantly increases the number of independent local control parameters (e.g.~using $n$ frequencies with $n$ independent  amplitudes per lattice site) and thereby presents a clear pathway to extending the full individual control over all tunnelling elements to larger 2D patches and possibly full 2D lattices.

%%%%%%%%%%%%%%%%%%%%%%%%%%%%%%%%%%%%%%%%%%%%%%%%%%%%%%%%%%%%%%%%%
\subsubsection{Full control over flux}
\label{SSec:ControlOverFlux2D}

\begin{figure}
\centering
    \includegraphics{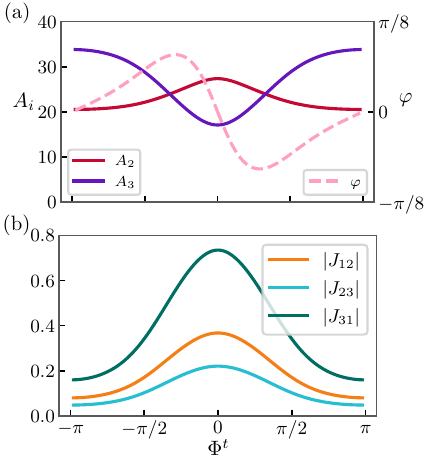}
    \caption{The results of the numerical optimisation algorithm for the target relative tunnelling strengths $(J_{\mathrm{mid}}/ J_{\mathrm{max}}, J_{\mathrm{min}}/J_{\mathrm{max}}) = (0.5, 0.3)$.
    (a) Driving parameters $A_2$, $A_3$ and $\varphi$ can be tuned continuously to realise any desired plaquette flux $\Phi^t$ while keeping the cost function~[Eq.~(\ref{Eq:CostFunc})] below $10^{-10}$. (b) The resulting tunneling amplitudes for driving parameters shown in (a). }
    \label{Fig:FindMinimumResults}
\end{figure}

In addition to controlling the ratios between tunnelling amplitudes, the local modulation scheme can simultaneously realise any desired target flux in our triangular plaquette. To show this, we numerically minimise \footnote{The numerical minimisation is performed using the FindMinimum function in Mathematica.} the cost function
\begin{equation}
    f(A_2/\omega_0, A_3/\omega_0, \varphi) := (x^t - x)^2 + (y^t - y)^2 + \frac{1}{20}(\Phi^t - \Phi)^2,
    \label{Eq:CostFunc}
\end{equation}
over the three independent control parameters $A_2/\omega_0$, $A_3/\omega_0$ and $\varphi$, where $( x^t, y^t)$ give the target ratios between tunnelling amplitudes and $\Phi^t \in [-\pi, \pi) $ is the desired target flux.
Here, $x,y,\Phi$ are calculated using Eqs.~(\ref{Eq:TriangleJ12Renormalised})-(\ref{eq:xy}) and we choose $\alpha=1$, $\beta=2$ . 

Figure~\ref{Fig:FindMinimumResults} shows the resulting control parameters as a function of target flux $\Phi^t$ for fixed target tunneling ratios $(x^t,y^t) = (0.5, 0.3)$. We indeed find that the plaquette flux can be continuously tuned by adjusting the driving parameters. Note that while the individual tunneling strengths vary significantly as a function of target flux, their ratios remain constant at the target value. 
Changing the relative phase $\varphi$ of the modulation to $-\varphi$ inverts the flux, i.e.~sends $\Phi \rightarrow -\Phi$, as discussed in  Appendix ~\ref{App:ReflectionofXiGivenReflectionofVarphi}.
%The reflection of $\Phi$ about zero when $\varphi \rightarrow -\varphi$ is a general feature as discussed in  Appendix ~\ref{App:ReflectionofXiGivenReflectionofVarphi}.

To check that this tunability is in fact generic, we chose $>100$ random sets of target $(x^t, y^t)$ values and could in all cases find driving parameters that keep the above cost function below $e^{-10}$ over the full range of $\Phi^t$. We show another example in Appendix ~\ref{App:Neighbourhood(0.9,0.9)}.

\subsection{Locally tunable flux in 2D}
\label{SubSec:LocalFlux}

\begin{figure}
\centering
    \centering\hspace{-4mm}
    \includegraphics{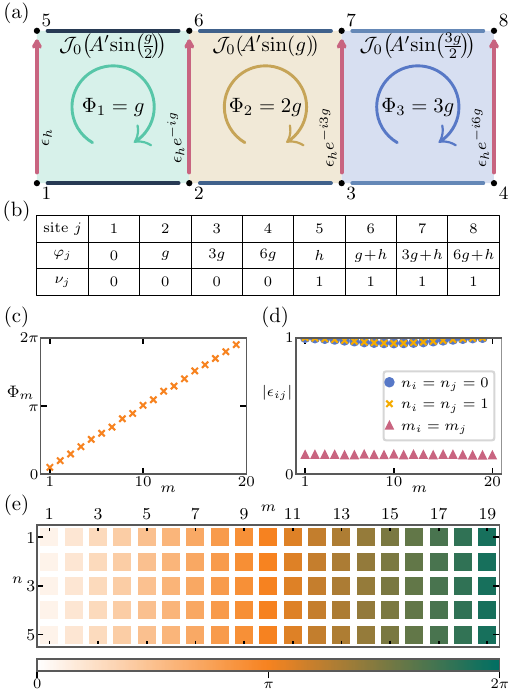}
    \caption{Independently tunable fluxes along one axis. (a) A $4\! \times\! 2$-site ladder showing effective tunnelling renormalisation factors $\epsilon_{ij}$ where each site $j$ is driven according to Eq.~\eqref{Eq:LocalDriveFuncWithStaticOffset} with phase $\varphi_j$ and static offset $\nu_j$ given in (b). 
    Tunnelling along the $x$-direction is given by Eq.~\eqref{Eq:LocalFluxLadderXRenorm} and is identical for the two horizontal legs. Complex tunnelling phases introduced along the rungs [Eq.~\eqref{Eq:LocalFluxLadderYRenorm}] are independently tunable by the $x$-dependent choices of $f_{m}$. (b) Driving phases $\varphi_j$ when $f_m$ is the $m^{\mathrm{th}}$ element of a list of triangular numbers $f = [0,1,3,6,10,...]$ -- a particular choice that generates linearly increasing flux values  with a gradient of $g$ per plaquette. (c) Linearly increasing plaquette flux values in a $20\! \times\! 2$-site ladder numerically calculated from the stroboscopic Hamiltonian $H_S^{t_0}$, and (d) corresponding magnitudes of $|\epsilon_{ij}|$, for $\omega =  25$, $A = 5$, $g = \pi/10$, $h = \pi/2$. % Tunnellings along the legs are modulated very weakly.
    % but is identical between the two legs. The magnitude of the     Tunnelling magnitudes on the rungs, given theoretically by Eq.~\eqref{Eq:LocalFluxLadderYRenorm}, are identical everywhere. 
    The horizontal and vertical tunnellings can be made approximately equal everywhere by choosing an anisotropic static lattice.
    %Vertically aligned $\epsilon_{ij}|_{n_i = n_j}$ are identical. 
%The value of $|\epsilon_{ij}|_{m_i = m_j}$ given theoretically by Eq.~\eqref{Eq:LocalFluxLadderYRenorm} is the same for all vertical tunnellings along the ladder.
    % Setting different bare tunnelling $J_x$ and $J_y$ allows for homogeneous tunnelling magnitudes across the lattice. 
    (e) Plaquette flux values (shown by the colourbar) for a larger $20\! \times\! 6$ system calculated numerically from the stroboscopic Hamiltonian for the same driving parameters as in (c), generating a 2D system with an artificial magnetic field gradient along the $x$-direction.
    }
    \label{Fig:GradientFlux}
\end{figure}

We now demonstrate how the local driving scheme can be employed in larger 2D settings to realise bespoke local flux configurations by combining local periodic driving with additional static offsets. In this case, each site $j = (m_j,n_j)$ is driven as
\begin{equation}
    W_j(t) = A \cos (\omega t + \varphi_j) + \nu_j \omega,
    \label{Eq:LocalDriveFuncWithStaticOffset}
\end{equation}
where the second term describes a site-dependent static offset. All sites are modulated with the same amplitude $A$ and frequency $\omega$, but with site-dependent phases. The effective Hamiltonian is given by Eq.~\eqref{Eq:EffectiveGFramework} with 
\begin{equation}
     \epsilon_{ij} = \mathcal{J}_{\nu_{ij}} \left( \frac{A_{ij}}{\omega} \right) e^{-i \nu_{ij} \xi_{ij} }
    \label{Eq:LocalDrivewStaticOffsetRenormFactor}
\end{equation}
where $\nu_{ij} = \nu_i - \nu_j$ and  $ A_{ij} = 2 A \sin \left( \frac{\varphi_i - \varphi_j}{2}\right)$ and $\xi_{ij} = (\varphi_j + \varphi_i - \pi)/2 $, as shown in Appendix \ref{Appx:GradientFluxCalculations} in detail and in \cite{eckardt_colloquium_2017}. Importantly, Eq.~\eqref{Eq:LocalDrivewStaticOffsetRenormFactor} demonstrates how non-zero static relative offsets $\nu_{ij}$ together with resonant modulation can be used to engineer complex tunnelling phases, while only requiring global reflection symmetry  to be broken ($\varphi_i - \varphi_j \not\in \{0,\pi \}$) \cite{eckardt_colloquium_2017} . This idea has been realised experimentally with periodic running lattices to engineer homogeneous plaquette flux magnitudes throughout the lattice 
\cite{Aidelsburger-Bloch-ExperimentalRealizationStrong-2011, Aidelsburger-Bloch-RealizationHofstadterHamiltonian-2013, Aidelsburger-Goldman-MeasuringChernNumber-2015, Atala-Bloch-ObservationChiralCurrents-2014, Miyake-Ketterle-RealizingHarperHamiltonian-2013, Kennedy-Ketterle-ObservationBoseEinstein-2015, Tai-Greiner-MicroscopyInteractingHarper-2017}. Here, we demonstrate how flexibility in local offset phases $\varphi_j$ in Eq.~\eqref{Eq:LocalDriveFuncWithStaticOffset} generated by our scheme allows for locally tunable flux values, an idea first proposed to simulate optical solenoids \cite{wang_floquet_2018}.

As an example, we demonstrate how to engineer a linear gradient in plaquette fluxes along one direction simulating a system experiencing a magnetic field gradient, see Fig.~\ref{Fig:GradientFlux} where we implement this analytically and numerically for different system sizes. We consider a 2D square lattice where sites experience a static, linear energy offset in the $y$-direction with a gradient corresponding to the modulation frequency  $\omega$ i.e., $\nu_j = n_j$ in Eq.~\eqref{Eq:LocalDriveFuncWithStaticOffset}.
While the tunnelling amplitudes along the $x$-direction are renormalised by Bessel functions $\mathcal{J}_0$ as before, the amplitudes in the $y$-direction are controlled by $\mathcal{J}_1$ with additional complex phases.  
% We exploit this to achieve local plaquette fluxes that are fully tunable along one axis while the tunnelling magnitudes are only weakly modulated, see Fig.~\ref{Fig:GradientFlux}e. 
We parameterise the local driving phases on each site as $\varphi_{j} = hn_j + f_{m_j} g$ where $h,g$ are real constants and $f_{m_j}$ is an $x$-dependent factor.
The tunnellings along the $x$- an $y$-directions are renormalised with respective factors
% \begin{align}
%     \epsilon_{ij}^{x} &= \mathcal{J}_0 \left(  A' \sin \left( \frac{g}{2} |f_{m_i} - f_{m_j}| \right)\right), \label{Eq:LocalFluxLadderXRenorm}\\
%     \epsilon_{ij}^{y} &=\epsilon_h e^{-i g \nu_{ij} f_{m_{ij}}} \label{Eq:LocalFluxLadderYRenorm}
% \end{align}
\begin{align}
    \epsilon_{ij}\> |_{n_i = n_j} &= \mathcal{J}_0 \left(  A' \sin \left( \frac{g}{2} |f_{m_i} - f_{m_j}| \right)\right), \label{Eq:LocalFluxLadderXRenorm}\\
    \epsilon_{ij} |_{m_i = m_j} &=\epsilon_h e^{-i g \nu_{ij} f_{m_{i}}} \label{Eq:LocalFluxLadderYRenorm}
\end{align}
where $A'= \frac{2A}{\omega}$ and $\epsilon_h =  \mathcal{J}_1 \left( A' \sin\left( \frac{h}{2} \right) \right) e^{-i \nu_{ij} \left(\frac{h - \pi}{2}\right)}$.

% In Fig.~\ref{Fig:GradientFlux}, we implement this scheme to achieve flux values that linearly increase along the $x$-direction, generating a magnetic field gradient.
Considering a two-leg ladder example in Fig.~\ref{Fig:GradientFlux}(a), we show the effective tunnelling renormalisation factors $\epsilon_{ij}$ for both horizontal and vertical tunnelling 
with constants $\varphi_j$ and $\nu_j$ set explicitly by the table in Fig.~\ref{Fig:GradientFlux}(b). This example realises a flux that is linearly increasing  with gradient $g$ per plaquette along the horizontal direction. It is implemented by choosing $f_m$ to be the $m^{\mathrm{th}}$ element of a list of triangular numbers $f = [0,1,3,6,10,...]$. We note that any desired array of local fluxes can be achieved by utilising different choices of $f_m$.  Tunnellings along the two horizontal legs of the ladder are identical e.g. $\epsilon_{12} = \epsilon_{56} = \mathcal{J}_0 \left( A' \sin \left(\frac{g}{2} \right) \right)$ and all $x$-tunnellings will be approximately equal for small $A'$ given $\mathcal{J}_0(\delta) \approx 1$ for small $|\delta|$. Tunnelling magnitudes along the vertical rungs of the ladder are identical everywhere.

Figure~\ref{Fig:GradientFlux}(c) shows the $m$-dependent flux for a $20 \times 2$ site ladder extracted from the stroboscopic Hamiltonian $H_S^{t_0}$, featuring local flux values that increase linearly across the horizontal direction. Figure~\ref{Fig:GradientFlux}(d) shows that horizontal ($n_i = n_j$) and vertical ($m_i = m_j$) tunnelling renormalisation factors $\epsilon_{ij}$ stay approximately constant across the ladder, with fluctuations of  $\sim 0.04$ and $0.007$ respectively given $A'=2/5$ is sufficiently small. 
By considering different bare tunnelling values $J_x$, $J_y$ along the $x$- and $y$-directions of the underlying lattice, we can choose global tunnelling magnitudes to be approximately homogeneous across the system. This scheme can easily be generalised to larger 2D systems by extending the linear static offsets; Fig.~\ref{Fig:GradientFlux}(e) shows plaquette fluxes numerically generated from the stroboscopic Hamiltonian for a $20\times6$ site system achieving a smooth linear flux gradient in the $x$-direction.

\section{Physical implementation}
\label{Sec:PhysImplement}
The key requirement for implementing this proposal is the ability to provide local time-dependent potentials to the sites of an optical lattice. Here we demonstrate that this is achievable with present-day  technology and discuss some details and limitations.

For simplicity, we have so far assumed that the modulation of the on-site energy takes the form $W_j(t)=A_j\cos(\omega_j t)$, which oscillates symmetrically around zero. This can be achieved in different ways, with the conceptually simplest configuration being to use both blue-detuned and red-detuned beams whose intensity-modulation is out of phase. In practice it will however typically be easier to apply a common offset $A_0$ to all sites and modulate around this, i.e.~$W_j(t)=A_0+A_j\cos(\omega_j t)$, such that only a single wavelength is required.

For typical (moderate) system sizes, the most straightforward way of implementing such a modulation is to use acousto-optical deflectors (AODs) in an arrangement similar to current tweezer array technologies~\cite{Endres-Lukin-AtombyatomAssemblyDefectfree-2016, Barredo-Browaeys-AtombyatomAssemblerDefectfree-2016, Bernien-Lukin-ProbingManybodyDynamics-2017, Chisholm-Kjaergaard-ThreedimensionalSteerableOptical-2018,  Omran-Lukin-GenerationManipulationSchrodinger-2019,Bluvstein-Lukin-QuantumProcessorBased-2022,Spar-Bakr-RealizationFermiHubbardOptical-2022}. Here, every beam (tweezer) can be focused onto an individual lattice site and provides independent time-resolved control over the corresponding on-site energy. The modulation bandwidth of AODs vastly exceeds the required modulation frequencies in the range of $\omega\sim 20-100\,J\leq 2\pi\times100\,$kHz.

For large SSH chains and other periodic arrangements, a convenient  alternative will be to split the modulation into parts by using a combination of acousto-optical modulators (AOMs) and spatial light modulators (SLMs) such as e.g.~digital mirror devices (DMDs) or liquid crystal modulators. In such a configuration, each required frequency or phase configuration, such as the two out-of-phase components in Sections~\ref{Sec:SuSchriefferHeegerModel} and \ref{Sec:SSH3}, is provided by an independent beam path $\gamma\in\{1,2\}$ with the common time-dependent modulation ($1+\cos(\omega_\gamma t+\varphi_\gamma)$) provided by the AOMs. The subsequent SLMs provide the required spatial intensity maps ($A_j^\gamma$) to address different subsets of sites. 
This combination results in potentials of the form $W_j(t)=\sum_\gamma A_j ^\gamma (1+\cos(\omega_\gamma t+\varphi_\gamma))$, where the time-averaged offset energy per site is $\sum_\gamma A_j^\gamma$  and can be cancelled using a static beam of the opposite detuning with matching spatial intensity map.

\begin{figure}[t]
	\includegraphics[width=0.9\linewidth]{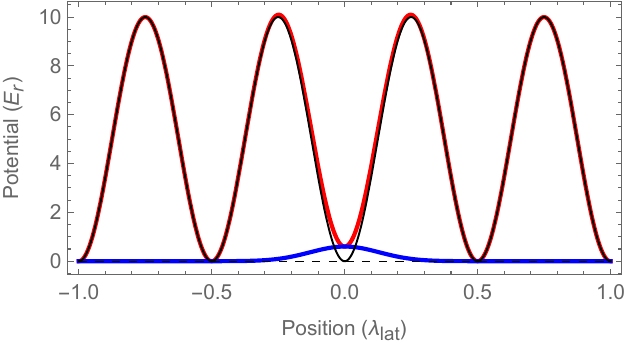}
	\caption{The black line shows a typical 1D lattice of depth $V_0=10\,E_r$ formed by a wavelength of $\lambda_{lat}=1064\,$nm. The blue line shows the modulation beam at $\lambda_{mod}=532\,$nm (amplitude $A=40$) focused by a diffraction-limited objective with NA$=0.68$ and the red line shows the total resulting potential.}
	\label{Fig:Implementation}
\end{figure}

In all cases, the potentials ideally have minimal cross-talk between neighbouring sites, i.e.~the beams should be tightly focused to achieve an effective modulation of only the targeted sites. Typical quantum gas microscopes with their high-numerical-aperture lenses can provide the required resolution, as demonstrated by their ability to image individual lattice sites~\cite{bakr_quantum_2009}.  
%Repurposing such a system to focus a laser to a similar resolution would provide a method of manipulating the onsite energies of single sites as required by our proposal. 
For example, the NA=0.68 system used by the Munich group~\cite{Choi-Gross-ExploringManybodyLocalization-2016} in one of the first quantum gas microscopes enabled them to demonstrate a correlation length of a projected potential of $0.6d$, where  $d=532\,$nm is the lattice constant and $\lambda_{mod}=787.55\,$nm.  Figure .\ref{Fig:Implementation} demonstrates that if the focused potentials instead use the common wavelength of $\lambda_{mod}=532\,$nm, which is significantly shorter than the typical lattice wavelength $\lambda_{lat}=1064\,$nm, the modulation is well-focused on the target site.  Due to the small amplitudes of the required modulation compared to the lattice depth (typically $<5\%$), the direct effect of the focused light on the bare tunnelling is minimal. Also, the small modulation frequencies of typically $\omega\approx8-30 J$ mean that interband excitations due to the modulation can be excluded \cite{Reitter2017}.

Importantly, the proposed scheme can be brought to fruition even at lower optical resolutions. In this case there will be an unwanted modulation leakage to neighbouring sites, but this can be efficiently cancelled by an additional opposite modulation to that neighbouring site. Numerical simulations show that even for $w/d=2$, where $w$ denotes the waist of the used beams (corresponding to a modest required NA$<0.2$), perfect cancellation can be achieved at the cost of only moderate (factor of 4) increases in the required laser intensity.  These modest resolution requirements represent a significant advantage compared to schemes based on directly changing the tunnelling barrier between sites by focusing static potentials there, as these require sub-lattice-site optical resolution and stability.

In addition, the proposed schemes directly incorporate the ability to also control the on-site energies by adjusting the static components of the modulation.
%. This can be done through the use of a locally tunable digital micromirror device (DMD) programmed to generate a static potential on each site that is equal and opposite to the effective on-site potential generated through driving.
As with every Floquet scheme, the practical limitations will likely come in the form of Floquet heating~\cite{eckardt_colloquium_2017}. An advantage of the presented schemes are their flexibility, for instance the possibility to choose modulation frequencies that avoid single-particle multi-photon resonances~\cite{Weinberg2015, Reitter2017}.

%Another option is to use liquid-crystal display (LCD) as a spatial light modulator \cite{bergamini_holographic_2004} . These devices generate arrays of microtraps holographically. Holographic optical tweezers use a computer-designed diffractive optical element to split a single collimated beam into several beams, which are then focused by a high numerical aperture lens into an array of tweezers. This generation of multitrap arrays for single atoms, allowing for atoms on the single site level to be addressed. Sites were spaced \(5\mathrm{\text\textmu m}\).
%\cite{bergamini_holographic_2004} per site, address individual sites, and measure the locations of trapped atoms within the lattice. The setup allows storing of a single atom per trap and addressing and manipulation of individual trapping sites.

Finally, we point out that the use of focused beams that affect individual lattice sites provides even more general opportunities if the used light is not too far detuned from the atomic transitions, but rather produces spin-dependent potentials~\cite{Mandel_SpinDep}. For instance, a modulated spin-dependent light shift will result in spin-dependent renormalised tunnellings, opening up further possibilities for mobile particles, as well as enabling local control over the effective spin Hamiltonian of a two-component Hubbard model in the Mott regime~\cite{Duan-Lukin-ControllingSpinExchange-2003,Garcia-Ripoll-Cirac-SpinDynamicsBosons-2003}. Another alternative, which is independent of the Floquet physics discussed in this work, is the use of two light frequencies within a given beam that can couple different spin states via local Raman transitions, thereby allowing independent control over local fields of the form $\sum_j \mathbf{h}_j\cdot\mathbf{\sigma_j}$, where the $\mathbf{\sigma}_j$ represent Pauli spin operators on site $j$.

%%%%%%%%%%%%%%%%%%%%%%%%%%%%%%%%%%%%%%%%%%%%%%%%%%%%%%%%%%
\section{Conclusion}
\label{Sec:Conclusion}
Local periodic driving enables individually tunable tunnelling amplitudes in optical lattices.  In 1D, we have demonstrated that full, independent control of all nearest-neighbour tunnelling coefficients can be achieved by controlling the amplitudes of a single-frequency modulation. By exploring a variety of examples, we have demonstrated the efficiency and generality of our scheme in generating 1D Hamiltonians. 

%Crucially, Floquet engineering allows the creation of complex tunnelling phases. 
We further extended our consideration to two dimensions. By modulating specific sites of a Lieb lattice, we can engineer a fully programmable square 2D network with flexible connectivity allowing for e.g. the simulation of quantum walks on bespoke graphs. In 2D, crucially, complex tunnelling amplitudes can give rise to local Berry fluxes resulting for instance in more involved topological phenomena~\cite{OkaAoki09_PRB,Miyake-Ketterle-RealizingHarperHamiltonian-2013, Jotzu14_Nat,Aidelsburger-Goldman-MeasuringChernNumber-2015,Tarnowski19_NatCom, Asteria19_NatPhys,Wintersperger20_NatPhys, Slager22_arXivAnomEuFloq} by breaking time-reversal symmetry in the effective Hamiltonian. By considering a single triangular plaquette, we showed that local periodic driving enables full simultaneous control over the gauge-invariant plaquette flux as well as all relative tunnelling strengths by employing numerical optimisation methods.

%Enhancing the experimental flexibility of our scheme, we demonstrate that a number of distinct driving parameter sets can be employed to achieve the same effective Hamiltonian. This facilitates flexible experimental implementation which can be tailored to the practicalities of different set-ups. 

The significant number of local control parameters, namely $n$ amplitudes and phases per lattice site when using $n$ modulation frequencies plus one static offset, suggests that high levels of individual control should be possible also for larger 2D systems. However, further work is needed to extend our scheme to infinite 2D optical lattices. 

Complex phases can also be engineered by combining local modulations with non-zero potential offsets. %This requires fewer constraints on the driving potential for the creation of tunnelling phases as only global reflection symmetry needs to be broken, which occurs when sites are driven out of phase. 
We utilise this in a square 2D lattice to generate fully tunable flux values along one direction while tunnelling magnitudes remain approximately constant. This can be used in particular to simulate a magnetic field gradient. 

The presented modulation scheme comes with rather modest resolution and positional stability requirements that are  easily exceeded by quantum gas microscopes. It hence provides a more stable and experimentally less demanding alternative compared to directly changing the tunnelling barriers by focusing static potentials on the barriers, as these would require sub-lattice-site optical resolution and stability. Furthermore, the ability to change the sign of the hopping elements or imbue them with complex phases makes the presented schemes more powerful and versatile than directly addressing the barrier.

The ability to encode and study Hamiltonians with arbitrary local tunnelling coefficients will allow optical lattice experiments to simulate an incredibly diverse range of problems and quantum phenomena well beyond solid state settings.
So far, studies of universal 2D Hamiltonians and complexity classes in quantum simulations have often focused on spin models: results include the universality of a locally tunable 2D Heisenberg model, which remarkably is shown to be capable of replicating, to any desired accuracy, the entire physics of any other quantum many-body system~\cite{Cubitt-Piddock-UniversalQuantumHamiltonians-2018}.  Moving beyond quantum phenomena, there exist explicit polynomial mappings between locally tunable Ising models and many important classical discrete optimisation problems such as the travelling salesman and graph colouring~\cite{Lucas-IsingFormulationsMany-2014}.
Indeed, the methods discussed in Section~\ref{Sec:PhysImplement} detailing how to generate independently tunable, spin-dependent tunnelling also enable local control over emerging spin models~\cite{Garcia-Ripoll-Cirac-SpinDynamicsBosons-2003, Duan-Lukin-ControllingSpinExchange-2003}.

However, spin models, such as the Heisenberg model, represent only limiting cases of more general models for mobile particles, such as Hubbard models. While spin models can more easily be mapped onto digital quantum computers, it will now be fascinating to investigate the necessarily larger possibilities of local programmable Hubbard models.

\begin{acknowledgments}
We are grateful to Bo Song, Tiffany Harte, Yao-Chih Kuo, Callum Duncan, Luca Donini, Kimberly Tkalcec and Emmanuel Gottlob for helpful and encouraging discussions. G.N. acknowledges funding from the Royal Society Te Apārangi and the Cambridge Trust. 
F.N.\"U.~acknowledges funding from a Royal Society Newton International Fellowship, the Marie Sk{\l}odowska-Curie programme of the European Commission Grant No 893915 and Trinity College Cambridge. This work was partially supported by a grant from the Sloan Foundation and F.N.\"U.~thanks Aspen Center for Physics for their hospitality. 
This work was partly funded by the European Commission ERC Starting Grant QUASICRYSTAL, the EPSRC Grant No. EP/R044627/1 and EP/X032795/1, and Programme Grant DesOEQ (Grant No. EP/P009565/1) as well as the EPSRC Hub in Quantum Computing and Simulation EP/T001062/1. 
\end{acknowledgments}

\section{Data Availability Statement}
All data accompanying this publication are directly available within the publication.

\newpage

%%%%%%%%%%%%%%%%%%%%%%%%%%%%%%%%%%%%%%%%%%%%%%%%%%%%%%%%%%%%%%%%%%%%%%%%%%%%%%%%%%%%%%%%%%%%%%%%%%%%%%%%%%%%%%%%%%%%%%%%%%%%%%%%%%
\appendix

\section{Vanishing first-order correction in high-frequency expansion }
\label{App:VanishingSecondOrderHFExpansionTerm}
We here show that the second term of the high-frequency expansion of the effective Hamiltonian, given by Eq.~(\ref{Eq:HighFreqencyExpansionSecondTerm}), vanishes for sinusoidal driving of onsite energies in the form of Eq.~(\ref{Eq:PotentialTermSinusoidalDrivingOneDimension}). Fourier components of the lattice-frame Hamiltonian are given by
\begin{equation}
\tilde{H}_m = -J \sum_{\langle i,j \rangle} \hat{a}_i^{\dagger} \hat{a}_j \mathcal{J}_m \left( \frac{A_i - A_j}{\omega} \right),
\end{equation}
where $\mathcal{J}_m$ is the Bessel function of the $m^{\mathrm{th}}$ kind. Noting that $\mathcal{J}_{-m}(x) = (-1)^m \mathcal{J}_m(x)$, the commutator between Fourier components $H_m$ and $H_{-m}$ is then
\begin{widetext}
\begin{align}
% \resizebox{.45\textwidth}{!} {
[H_m, H_{-m}] &= J^2 \left[ \sum_{\langle ij \rangle} \mathcal{J}_m \left( \frac{A_i - A_j}{ \omega} \right) \hat{a}_i^{\dagger} \hat{a}_j, (-1)^m \sum_{\langle ij \rangle} \mathcal{J}_{m} \left( \frac{A_i - A_j}{\omega} \right) \hat{a}_i^{\dagger} \hat{a}_j\right] \\
% &= J^2 (-1)^m \sum_{\substack{\langle ij \rangle\\ \langle kl \rangle}} \mathcal{J}_m \left( \frac{A_i - A_j}{\omega} \right) \mathcal{J}_{m} \left( \frac{A_k - A_l}{\omega} \right) \left[\hat{a}_i^{\dagger} \hat{a}_j, \hat{a}_k^{\dagger} \hat{a}_l \right] \\
&= J^2 (-1)^m \sum_{\substack{\langle ij \rangle\\ \langle kl \rangle}} \mathcal{J}_m \left( \frac{A_i - A_j}{ \omega} \right) \mathcal{J}_{m} \left( \frac{A_k - A_l}{ \omega} \right) \left( \hat{a}_i^{\dagger} \hat{a}_l^{} \delta_{jk}- \hat{a}_k^{\dagger} \hat{a}_j^{} \delta_{il} \right)  \\
&= J^2 (-1)^m \sum_j \left( \mathcal{J}_m\left( \frac{A_j - A_{j+1}}{\omega} \right)  \mathcal{J}_m \left( \frac{A_{j+1} - A_{j+2}}{ \omega} \right) \hat{a}_j^{\dagger}  \hat{a}_{j+2}^{} \right.  \nonumber \\
&\quad \quad \quad + \mathcal{J}_m\left( \frac{A_j - A_{j+1}}{ \omega} \right)  \mathcal{J}_m \left( \frac{A_{j+1} - A_{j}}{ \omega} \right)\left( \hat{a}_j^{\dagger}  \hat{a}_{j}^{} -   \hat{a}_{j+1}^{\dagger}  \hat{a}_{j+1}^{} \right) \nonumber \\
&\quad \quad \quad +  \mathcal{J}_m\left( \frac{A_{j+2} - A_{j+1}}{ \omega} \right)  \mathcal{J}_m \left( \frac{A_{j+1} - A_{j}}{\omega} \right) \hat{a}_{j+2}^{\dagger}  \hat{a}_{j}^{}  \nonumber \\
&\quad \quad \quad + \mathcal{J}_m\left( \frac{A_{j+1} - A_{j}}{ \omega} \right)  \mathcal{J}_m \left( \frac{A_{j} - A_{j+1}}{ \omega} \right)\left( \hat{a}_{j+1}^{\dagger}  \hat{a}_{j+1}^{} -   \hat{a}_{j}^{\dagger}  \hat{a}_{j}^{} \right) \nonumber \\
&\quad \quad \quad +  \mathcal{J}_m\left( \frac{A_{j+1} - A_{j}}{\omega} \right)  \mathcal{J}_m \left( \frac{A_{j+2} - A_{j+1}}{\omega} \right) \left(- \hat{a}_{j+2}^{\dagger}  \hat{a}_{j}^{}\right) \nonumber \\
&\quad \quad \quad + \left. \mathcal{J}_m\left( \frac{A_{j+1} - A_{j+2}}{ \omega} \right)  \mathcal{J}_m \left( \frac{A_{j} - A_{j+1}}{ \omega} \right) \left(- \hat{a}_{j}^{\dagger}  \hat{a}_{j+2}^{}\right) \right) \nonumber \\
&= 0.
% }
\end{align}
\end{widetext}

Therefore, corrections to the lowest (zeroth) order term describing the effective Hamiltonian Eq.~(\ref{Eq:FirstOrderHFTermOneDSinusoidalDrive}) then come in the order of $1/\omega^2$.

\section{Numerical calculation of Stroboscopic and Effective Hamiltonian}
\label{App:NumericalCalculationStrobHam}
As discussed in Sec.~\ref{Sec:FloquetEngineering}, Floquet-engineered systems can be analysed using the effective Hamiltonian $H_{\mathrm{eff}}$. This representation of the Floquet Hamiltonian is Floquet gauge invariant such that the initial time $t_0$-dependence is transferred to the micromotion operator \cite{bukov_universal_2015}. Under moderate frequencies, the high-frequency expansion of the effective Hamiltonian can be employed to analytically study the dynamics~\cite{goldman_periodically_2014,eckardt_colloquium_2017}. 

To extend our analysis, we can also numerically calculate the $t_0$-dependent stroboscopic Hamiltonian $H_S^{t_0}$ which captures the effect of the drive without truncation of higher-order contributions. 
%We briefly describe how to numerically calculate the stroboscopic Hamiltonian $H_S^{t_0}$. 
This can be done by simulating the time evolution by numerically solving the Schr\"{o}dinger equation. After one period, we diagonalise the time evolution operator $U(t_0 + T, t_0)|\psi_n(t_0)\rangle=u_n |\psi_n(t_0)\rangle$. The quasienergy $\{ \epsilon_n \}$ is given by the natural logarithm $\epsilon_n = \frac{i}{T} \log (u_n )$, defining the stroboscopic Hamiltonian via $H_S^{t_0} = \sum_n \epsilon_n |\psi_n(t_0)\rangle\langle\psi_n(t_0)|$. 

\begin{figure}[t]
	\includegraphics[width=0.9\linewidth]{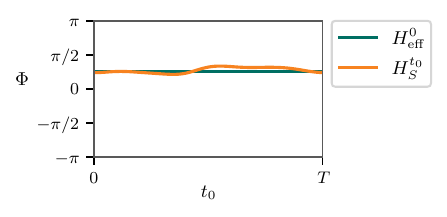}
	\caption{The plaquette flux $\Phi$ calculated using the lowest order term of the effective Hamiltonian $H_{\mathrm{eff}}^0$ and the stroboscopic Hamiltonian $H_S^{t_0}$ as a function of $t_0 \in [0,T)$. Here, we set $A_2 = A_3 = 15$, $\alpha=1$, $\beta=2$, $\omega=8$, $\varphi=0.2 \pi$. The variation in plaquette flux resulting from the Floquet gauge remains small ($0.1\pi$).   }
	\label{Fig:StroboscopicFluctuations}
\end{figure}

%\section{Floquet gauge and fluctuations from initial time}
\label{App:FluctuationsFromInitialTime}

% As discussed in Sec.~\ref{Sec:FloquetEngineering}, it can also be convenient to analyse Floquet-engineered systems using the effective Hamiltonian $H_{\mathrm{eff}}$. This representation of the Floquet Hamiltonian is Floquet gauge invariant such that the initial time $t_0$-dependence is transferred to the micromotion operator \cite{bukov_universal_2015}. Under moderate frequencies, the high-frequency expansion of the effective Hamiltonian can be employed to analytically study the dynamics. To extend our analysis, we can also numerically calculate the $t_0$-dependent stroboscopic Hamiltonian $H_S^{t_0}$ which captures the effect of the drive without truncation of higher-order contributions. 

Fig.~\ref{Fig:StroboscopicFluctuations} shows the flux values calculated using both approaches for a specific  set of driving parameters. We numerically evaluate the stroboscopic Hamiltonian for initial times $t_0 \in [0, T]$ and extract the flux (orange line) which depends on the Floquet gauge $t_0$ as expected. We compare this to the flux calculated from the lowest order effective Hamiltonian as implemented in the main text (green line), where the stroboscopic Hamiltonian $t_0$ values fluctuate around the high-frequency expansion value by less than $10\%$. 
%The amount of this fluctuation can be minimised by choice of driving parameters, which can be also used as a resource in realising a specific flux value by adjusting the initial time in experiments.

%%%%%%%%%%%%%%%%%%%%%%%%%%%%%%%%%%%%%%%%%%%%%
\section{Relative phase and the induced plaquette flux}
\label{Sec:2DZeroRelativePhaseRemovesPlaquetteFlux}

We here analyse some relations between the relative phase $\varphi$ between the local drives applied to the two sites in our triangle plaquette and the induced flux values $\Phi$. The imaginary part of the renormalised tunnelling element $J^0_{23}$ is given by
\begin{multline}
\mathrm{Imag}\left(J^0_{23}\right) = \frac{J}{T} \int_{0}^{T} \mathrm{d} t \sin \left( \frac{A_3}{ \beta \omega_0} \sin (\beta \omega_0 t + \varphi )  \right. \\
- \left. \frac{A_2}{ \alpha \omega_0} \sin (\alpha \omega_0 t)  \right) .
\label{Eq:AppImagJ23}
\end{multline}
Certain symmetries of the driving function, such as the (time-)reflection symmetry within a period~\cite{eckardt_colloquium_2017}, can force this imaginary part to vanish. We can demonstrate this by analysing Eq.(\ref{Eq:AppImagJ23}) with respect to a time instance $\tau \in [0,T]$.  We separate the integral into two parts from $\int_0^{\tau}$ and $\int_{\tau}^{T}$ and make a change of variables $t'=-t +T$ in the second term giving
\begin{multline}
     \mathrm{Imag}(J^0_{23}) = \frac{J}{T} \int_{0}^{\tau}  \mathrm{d}t  \sin \left( \frac{A_3}{ \beta \omega_0} \sin (\beta \omega_0 t + \varphi ) \right. \\
- \left. \frac{A_2}{ \alpha \omega_0} \sin (\alpha \omega_0 t) \right) \\
- \frac{J}{T} \int_{0}^{T - \tau}   \mathrm{d}t'  \sin \bigg( \frac{A_3}{ \beta \omega_0} \sin (\beta \omega_0 t' - \varphi ) \\
- \bigg. \frac{A_2}{ \alpha \omega_0} \sin (\alpha \omega_0 t') \bigg). 
\label{Eq:ImagJ23TwoTermsTau}
\end{multline}
For $\varphi \in \{ 0, \pi \}$ %and $\alpha \neq \beta$, 
this integral (\ref{Eq:ImagJ23TwoTermsTau}) vanishes for $\tau=T/2$ as the driving function is odd. We therefore require a finite phase shift $\varphi\notin\{0,\pi\}$ to obtain a complex tunneling amplitude contributing to a finite flux in Section \ref{SSec:ControlOverFlux2D}.

Secondly, we show that $J_{23}^0$ is purely real, given by a Bessel function, when $\alpha=\beta$ for any $\varphi$. In this case, setting $\omega = \alpha \omega_0 = \beta \omega_0$, and using a double angle formula, we can express $J_{23}^0$ given by Eq.~(\ref{Eq:TriangleJ23Renormalised}) as
\begin{equation}
J^0_{23}= \frac{J}{T} \int_{-T/2}^{T/2} \mathrm{d} t  \> \mathrm{e}^{i \left( \left[ \frac{A_3}{\omega} \cos \varphi - \frac{A_2}{\omega} \right] \sin (\omega t) +  \frac{A_3}{ \omega} \sin \varphi \cos (\omega t)  \right)}.
\label{Eq:AppImagJ23alpha=beta}
\end{equation}
Using that any sum of trigonometric functions $(a \sin x + b \cos x)$ can be written as $R \sin (x + \gamma)$ where $R = \sqrt{a^2 + b^2}$ and $\tan \gamma = b/a$, we can rewrite Eq.~(\ref{Eq:AppImagJ23alpha=beta}) as
\begin{equation}
  J^0_{23} = \frac{J}{T} \int_{-T/2}^{T/2}  \mathrm{d}t \>  \mathrm{e}^{ i  R \sin (\omega t + \gamma )} = J \mathcal{J}_0 \left( R \right),
\end{equation}
  where $R = \frac{1}{\omega_0} \sqrt{A_2^2 + A_3^2 - 2 A_2 A_3 \cos \varphi }$ and $\tan \gamma = (A_3 \cos \varphi - A_2) / A_3 \sin \varphi$.

%Using a trigonometric double-angle formula, Eq.~(\ref{Eq:AppImagJ23}) is proportional to 
%\begin{multline}
%      \int_{-T/2}^{T/2} \mathrm{d}t \sin \left( \frac{A_3}{ \beta \omega_0} \sin (\beta \omega_0 t) \right) \cos \left( \frac{A_2}{\alpha \omega_0} \sin (\alpha \omega_0 t) \right)  \\
%  - \cos \left( \frac{A_3}{ \beta \omega_0} \sin (\beta \omega_0 t) \right) \sin \left( \frac{A_2}{ \alpha \omega_0} \sin (\alpha \omega_0 t) \right) .  \label{EqApp:ImaginaryJ23}
%\end{multline}
%Equation~(\ref{EqApp:ImaginaryJ23}) is necessarily zero given its integrand is an odd function.

%\section{Reflection of $\xi$ about zero when $\varphi \rightarrow -\varphi $} 
\label{App:ReflectionofXiGivenReflectionofVarphi}

Thirdly, we study the reflection symmetry of renormalised tunneling parameters upon reversing the sign of the relative phase, $\varphi \rightarrow -\varphi $.
The phase of a complex number $z=C e^{i \xi}$ is given by $\xi = \tan^{-1} (\mathrm{Im}(z)/ \mathrm{Re}(z))$. Utilising trigonometric angle sum identities and the fact that the integrand is an odd function around zero, we find that the phase of  $J^0_{23}$ given by Eq.~(\ref{Eq:TriangleJ23Renormalised}) is
% \begin{equation}
% \resizebox{.45\textwidth}{!} {
% \xi_{J^0_{23}} = \tan^{-1} \left(\frac{\int_{T/2}^{T/2} \sin \left( \frac{A_3}{\beta \omega_0} \sin(\beta \omega_0 t + \varphi) - \frac{A_2}{\alpha \omega_0} \sin (\alpha \omega_0 t)\right)}{\int_{T/2}^{T/2} \cos \left( \frac{A_3}{\beta \omega_0} \sin(\beta \omega_0 t + \varphi) - \frac{A_2}{\alpha \omega_0} \sin (\alpha \omega_0 t)\right)}\right)}
% \end{equation}

% Using trigonometric angle sum and difference identities, and utilising that the integral of an odd function across an  interval symmetric around zero vanishes, we find that
\begin{widetext}
\begin{equation}
    \xi_{J^0_{23}} = \tan^{-1} \left(\frac{\int_{-T/2}^{T/2} \sin \left( \frac{A_3}{\beta \omega_0} \sin(\varphi) \cos (\beta \omega_0 t) \right) \cos \left( \frac{A_3}{\beta \omega} \sin (\beta \omega t) \cos (\varphi ) \right)  \cos \left( \frac{A_2}{\alpha \omega_0} \sin(\alpha \omega_0 t) \right)}{\int_{-T/2}^{T/2} \cos \left( \frac{A_3}{\beta \omega_0} \cos(\varphi) \sin (\beta \omega_0 t) \right) \cos \left( \frac{A_3}{\beta \omega} \cos (\beta \omega t) \sin (\varphi ) \right)  \cos \left( \frac{A_2}{\alpha \omega_0} \sin(\alpha \omega_0 t) \right)}\right). \label{EqApp:Tan}
\end{equation}
\end{widetext}
Sending $\varphi \rightarrow -\varphi$  then changes the sign of the argument of the inverse tangent function in Eq.~(\ref{EqApp:Tan}) and thereby also the sign of $\xi_{J_{23}}$. The other two tunnelling elements $J_{12}$ and $J_{31}$ are unaffected, therefore, the sign of the total plaquette flux $\Phi$ is inverted. 

%\section{Numerical Optimisation Routine} // or footnote \label{App:NumericalOptimisationRoutine}
%We utilise the "FindMinimum" function from Mathematica to minimise the cost function in Eq.~(\ref{Eq:CostFunc}). We set initial conditions based on the butterfly. 

\section{Tuning flux values for the fixed relative tunnelling ratio  \texorpdfstring{$(x,y) = (0.9,0.9)$}{Lg}}
\label{App:Neighbourhood(0.9,0.9)}

%Similar to Section \ref{SSec:ControlOverFlux2D}, we here demonstrate how to tune the flux values from $-\pi$ to $\pi$ while maintaining a tunnelling ratio $(x^t,y^t)=(0.9,0.9)$.  
We here demonstrate another example illustrating how the plaquette flux can be tuned for the target tunnelling ratio $(x^t,y^t)=(0.9,0.9)$.
%results of the numerical minimisation routine, detailing how $A_2$, $A_3$ and $\varphi$ values can be tuned to achieve various flux values while fixing $\omega_0=8$.
%This example captures an instance where the plaquette flux is zero for some $\varphi \notin \{0, \pi \}$. 
The target flux $\Phi^t$ can be continuously changed while keeping the tunnelling ratios constant. %As is shown here, there is a discontinuity in $\varphi$ as the target plaquette flux is tuned across zero. 
\begin{figure}
    \includegraphics[]{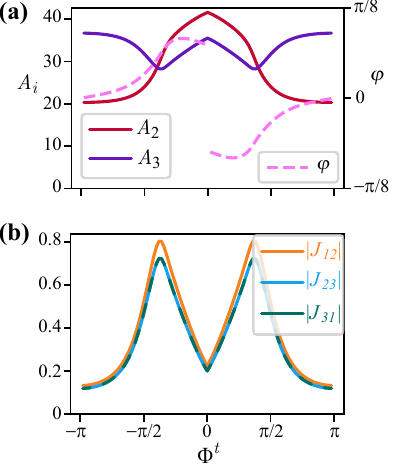}
    \caption{(a) Driving parameters $A_2$, $A_3$ and $\varphi$ used to achieve a target tunnelling ratio $(x^t,y^t)=(0.9,0.9)$ and flux $\Phi^t \in [-\pi, \pi)$. (b) The magnitude of the three renormalised tunnelling elements as $\Phi^t$ is tuned to various values. Their ratio remains constant at $(x^t,y^t)=(0.9,0.9)$ . }
    \label{Fig:TuningFluxVals(x,y)=(0.9,0.9)}
\end{figure}

\section{Peierls phases from local static offsets}
\label{Appx:GradientFluxCalculations}
Here, we show how renormalisation factors in Eq.~\eqref{Eq:LocalDrivewStaticOffsetRenormFactor} are generated by driving sites via Eq.~\eqref{Eq:LocalDriveFuncWithStaticOffset}. Similarly to Sec.~\ref{Sec:FloquetEngineering}, the Hamiltonian in the lattice frame takes the form of Eq.~\eqref{Eq:HTildeGeneral} with Peierls phases
\begin{align}
    \phi_{ij} (t) &= - \int_{t_0}^t \left[ W_j(\tau) - W_i(\tau) \right] d \tau - \phi_{ij}^0 \\
    &= -\frac{A}{\omega} \left[ \sin (\omega t + \varphi_j) - \sin(\omega t + \varphi_i )\right] + \nu_{ij} \omega t 
\end{align}
where $\nu_{ij} = \nu_i - \nu_j$ and we set the gauge constant $\phi_{ij}^0 =  -\nu_{ij} \omega t_0 + \frac{A}{\omega} \left[ \sin(\omega t_0 + \varphi_j) - \sin(\omega t_0 + \varphi_i) \right]$. Using the first term of the high-frequency expansion (see Sec.~\ref{SSec:DrivingASingleSite}), the effective Hamiltonian is given by Eq.~\eqref{Eq:EffectiveGFramework}  with
% \begin{align}
%     \epsilon_{ij}^{\mathrm{eff}} &= \frac{1}{T} \int_0^T e^{i (\phi_j(t) - \phi_i (t))} dt \\
%     &= \frac{J}{T}\int_0^T e^{i \left[ -\frac{A}{\omega} \sin (\omega t + \varphi_j)  + \frac{A}{\omega} \sin (\omega t + \varphi_i) + \nu_{ij} \omega t \right]} dt  
% \end{align}
% where . Merging the sin functions 
\begin{equation}
    \epsilon_{ij} = \frac{1}{T}\int_{0}^{ T} e^{i \left[ \frac{-A_{ij}}{\omega} \sin (\omega t + \xi_{ij} )  + \nu_{ij} \omega t  \right]} dt,
\end{equation}
where $ A_{ij} = 2 A \sin \left( \frac{\varphi_i - \varphi_j}{2}\right)$ and $\xi_{ij} = \frac{\varphi_j + \varphi_i - \pi}{2} $. Making a change of variables $t = t' - \delta$ where $\delta \omega = \xi_{ij}$, 
\begin{equation}
     \epsilon_{ij}= \frac{J}{T}\int_{\delta }^{\delta + T} e^{i \left[ \frac{-A_{ij}}{\omega} \sin (\omega t') + \nu_{ij} \omega t'\right]}  dt' \> e^{-i \nu_{ij} \xi_{ij}}.
\end{equation}
This amounts to the integral representation of a Bessel function multiplied by a phase factor giving
\begin{equation}
     \epsilon_{ij} =  \mathcal{J}_{\nu_{ij}} \left( \frac{A_{ij} }{\omega} \right) e^{-i \nu_{ij} \xi_{ij} }.
\end{equation}

\clearpage
\newpage
\bibliography{Paper}

%apsrev4-2.bst 2019-01-14 (MD) hand-edited version of apsrev4-1.bst
%Control: key (0)
%Control: author (72) initials jnrlst
%Control: editor formatted (1) identically to author
%Control: production of article title (-1) disabled
%Control: page (0) single
%Control: year (1) truncated
%Control: production of eprint (0) enabled
\begin{thebibliography}{93}%
\makeatletter
\providecommand \@ifxundefined [1]{%
 \@ifx{#1\undefined}
}%
\providecommand \@ifnum [1]{%
 \ifnum #1\expandafter \@firstoftwo
 \else \expandafter \@secondoftwo
 \fi
}%
\providecommand \@ifx [1]{%
 \ifx #1\expandafter \@firstoftwo
 \else \expandafter \@secondoftwo
 \fi
}%
\providecommand \natexlab [1]{#1}%
\providecommand \enquote  [1]{``#1''}%
\providecommand \bibnamefont  [1]{#1}%
\providecommand \bibfnamefont [1]{#1}%
\providecommand \citenamefont [1]{#1}%
\providecommand \href@noop [0]{\@secondoftwo}%
\providecommand \href [0]{\begingroup \@sanitize@url \@href}%
\providecommand \@href[1]{\@@startlink{#1}\@@href}%
\providecommand \@@href[1]{\endgroup#1\@@endlink}%
\providecommand \@sanitize@url [0]{\catcode `\\12\catcode `\$12\catcode
  `\&12\catcode `\#12\catcode `\^12\catcode `\_12\catcode `\%12\relax}%
\providecommand \@@startlink[1]{}%
\providecommand \@@endlink[0]{}%
\providecommand \url  [0]{\begingroup\@sanitize@url \@url }%
\providecommand \@url [1]{\endgroup\@href {#1}{\urlprefix }}%
\providecommand \urlprefix  [0]{URL }%
\providecommand \Eprint [0]{\href }%
\providecommand \doibase [0]{https://doi.org/}%
\providecommand \selectlanguage [0]{\@gobble}%
\providecommand \bibinfo  [0]{\@secondoftwo}%
\providecommand \bibfield  [0]{\@secondoftwo}%
\providecommand \translation [1]{[#1]}%
\providecommand \BibitemOpen [0]{}%
\providecommand \bibitemStop [0]{}%
\providecommand \bibitemNoStop [0]{.\EOS\space}%
\providecommand \EOS [0]{\spacefactor3000\relax}%
\providecommand \BibitemShut  [1]{\csname bibitem#1\endcsname}%
\let\auto@bib@innerbib\@empty
%</preamble>
\bibitem [{\citenamefont {Cirac}\ and\ \citenamefont
  {Zoller}(2012)}]{Cirac-Zoller-GoalsOpportunitiesQuantum-2012}%
  \BibitemOpen
  \bibfield  {author} {\bibinfo {author} {\bibfnamefont {J.~I.}\ \bibnamefont
  {Cirac}}\ and\ \bibinfo {author} {\bibfnamefont {P.}~\bibnamefont {Zoller}},\
  }\href {https://doi.org/10.1038/nphys2275} {\bibfield  {journal} {\bibinfo
  {journal} {Nature Physics}\ }\textbf {\bibinfo {volume} {8}},\ \bibinfo
  {pages} {264} (\bibinfo {year} {2012})}\BibitemShut {NoStop}%
\bibitem [{\citenamefont {Georgescu}\ \emph {et~al.}(2014)\citenamefont
  {Georgescu}, \citenamefont {Ashhab},\ and\ \citenamefont
  {Nori}}]{Georgescu14_RMP}%
  \BibitemOpen
  \bibfield  {author} {\bibinfo {author} {\bibfnamefont {I.~M.}\ \bibnamefont
  {Georgescu}}, \bibinfo {author} {\bibfnamefont {S.}~\bibnamefont {Ashhab}},\
  and\ \bibinfo {author} {\bibfnamefont {F.}~\bibnamefont {Nori}},\ }\href
  {https://doi.org/10.1103/RevModPhys.86.153} {\bibfield  {journal} {\bibinfo
  {journal} {Reviews of Modern Physics}\ }\textbf {\bibinfo {volume} {86}},\
  \bibinfo {pages} {153} (\bibinfo {year} {2014})}\BibitemShut {NoStop}%
\bibitem [{\citenamefont {Daley}\ \emph {et~al.}(2022)\citenamefont {Daley},
  \citenamefont {Bloch}, \citenamefont {Kokail}, \citenamefont {Flannigan},
  \citenamefont {Pearson}, \citenamefont {Troyer},\ and\ \citenamefont
  {Zoller}}]{Daley-Zoller-PracticalQuantumAdvantage-2022}%
  \BibitemOpen
  \bibfield  {author} {\bibinfo {author} {\bibfnamefont {A.~J.}\ \bibnamefont
  {Daley}}, \bibinfo {author} {\bibfnamefont {I.}~\bibnamefont {Bloch}},
  \bibinfo {author} {\bibfnamefont {C.}~\bibnamefont {Kokail}}, \bibinfo
  {author} {\bibfnamefont {S.}~\bibnamefont {Flannigan}}, \bibinfo {author}
  {\bibfnamefont {N.}~\bibnamefont {Pearson}}, \bibinfo {author} {\bibfnamefont
  {M.}~\bibnamefont {Troyer}},\ and\ \bibinfo {author} {\bibfnamefont
  {P.}~\bibnamefont {Zoller}},\ }\href
  {https://doi.org/10.1038/s41586-022-04940-6} {\bibfield  {journal} {\bibinfo
  {journal} {Nature}\ }\textbf {\bibinfo {volume} {607}},\ \bibinfo {pages}
  {667} (\bibinfo {year} {2022})}\BibitemShut {NoStop}%
\bibitem [{\citenamefont {Bloch}(2005)}]{Bloch-UltracoldQuantumGases-2005}%
  \BibitemOpen
  \bibfield  {author} {\bibinfo {author} {\bibfnamefont {I.}~\bibnamefont
  {Bloch}},\ }\href {https://doi.org/10.1038/nphys138} {\bibfield  {journal}
  {\bibinfo  {journal} {Nature Physics}\ }\textbf {\bibinfo {volume} {1}},\
  \bibinfo {pages} {23} (\bibinfo {year} {2005})}\BibitemShut {NoStop}%
\bibitem [{\citenamefont {Bloch}\ \emph {et~al.}(2008)\citenamefont {Bloch},
  \citenamefont {Dalibard},\ and\ \citenamefont
  {Zwerger}}]{Bloch-Zwerger-ManybodyPhysicsUltracold-2008}%
  \BibitemOpen
  \bibfield  {author} {\bibinfo {author} {\bibfnamefont {I.}~\bibnamefont
  {Bloch}}, \bibinfo {author} {\bibfnamefont {J.}~\bibnamefont {Dalibard}},\
  and\ \bibinfo {author} {\bibfnamefont {W.}~\bibnamefont {Zwerger}},\ }\href
  {https://doi.org/10.1103/RevModPhys.80.885} {\bibfield  {journal} {\bibinfo
  {journal} {Reviews of Modern Physics}\ }\textbf {\bibinfo {volume} {80}},\
  \bibinfo {pages} {885} (\bibinfo {year} {2008})}\BibitemShut {NoStop}%
\bibitem [{\citenamefont {Sch{\"a}fer}\ \emph {et~al.}(2020)\citenamefont
  {Sch{\"a}fer}, \citenamefont {Fukuhara}, \citenamefont {Sugawa},
  \citenamefont {Takasu},\ and\ \citenamefont
  {Takahashi}}]{Schafer-Takahashi-ToolsQuantumSimulation-2020}%
  \BibitemOpen
  \bibfield  {author} {\bibinfo {author} {\bibfnamefont {F.}~\bibnamefont
  {Sch{\"a}fer}}, \bibinfo {author} {\bibfnamefont {T.}~\bibnamefont
  {Fukuhara}}, \bibinfo {author} {\bibfnamefont {S.}~\bibnamefont {Sugawa}},
  \bibinfo {author} {\bibfnamefont {Y.}~\bibnamefont {Takasu}},\ and\ \bibinfo
  {author} {\bibfnamefont {Y.}~\bibnamefont {Takahashi}},\ }\href
  {https://doi.org/10.1038/s42254-020-0195-3} {\bibfield  {journal} {\bibinfo
  {journal} {Nature Reviews Physics}\ }\textbf {\bibinfo {volume} {2}},\
  \bibinfo {pages} {411} (\bibinfo {year} {2020})}\BibitemShut {NoStop}%
\bibitem [{\citenamefont {Gross}\ and\ \citenamefont
  {Bloch}(2017)}]{Gross-Bloch-QuantumSimulationsUltracold-2017}%
  \BibitemOpen
  \bibfield  {author} {\bibinfo {author} {\bibfnamefont {C.}~\bibnamefont
  {Gross}}\ and\ \bibinfo {author} {\bibfnamefont {I.}~\bibnamefont {Bloch}},\
  }\href {https://doi.org/10.1126/science.aal3837} {\bibfield  {journal}
  {\bibinfo  {journal} {Science}\ }\textbf {\bibinfo {volume} {357}},\ \bibinfo
  {pages} {995} (\bibinfo {year} {2017})}\BibitemShut {NoStop}%
\bibitem [{\citenamefont {Greiner}\ \emph {et~al.}(2002)\citenamefont
  {Greiner}, \citenamefont {Mandel}, \citenamefont {Esslinger}, \citenamefont
  {Hänsch},\ and\ \citenamefont {Bloch}}]{greiner_quantum_2002}%
  \BibitemOpen
  \bibfield  {author} {\bibinfo {author} {\bibfnamefont {M.}~\bibnamefont
  {Greiner}}, \bibinfo {author} {\bibfnamefont {O.}~\bibnamefont {Mandel}},
  \bibinfo {author} {\bibfnamefont {T.}~\bibnamefont {Esslinger}}, \bibinfo
  {author} {\bibfnamefont {T.~W.}\ \bibnamefont {Hänsch}},\ and\ \bibinfo
  {author} {\bibfnamefont {I.}~\bibnamefont {Bloch}},\ }\href
  {https://doi.org/10.1038/415039a} {\bibfield  {journal} {\bibinfo  {journal}
  {Nature}\ }\textbf {\bibinfo {volume} {415}},\ \bibinfo {pages} {39}
  (\bibinfo {year} {2002})}\BibitemShut {NoStop}%
\bibitem [{\citenamefont {Struck}\ \emph {et~al.}(2011)\citenamefont {Struck},
  \citenamefont {Ölschläger}, \citenamefont {Le~Targat}, \citenamefont
  {Soltan-Panahi}, \citenamefont {Eckardt}, \citenamefont {Lewenstein},
  \citenamefont {Windpassinger},\ and\ \citenamefont
  {Sengstock}}]{struck_quantum_2011}%
  \BibitemOpen
  \bibfield  {author} {\bibinfo {author} {\bibfnamefont {J.}~\bibnamefont
  {Struck}}, \bibinfo {author} {\bibfnamefont {C.}~\bibnamefont
  {Ölschläger}}, \bibinfo {author} {\bibfnamefont {R.}~\bibnamefont
  {Le~Targat}}, \bibinfo {author} {\bibfnamefont {P.}~\bibnamefont
  {Soltan-Panahi}}, \bibinfo {author} {\bibfnamefont {A.}~\bibnamefont
  {Eckardt}}, \bibinfo {author} {\bibfnamefont {M.}~\bibnamefont {Lewenstein}},
  \bibinfo {author} {\bibfnamefont {P.}~\bibnamefont {Windpassinger}},\ and\
  \bibinfo {author} {\bibfnamefont {K.}~\bibnamefont {Sengstock}},\ }\href
  {https://doi.org/10.1126/science.1207239} {\bibfield  {journal} {\bibinfo
  {journal} {Science}\ }\textbf {\bibinfo {volume} {333}},\ \bibinfo {pages}
  {996} (\bibinfo {year} {2011})}\BibitemShut {NoStop}%
\bibitem [{\citenamefont {Simon}\ \emph {et~al.}(2011)\citenamefont {Simon},
  \citenamefont {Bakr}, \citenamefont {Ma}, \citenamefont {Tai}, \citenamefont
  {Preiss},\ and\ \citenamefont {Greiner}}]{simon_quantum_2011}%
  \BibitemOpen
  \bibfield  {author} {\bibinfo {author} {\bibfnamefont {J.}~\bibnamefont
  {Simon}}, \bibinfo {author} {\bibfnamefont {W.~S.}\ \bibnamefont {Bakr}},
  \bibinfo {author} {\bibfnamefont {R.}~\bibnamefont {Ma}}, \bibinfo {author}
  {\bibfnamefont {M.~E.}\ \bibnamefont {Tai}}, \bibinfo {author} {\bibfnamefont
  {P.~M.}\ \bibnamefont {Preiss}},\ and\ \bibinfo {author} {\bibfnamefont
  {M.}~\bibnamefont {Greiner}},\ }\href {https://doi.org/10.1038/nature09994}
  {\bibfield  {journal} {\bibinfo  {journal} {Nature}\ }\textbf {\bibinfo
  {volume} {472}},\ \bibinfo {pages} {307} (\bibinfo {year}
  {2011})}\BibitemShut {NoStop}%
\bibitem [{\citenamefont {Jotzu}\ \emph {et~al.}(2014)\citenamefont {Jotzu},
  \citenamefont {Messer}, \citenamefont {Desbuquois}, \citenamefont {Lebrat},
  \citenamefont {Uehlinger}, \citenamefont {Greif},\ and\ \citenamefont
  {Esslinger}}]{Jotzu14_Nat}%
  \BibitemOpen
  \bibfield  {author} {\bibinfo {author} {\bibfnamefont {G.}~\bibnamefont
  {Jotzu}}, \bibinfo {author} {\bibfnamefont {M.}~\bibnamefont {Messer}},
  \bibinfo {author} {\bibfnamefont {R.}~\bibnamefont {Desbuquois}}, \bibinfo
  {author} {\bibfnamefont {M.}~\bibnamefont {Lebrat}}, \bibinfo {author}
  {\bibfnamefont {T.}~\bibnamefont {Uehlinger}}, \bibinfo {author}
  {\bibfnamefont {D.}~\bibnamefont {Greif}},\ and\ \bibinfo {author}
  {\bibfnamefont {T.}~\bibnamefont {Esslinger}},\ }\href
  {https://doi.org/10.1038/nature13915} {\bibfield  {journal} {\bibinfo
  {journal} {Nature}\ }\textbf {\bibinfo {volume} {515}},\ \bibinfo {pages}
  {237} (\bibinfo {year} {2014})}\BibitemShut {NoStop}%
\bibitem [{\citenamefont {Glaetzle}\ \emph {et~al.}(2015)\citenamefont
  {Glaetzle}, \citenamefont {Dalmonte}, \citenamefont {Nath}, \citenamefont
  {Gross}, \citenamefont {Bloch},\ and\ \citenamefont
  {Zoller}}]{glaetzle_designing_2015}%
  \BibitemOpen
  \bibfield  {author} {\bibinfo {author} {\bibfnamefont {A.~W.}\ \bibnamefont
  {Glaetzle}}, \bibinfo {author} {\bibfnamefont {M.}~\bibnamefont {Dalmonte}},
  \bibinfo {author} {\bibfnamefont {R.}~\bibnamefont {Nath}}, \bibinfo {author}
  {\bibfnamefont {C.}~\bibnamefont {Gross}}, \bibinfo {author} {\bibfnamefont
  {I.}~\bibnamefont {Bloch}},\ and\ \bibinfo {author} {\bibfnamefont
  {P.}~\bibnamefont {Zoller}},\ }\href
  {https://doi.org/10.1103/PhysRevLett.114.173002} {\bibfield  {journal}
  {\bibinfo  {journal} {Physical Review Letters}\ }\textbf {\bibinfo {volume}
  {114}},\ \bibinfo {pages} {173002} (\bibinfo {year} {2015})}\BibitemShut
  {NoStop}%
\bibitem [{\citenamefont {Mazurenko}\ \emph {et~al.}(2017)\citenamefont
  {Mazurenko}, \citenamefont {Chiu}, \citenamefont {Ji}, \citenamefont
  {Parsons}, \citenamefont {{Kan{\'a}sz-Nagy}}, \citenamefont {Schmidt},
  \citenamefont {Grusdt}, \citenamefont {Demler}, \citenamefont {Greif},\ and\
  \citenamefont {Greiner}}]{Mazurenko-Greiner-ColdatomFermiHubbard-2017}%
  \BibitemOpen
  \bibfield  {author} {\bibinfo {author} {\bibfnamefont {A.}~\bibnamefont
  {Mazurenko}}, \bibinfo {author} {\bibfnamefont {C.~S.}\ \bibnamefont {Chiu}},
  \bibinfo {author} {\bibfnamefont {G.}~\bibnamefont {Ji}}, \bibinfo {author}
  {\bibfnamefont {M.~F.}\ \bibnamefont {Parsons}}, \bibinfo {author}
  {\bibfnamefont {M.}~\bibnamefont {{Kan{\'a}sz-Nagy}}}, \bibinfo {author}
  {\bibfnamefont {R.}~\bibnamefont {Schmidt}}, \bibinfo {author} {\bibfnamefont
  {F.}~\bibnamefont {Grusdt}}, \bibinfo {author} {\bibfnamefont
  {E.}~\bibnamefont {Demler}}, \bibinfo {author} {\bibfnamefont
  {D.}~\bibnamefont {Greif}},\ and\ \bibinfo {author} {\bibfnamefont
  {M.}~\bibnamefont {Greiner}},\ }\href {https://doi.org/10.1038/nature22362}
  {\bibfield  {journal} {\bibinfo  {journal} {Nature}\ }\textbf {\bibinfo
  {volume} {545}},\ \bibinfo {pages} {462} (\bibinfo {year}
  {2017})}\BibitemShut {NoStop}%
\bibitem [{\citenamefont {Asteria}\ \emph {et~al.}(2019)\citenamefont
  {Asteria}, \citenamefont {Tran}, \citenamefont {Ozawa}, \citenamefont
  {Tarnowski}, \citenamefont {Rem}, \citenamefont {Flashner}, \citenamefont
  {Sengstock}, \citenamefont {Goldman},\ and\ \citenamefont
  {Weitenberg}}]{Asteria19_NatPhys}%
  \BibitemOpen
  \bibfield  {author} {\bibinfo {author} {\bibfnamefont {L.}~\bibnamefont
  {Asteria}}, \bibinfo {author} {\bibfnamefont {D.}~\bibnamefont {Tran}},
  \bibinfo {author} {\bibfnamefont {T.}~\bibnamefont {Ozawa}}, \bibinfo
  {author} {\bibfnamefont {M.}~\bibnamefont {Tarnowski}}, \bibinfo {author}
  {\bibfnamefont {B.~S.}\ \bibnamefont {Rem}}, \bibinfo {author} {\bibfnamefont
  {N.}~\bibnamefont {Flashner}}, \bibinfo {author} {\bibfnamefont
  {K.}~\bibnamefont {Sengstock}}, \bibinfo {author} {\bibfnamefont
  {N.}~\bibnamefont {Goldman}},\ and\ \bibinfo {author} {\bibfnamefont
  {C.}~\bibnamefont {Weitenberg}},\ }\href
  {https://doi.org/10.1038/s41567-019-0417-8} {\bibfield  {journal} {\bibinfo
  {journal} {Nature Physics}\ }\textbf {\bibinfo {volume} {15}},\ \bibinfo
  {pages} {449} (\bibinfo {year} {2019})}\BibitemShut {NoStop}%
\bibitem [{\citenamefont {Wintersperger}\ \emph {et~al.}(2020)\citenamefont
  {Wintersperger}, \citenamefont {Braun}, \citenamefont {{\"U}nal},
  \citenamefont {Eckardt}, \citenamefont {Liberto}, \citenamefont {Goldman},
  \citenamefont {Bloch},\ and\ \citenamefont
  {Aidelsburger}}]{Wintersperger20_NatPhys}%
  \BibitemOpen
  \bibfield  {author} {\bibinfo {author} {\bibfnamefont {K.}~\bibnamefont
  {Wintersperger}}, \bibinfo {author} {\bibfnamefont {C.}~\bibnamefont
  {Braun}}, \bibinfo {author} {\bibfnamefont {F.~N.}\ \bibnamefont {{\"U}nal}},
  \bibinfo {author} {\bibfnamefont {A.}~\bibnamefont {Eckardt}}, \bibinfo
  {author} {\bibfnamefont {M.~D.}\ \bibnamefont {Liberto}}, \bibinfo {author}
  {\bibfnamefont {N.}~\bibnamefont {Goldman}}, \bibinfo {author} {\bibfnamefont
  {I.}~\bibnamefont {Bloch}},\ and\ \bibinfo {author} {\bibfnamefont
  {M.}~\bibnamefont {Aidelsburger}},\ }\href
  {https://doi.org/10.1038/s41567-020-0949-y} {\bibfield  {journal} {\bibinfo
  {journal} {Nature Physics}\ }\textbf {\bibinfo {volume} {16}},\ \bibinfo
  {pages} {1058} (\bibinfo {year} {2020})}\BibitemShut {NoStop}%
\bibitem [{\citenamefont {\"Unal}\ \emph {et~al.}(2020)\citenamefont {\"Unal},
  \citenamefont {Bouhon},\ and\ \citenamefont {Slager}}]{Unal20_PRL}%
  \BibitemOpen
  \bibfield  {author} {\bibinfo {author} {\bibfnamefont {F.~N.}\ \bibnamefont
  {\"Unal}}, \bibinfo {author} {\bibfnamefont {A.}~\bibnamefont {Bouhon}},\
  and\ \bibinfo {author} {\bibfnamefont {R.-J.}\ \bibnamefont {Slager}},\
  }\href {https://doi.org/10.1103/PhysRevLett.125.053601} {\bibfield  {journal}
  {\bibinfo  {journal} {Physical Review Letters}\ }\textbf {\bibinfo {volume}
  {125}},\ \bibinfo {pages} {053601} (\bibinfo {year} {2020})}\BibitemShut
  {NoStop}%
\bibitem [{\citenamefont {Zhao}\ \emph {et~al.}(2022)\citenamefont {Zhao},
  \citenamefont {Yang}, \citenamefont {Jiang}, \citenamefont {Mao},
  \citenamefont {Guo}, \citenamefont {Qiu}, \citenamefont {Wang}, \citenamefont
  {Yao}, \citenamefont {He}, \citenamefont {Zhou}, \citenamefont {Xu},\ and\
  \citenamefont {Duan}}]{Zhao22_CommPhys_EuQuench}%
  \BibitemOpen
  \bibfield  {author} {\bibinfo {author} {\bibfnamefont {W.}~\bibnamefont
  {Zhao}}, \bibinfo {author} {\bibfnamefont {Y.-B.}\ \bibnamefont {Yang}},
  \bibinfo {author} {\bibfnamefont {Y.}~\bibnamefont {Jiang}}, \bibinfo
  {author} {\bibfnamefont {Z.}~\bibnamefont {Mao}}, \bibinfo {author}
  {\bibfnamefont {W.}~\bibnamefont {Guo}}, \bibinfo {author} {\bibfnamefont
  {L.}~\bibnamefont {Qiu}}, \bibinfo {author} {\bibfnamefont {G.}~\bibnamefont
  {Wang}}, \bibinfo {author} {\bibfnamefont {L.}~\bibnamefont {Yao}}, \bibinfo
  {author} {\bibfnamefont {L.}~\bibnamefont {He}}, \bibinfo {author}
  {\bibfnamefont {Z.}~\bibnamefont {Zhou}}, \bibinfo {author} {\bibfnamefont
  {Y.}~\bibnamefont {Xu}},\ and\ \bibinfo {author} {\bibfnamefont
  {L.}~\bibnamefont {Duan}},\ }\href
  {https://doi.org/10.1038/s42005-022-01001-2} {\bibfield  {journal} {\bibinfo
  {journal} {Communications Physics}\ }\textbf {\bibinfo {volume} {5}},\
  \bibinfo {pages} {223} (\bibinfo {year} {2022})}\BibitemShut {NoStop}%
\bibitem [{\citenamefont {Deissler}\ \emph {et~al.}(2010)\citenamefont
  {Deissler}, \citenamefont {Zaccanti}, \citenamefont {Roati}, \citenamefont
  {D’Errico}, \citenamefont {Fattori}, \citenamefont {Modugno}, \citenamefont
  {Modugno},\ and\ \citenamefont {Inguscio}}]{Deissler10_NatPhys}%
  \BibitemOpen
  \bibfield  {author} {\bibinfo {author} {\bibfnamefont {B.}~\bibnamefont
  {Deissler}}, \bibinfo {author} {\bibfnamefont {M.}~\bibnamefont {Zaccanti}},
  \bibinfo {author} {\bibfnamefont {G.}~\bibnamefont {Roati}}, \bibinfo
  {author} {\bibfnamefont {C.}~\bibnamefont {D’Errico}}, \bibinfo {author}
  {\bibfnamefont {M.}~\bibnamefont {Fattori}}, \bibinfo {author} {\bibfnamefont
  {M.}~\bibnamefont {Modugno}}, \bibinfo {author} {\bibfnamefont
  {G.}~\bibnamefont {Modugno}},\ and\ \bibinfo {author} {\bibfnamefont
  {M.}~\bibnamefont {Inguscio}},\ }\href
  {https://doi.org/https://doi.org/10.1038/nphys1635} {\bibfield  {journal}
  {\bibinfo  {journal} {Nature Physics}\ }\textbf {\bibinfo {volume} {6}},\
  \bibinfo {pages} {354} (\bibinfo {year} {2010})}\BibitemShut {NoStop}%
\bibitem [{\citenamefont {Schreiber}\ \emph {et~al.}(2015)\citenamefont
  {Schreiber}, \citenamefont {Hodgman}, \citenamefont {Bordia}, \citenamefont
  {L{\"u}schen}, \citenamefont {Fischer}, \citenamefont {Vosk}, \citenamefont
  {Altman}, \citenamefont {Schneider},\ and\ \citenamefont
  {Bloch}}]{Schreiber-Bloch-ObservationManybodyLocalization-2015}%
  \BibitemOpen
  \bibfield  {author} {\bibinfo {author} {\bibfnamefont {M.}~\bibnamefont
  {Schreiber}}, \bibinfo {author} {\bibfnamefont {S.~S.}\ \bibnamefont
  {Hodgman}}, \bibinfo {author} {\bibfnamefont {P.}~\bibnamefont {Bordia}},
  \bibinfo {author} {\bibfnamefont {H.~P.}\ \bibnamefont {L{\"u}schen}},
  \bibinfo {author} {\bibfnamefont {M.~H.}\ \bibnamefont {Fischer}}, \bibinfo
  {author} {\bibfnamefont {R.}~\bibnamefont {Vosk}}, \bibinfo {author}
  {\bibfnamefont {E.}~\bibnamefont {Altman}}, \bibinfo {author} {\bibfnamefont
  {U.}~\bibnamefont {Schneider}},\ and\ \bibinfo {author} {\bibfnamefont
  {I.}~\bibnamefont {Bloch}},\ }\href {https://doi.org/10.1126/science.aaa7432}
  {\bibfield  {journal} {\bibinfo  {journal} {Science}\ }\textbf {\bibinfo
  {volume} {349}},\ \bibinfo {pages} {842} (\bibinfo {year}
  {2015})}\BibitemShut {NoStop}%
\bibitem [{\citenamefont {Kondov}\ \emph {et~al.}(2015)\citenamefont {Kondov},
  \citenamefont {McGehee}, \citenamefont {Xu},\ and\ \citenamefont
  {DeMarco}}]{Kondov-DeMarco-DisorderInducedLocalizationStrongly-2015}%
  \BibitemOpen
  \bibfield  {author} {\bibinfo {author} {\bibfnamefont {S.~S.}\ \bibnamefont
  {Kondov}}, \bibinfo {author} {\bibfnamefont {W.~R.}\ \bibnamefont {McGehee}},
  \bibinfo {author} {\bibfnamefont {W.}~\bibnamefont {Xu}},\ and\ \bibinfo
  {author} {\bibfnamefont {B.}~\bibnamefont {DeMarco}},\ }\href
  {https://doi.org/10.1103/PhysRevLett.114.083002} {\bibfield  {journal}
  {\bibinfo  {journal} {Physical Review Letters}\ }\textbf {\bibinfo {volume}
  {114}},\ \bibinfo {pages} {083002} (\bibinfo {year} {2015})}\BibitemShut
  {NoStop}%
\bibitem [{\citenamefont {Choi}\ \emph {et~al.}(2016)\citenamefont {Choi},
  \citenamefont {Hild}, \citenamefont {Zeiher}, \citenamefont {Schau{\ss}},
  \citenamefont {{Rubio-Abadal}}, \citenamefont {Yefsah}, \citenamefont
  {Khemani}, \citenamefont {Huse}, \citenamefont {Bloch},\ and\ \citenamefont
  {Gross}}]{Choi-Gross-ExploringManybodyLocalization-2016}%
  \BibitemOpen
  \bibfield  {author} {\bibinfo {author} {\bibfnamefont {J.-Y.}\ \bibnamefont
  {Choi}}, \bibinfo {author} {\bibfnamefont {S.}~\bibnamefont {Hild}}, \bibinfo
  {author} {\bibfnamefont {J.}~\bibnamefont {Zeiher}}, \bibinfo {author}
  {\bibfnamefont {P.}~\bibnamefont {Schau{\ss}}}, \bibinfo {author}
  {\bibfnamefont {A.}~\bibnamefont {{Rubio-Abadal}}}, \bibinfo {author}
  {\bibfnamefont {T.}~\bibnamefont {Yefsah}}, \bibinfo {author} {\bibfnamefont
  {V.}~\bibnamefont {Khemani}}, \bibinfo {author} {\bibfnamefont {D.~A.}\
  \bibnamefont {Huse}}, \bibinfo {author} {\bibfnamefont {I.}~\bibnamefont
  {Bloch}},\ and\ \bibinfo {author} {\bibfnamefont {C.}~\bibnamefont {Gross}},\
  }\href {https://doi.org/10.1126/science.aaf883} {\bibfield  {journal}
  {\bibinfo  {journal} {Science}\ }\textbf {\bibinfo {volume} {352}},\ \bibinfo
  {pages} {1547} (\bibinfo {year} {2016})}\BibitemShut {NoStop}%
\bibitem [{\citenamefont {Yu}\ \emph {et~al.}(2023)\citenamefont {Yu},
  \citenamefont {Bhave}, \citenamefont {Reeve}, \citenamefont {Song},\ and\
  \citenamefont {Schneider}}]{Yu23_arXiv_BoseGlass}%
  \BibitemOpen
  \bibfield  {author} {\bibinfo {author} {\bibfnamefont {J.-C.}\ \bibnamefont
  {Yu}}, \bibinfo {author} {\bibfnamefont {S.}~\bibnamefont {Bhave}}, \bibinfo
  {author} {\bibfnamefont {L.}~\bibnamefont {Reeve}}, \bibinfo {author}
  {\bibfnamefont {B.}~\bibnamefont {Song}},\ and\ \bibinfo {author}
  {\bibfnamefont {U.}~\bibnamefont {Schneider}},\ }\bibfield  {journal}
  {\bibinfo  {journal} {arXiv preprint}\ }\href
  {https://doi.org/10.48550/arXiv.2303.00737} {10.48550/arXiv.2303.00737}
  (\bibinfo {year} {2023})\BibitemShut {NoStop}%
\bibitem [{\citenamefont {Viermann}\ \emph {et~al.}(2022)\citenamefont
  {Viermann}, \citenamefont {Sparn}, \citenamefont {Liebster}, \citenamefont
  {Hans}, \citenamefont {Kath}, \citenamefont {{Parra-L{\'o}pez}},
  \citenamefont {{Tolosa-Sime{\'o}n}}, \citenamefont {{S{\'a}nchez-Kuntz}},
  \citenamefont {Haas}, \citenamefont {Strobel}, \citenamefont {Floerchinger},\
  and\ \citenamefont
  {Oberthaler}}]{Viermann-Oberthaler-QuantumFieldSimulator-2022}%
  \BibitemOpen
  \bibfield  {author} {\bibinfo {author} {\bibfnamefont {C.}~\bibnamefont
  {Viermann}}, \bibinfo {author} {\bibfnamefont {M.}~\bibnamefont {Sparn}},
  \bibinfo {author} {\bibfnamefont {N.}~\bibnamefont {Liebster}}, \bibinfo
  {author} {\bibfnamefont {M.}~\bibnamefont {Hans}}, \bibinfo {author}
  {\bibfnamefont {E.}~\bibnamefont {Kath}}, \bibinfo {author} {\bibfnamefont
  {{\'A}.}~\bibnamefont {{Parra-L{\'o}pez}}}, \bibinfo {author} {\bibfnamefont
  {M.}~\bibnamefont {{Tolosa-Sime{\'o}n}}}, \bibinfo {author} {\bibfnamefont
  {N.}~\bibnamefont {{S{\'a}nchez-Kuntz}}}, \bibinfo {author} {\bibfnamefont
  {T.}~\bibnamefont {Haas}}, \bibinfo {author} {\bibfnamefont {H.}~\bibnamefont
  {Strobel}}, \bibinfo {author} {\bibfnamefont {S.}~\bibnamefont
  {Floerchinger}},\ and\ \bibinfo {author} {\bibfnamefont {M.~K.}\ \bibnamefont
  {Oberthaler}},\ }\href {https://doi.org/10.1038/s41586-022-05313-9}
  {\bibfield  {journal} {\bibinfo  {journal} {Nature}\ }\textbf {\bibinfo
  {volume} {611}},\ \bibinfo {pages} {260} (\bibinfo {year}
  {2022})}\BibitemShut {NoStop}%
\bibitem [{\citenamefont {Song}\ \emph {et~al.}(2022)\citenamefont {Song},
  \citenamefont {Dutta}, \citenamefont {Bhave}, \citenamefont {Yu},
  \citenamefont {Carter}, \citenamefont {Cooper},\ and\ \citenamefont
  {Schneider}}]{Song-Schneider-RealizingDiscontinuousQuantum-2022}%
  \BibitemOpen
  \bibfield  {author} {\bibinfo {author} {\bibfnamefont {B.}~\bibnamefont
  {Song}}, \bibinfo {author} {\bibfnamefont {S.}~\bibnamefont {Dutta}},
  \bibinfo {author} {\bibfnamefont {S.}~\bibnamefont {Bhave}}, \bibinfo
  {author} {\bibfnamefont {J.-C.}\ \bibnamefont {Yu}}, \bibinfo {author}
  {\bibfnamefont {E.}~\bibnamefont {Carter}}, \bibinfo {author} {\bibfnamefont
  {N.}~\bibnamefont {Cooper}},\ and\ \bibinfo {author} {\bibfnamefont
  {U.}~\bibnamefont {Schneider}},\ }\href
  {https://doi.org/10.1038/s41567-021-01476-w} {\bibfield  {journal} {\bibinfo
  {journal} {Nature Physics}\ }\textbf {\bibinfo {volume} {18}},\ \bibinfo
  {pages} {259} (\bibinfo {year} {2022})}\BibitemShut {NoStop}%
\bibitem [{\citenamefont {Zohar}\ \emph {et~al.}(2015)\citenamefont {Zohar},
  \citenamefont {Cirac},\ and\ \citenamefont
  {Reznik}}]{Zohar-Reznik-QuantumSimulationsLattice-2015}%
  \BibitemOpen
  \bibfield  {author} {\bibinfo {author} {\bibfnamefont {E.}~\bibnamefont
  {Zohar}}, \bibinfo {author} {\bibfnamefont {J.~I.}\ \bibnamefont {Cirac}},\
  and\ \bibinfo {author} {\bibfnamefont {B.}~\bibnamefont {Reznik}},\ }\href
  {https://doi.org/10.1088/0034-4885/79/1/014401} {\bibfield  {journal}
  {\bibinfo  {journal} {Reports on Progress in Physics}\ }\textbf {\bibinfo
  {volume} {79}},\ \bibinfo {pages} {014401} (\bibinfo {year}
  {2015})}\BibitemShut {NoStop}%
\bibitem [{\citenamefont {{Arg{\"u}ello-Luengo}}\ \emph
  {et~al.}(2019)\citenamefont {{Arg{\"u}ello-Luengo}}, \citenamefont
  {{Gonz{\'a}lez-Tudela}}, \citenamefont {Shi}, \citenamefont {Zoller},\ and\
  \citenamefont {Cirac}}]{Arguello-Luengo-Cirac-AnalogueQuantumChemistry-2019}%
  \BibitemOpen
  \bibfield  {author} {\bibinfo {author} {\bibfnamefont {J.}~\bibnamefont
  {{Arg{\"u}ello-Luengo}}}, \bibinfo {author} {\bibfnamefont {A.}~\bibnamefont
  {{Gonz{\'a}lez-Tudela}}}, \bibinfo {author} {\bibfnamefont {T.}~\bibnamefont
  {Shi}}, \bibinfo {author} {\bibfnamefont {P.}~\bibnamefont {Zoller}},\ and\
  \bibinfo {author} {\bibfnamefont {J.~I.}\ \bibnamefont {Cirac}},\ }\href
  {https://doi.org/10.1038/s41586-019-1614-4} {\bibfield  {journal} {\bibinfo
  {journal} {Nature}\ }\textbf {\bibinfo {volume} {574}},\ \bibinfo {pages}
  {215} (\bibinfo {year} {2019})}\BibitemShut {NoStop}%
\bibitem [{\citenamefont {Eckardt}(2017)}]{eckardt_colloquium_2017}%
  \BibitemOpen
  \bibfield  {author} {\bibinfo {author} {\bibfnamefont {A.}~\bibnamefont
  {Eckardt}},\ }\href {https://doi.org/10.1103/RevModPhys.89.011004} {\bibfield
   {journal} {\bibinfo  {journal} {Reviews of Modern Physics}\ }\textbf
  {\bibinfo {volume} {89}},\ \bibinfo {pages} {011004} (\bibinfo {year}
  {2017})}\BibitemShut {NoStop}%
\bibitem [{\citenamefont {Cooper}\ \emph {et~al.}(2019)\citenamefont {Cooper},
  \citenamefont {Dalibard},\ and\ \citenamefont {Spielman}}]{Cooper19_RMP}%
  \BibitemOpen
  \bibfield  {author} {\bibinfo {author} {\bibfnamefont {N.~R.}\ \bibnamefont
  {Cooper}}, \bibinfo {author} {\bibfnamefont {J.}~\bibnamefont {Dalibard}},\
  and\ \bibinfo {author} {\bibfnamefont {I.~B.}\ \bibnamefont {Spielman}},\
  }\href {https://doi.org/10.1103/RevModPhys.91.015005} {\bibfield  {journal}
  {\bibinfo  {journal} {Reviews of Modern Physics}\ }\textbf {\bibinfo {volume}
  {91}},\ \bibinfo {pages} {015005} (\bibinfo {year} {2019})}\BibitemShut
  {NoStop}%
\bibitem [{\citenamefont {Weitenberg}\ and\ \citenamefont
  {Simonet}(2021)}]{Weitenberg-Simonet-TailoringQuantumGases-2021}%
  \BibitemOpen
  \bibfield  {author} {\bibinfo {author} {\bibfnamefont {C.}~\bibnamefont
  {Weitenberg}}\ and\ \bibinfo {author} {\bibfnamefont {J.}~\bibnamefont
  {Simonet}},\ }\href {https://doi.org/10.1038/s41567-021-01316-x} {\bibfield
  {journal} {\bibinfo  {journal} {Nature Physics}\ }\textbf {\bibinfo {volume}
  {17}},\ \bibinfo {pages} {1342} (\bibinfo {year} {2021})}\BibitemShut
  {NoStop}%
\bibitem [{\citenamefont {Lignier}\ \emph {et~al.}(2007)\citenamefont
  {Lignier}, \citenamefont {Sias}, \citenamefont {Ciampini}, \citenamefont
  {Singh}, \citenamefont {Zenesini}, \citenamefont {Morsch},\ and\
  \citenamefont {Arimondo}}]{lignier_dynamical_2007}%
  \BibitemOpen
  \bibfield  {author} {\bibinfo {author} {\bibfnamefont {H.}~\bibnamefont
  {Lignier}}, \bibinfo {author} {\bibfnamefont {C.}~\bibnamefont {Sias}},
  \bibinfo {author} {\bibfnamefont {D.}~\bibnamefont {Ciampini}}, \bibinfo
  {author} {\bibfnamefont {Y.}~\bibnamefont {Singh}}, \bibinfo {author}
  {\bibfnamefont {A.}~\bibnamefont {Zenesini}}, \bibinfo {author}
  {\bibfnamefont {O.}~\bibnamefont {Morsch}},\ and\ \bibinfo {author}
  {\bibfnamefont {E.}~\bibnamefont {Arimondo}},\ }\href
  {https://doi.org/10.1103/PhysRevLett.99.220403} {\bibfield  {journal}
  {\bibinfo  {journal} {Physical Review Letters}\ }\textbf {\bibinfo {volume}
  {99}},\ \bibinfo {pages} {220403} (\bibinfo {year} {2007})}\BibitemShut
  {NoStop}%
\bibitem [{\citenamefont {Eckardt}\ \emph {et~al.}(2009)\citenamefont
  {Eckardt}, \citenamefont {Holthaus}, \citenamefont {Lignier}, \citenamefont
  {Zenesini}, \citenamefont {Ciampini}, \citenamefont {Morsch},\ and\
  \citenamefont {Arimondo}}]{eckardt_exploring_2009}%
  \BibitemOpen
  \bibfield  {author} {\bibinfo {author} {\bibfnamefont {A.}~\bibnamefont
  {Eckardt}}, \bibinfo {author} {\bibfnamefont {M.}~\bibnamefont {Holthaus}},
  \bibinfo {author} {\bibfnamefont {H.}~\bibnamefont {Lignier}}, \bibinfo
  {author} {\bibfnamefont {A.}~\bibnamefont {Zenesini}}, \bibinfo {author}
  {\bibfnamefont {D.}~\bibnamefont {Ciampini}}, \bibinfo {author}
  {\bibfnamefont {O.}~\bibnamefont {Morsch}},\ and\ \bibinfo {author}
  {\bibfnamefont {E.}~\bibnamefont {Arimondo}},\ }\href
  {https://doi.org/10.1103/PhysRevA.79.013611} {\bibfield  {journal} {\bibinfo
  {journal} {Physical Review A}\ }\textbf {\bibinfo {volume} {79}},\ \bibinfo
  {pages} {013611} (\bibinfo {year} {2009})}\BibitemShut {NoStop}%
\bibitem [{\citenamefont {Creffield}\ \emph {et~al.}(2010)\citenamefont
  {Creffield}, \citenamefont {Sols}, \citenamefont {Ciampini}, \citenamefont
  {Morsch},\ and\ \citenamefont {Arimondo}}]{creffield_expansion_2010}%
  \BibitemOpen
  \bibfield  {author} {\bibinfo {author} {\bibfnamefont {C.~E.}\ \bibnamefont
  {Creffield}}, \bibinfo {author} {\bibfnamefont {F.}~\bibnamefont {Sols}},
  \bibinfo {author} {\bibfnamefont {D.}~\bibnamefont {Ciampini}}, \bibinfo
  {author} {\bibfnamefont {O.}~\bibnamefont {Morsch}},\ and\ \bibinfo {author}
  {\bibfnamefont {E.}~\bibnamefont {Arimondo}},\ }\href
  {https://doi.org/10.1103/PhysRevA.82.035601} {\bibfield  {journal} {\bibinfo
  {journal} {Physical Review A}\ }\textbf {\bibinfo {volume} {82}},\ \bibinfo
  {pages} {035601} (\bibinfo {year} {2010})}\BibitemShut {NoStop}%
\bibitem [{\citenamefont {Aidelsburger}\ \emph {et~al.}(2011)\citenamefont
  {Aidelsburger}, \citenamefont {Atala}, \citenamefont {Nascimb{\`e}ne},
  \citenamefont {Trotzky}, \citenamefont {Chen},\ and\ \citenamefont
  {Bloch}}]{Aidelsburger-Bloch-ExperimentalRealizationStrong-2011}%
  \BibitemOpen
  \bibfield  {author} {\bibinfo {author} {\bibfnamefont {M.}~\bibnamefont
  {Aidelsburger}}, \bibinfo {author} {\bibfnamefont {M.}~\bibnamefont {Atala}},
  \bibinfo {author} {\bibfnamefont {S.}~\bibnamefont {Nascimb{\`e}ne}},
  \bibinfo {author} {\bibfnamefont {S.}~\bibnamefont {Trotzky}}, \bibinfo
  {author} {\bibfnamefont {Y.-A.}\ \bibnamefont {Chen}},\ and\ \bibinfo
  {author} {\bibfnamefont {I.}~\bibnamefont {Bloch}},\ }\href
  {https://doi.org/10.1103/PhysRevLett.107.255301} {\bibfield  {journal}
  {\bibinfo  {journal} {Physical Review Letters}\ }\textbf {\bibinfo {volume}
  {107}},\ \bibinfo {pages} {255301} (\bibinfo {year} {2011})}\BibitemShut
  {NoStop}%
\bibitem [{\citenamefont {Aidelsburger}\ \emph {et~al.}(2013)\citenamefont
  {Aidelsburger}, \citenamefont {Atala}, \citenamefont {Lohse}, \citenamefont
  {Barreiro}, \citenamefont {Paredes},\ and\ \citenamefont
  {Bloch}}]{Aidelsburger-Bloch-RealizationHofstadterHamiltonian-2013}%
  \BibitemOpen
  \bibfield  {author} {\bibinfo {author} {\bibfnamefont {M.}~\bibnamefont
  {Aidelsburger}}, \bibinfo {author} {\bibfnamefont {M.}~\bibnamefont {Atala}},
  \bibinfo {author} {\bibfnamefont {M.}~\bibnamefont {Lohse}}, \bibinfo
  {author} {\bibfnamefont {J.~T.}\ \bibnamefont {Barreiro}}, \bibinfo {author}
  {\bibfnamefont {B.}~\bibnamefont {Paredes}},\ and\ \bibinfo {author}
  {\bibfnamefont {I.}~\bibnamefont {Bloch}},\ }\href
  {https://doi.org/10.1103/PhysRevLett.111.185301} {\bibfield  {journal}
  {\bibinfo  {journal} {Physical Review Letters}\ }\textbf {\bibinfo {volume}
  {111}},\ \bibinfo {pages} {185301} (\bibinfo {year} {2013})}\BibitemShut
  {NoStop}%
\bibitem [{\citenamefont {Miyake}\ \emph {et~al.}(2013)\citenamefont {Miyake},
  \citenamefont {Siviloglou}, \citenamefont {Kennedy}, \citenamefont {Burton},\
  and\ \citenamefont
  {Ketterle}}]{Miyake-Ketterle-RealizingHarperHamiltonian-2013}%
  \BibitemOpen
  \bibfield  {author} {\bibinfo {author} {\bibfnamefont {H.}~\bibnamefont
  {Miyake}}, \bibinfo {author} {\bibfnamefont {G.~A.}\ \bibnamefont
  {Siviloglou}}, \bibinfo {author} {\bibfnamefont {C.~J.}\ \bibnamefont
  {Kennedy}}, \bibinfo {author} {\bibfnamefont {W.~C.}\ \bibnamefont
  {Burton}},\ and\ \bibinfo {author} {\bibfnamefont {W.}~\bibnamefont
  {Ketterle}},\ }\href {https://doi.org/10.1103/PhysRevLett.111.185302}
  {\bibfield  {journal} {\bibinfo  {journal} {Physical Review Letters}\
  }\textbf {\bibinfo {volume} {111}},\ \bibinfo {pages} {185302} (\bibinfo
  {year} {2013})}\BibitemShut {NoStop}%
\bibitem [{\citenamefont {Aidelsburger}\ \emph {et~al.}(2015)\citenamefont
  {Aidelsburger}, \citenamefont {Lohse}, \citenamefont {Schweizer},
  \citenamefont {Atala}, \citenamefont {Barreiro}, \citenamefont
  {Nascimb{\`e}ne}, \citenamefont {Cooper}, \citenamefont {Bloch},\ and\
  \citenamefont {Goldman}}]{Aidelsburger-Goldman-MeasuringChernNumber-2015}%
  \BibitemOpen
  \bibfield  {author} {\bibinfo {author} {\bibfnamefont {M.}~\bibnamefont
  {Aidelsburger}}, \bibinfo {author} {\bibfnamefont {M.}~\bibnamefont {Lohse}},
  \bibinfo {author} {\bibfnamefont {C.}~\bibnamefont {Schweizer}}, \bibinfo
  {author} {\bibfnamefont {M.}~\bibnamefont {Atala}}, \bibinfo {author}
  {\bibfnamefont {J.~T.}\ \bibnamefont {Barreiro}}, \bibinfo {author}
  {\bibfnamefont {S.}~\bibnamefont {Nascimb{\`e}ne}}, \bibinfo {author}
  {\bibfnamefont {N.~R.}\ \bibnamefont {Cooper}}, \bibinfo {author}
  {\bibfnamefont {I.}~\bibnamefont {Bloch}},\ and\ \bibinfo {author}
  {\bibfnamefont {N.}~\bibnamefont {Goldman}},\ }\href
  {https://doi.org/10.1038/nphys3171} {\bibfield  {journal} {\bibinfo
  {journal} {Nature Physics}\ }\textbf {\bibinfo {volume} {11}},\ \bibinfo
  {pages} {162} (\bibinfo {year} {2015})}\BibitemShut {NoStop}%
\bibitem [{\citenamefont {Tarnowski}\ \emph {et~al.}(2019)\citenamefont
  {Tarnowski}, \citenamefont {{\"U}nal}, \citenamefont {Fl{\"a}schner},
  \citenamefont {Rem}, \citenamefont {Eckardt}, \citenamefont {Sengstock},\
  and\ \citenamefont {Weitenberg}}]{Tarnowski19_NatCom}%
  \BibitemOpen
  \bibfield  {author} {\bibinfo {author} {\bibfnamefont {M.}~\bibnamefont
  {Tarnowski}}, \bibinfo {author} {\bibfnamefont {F.~N.}\ \bibnamefont
  {{\"U}nal}}, \bibinfo {author} {\bibfnamefont {N.}~\bibnamefont
  {Fl{\"a}schner}}, \bibinfo {author} {\bibfnamefont {B.~S.}\ \bibnamefont
  {Rem}}, \bibinfo {author} {\bibfnamefont {A.}~\bibnamefont {Eckardt}},
  \bibinfo {author} {\bibfnamefont {K.}~\bibnamefont {Sengstock}},\ and\
  \bibinfo {author} {\bibfnamefont {C.}~\bibnamefont {Weitenberg}},\ }\href
  {https://www.nature.com/articles/s41467-019-09668-y} {\bibfield  {journal}
  {\bibinfo  {journal} {Nature Communications}\ }\textbf {\bibinfo {volume}
  {10}} (\bibinfo {year} {2019})}\BibitemShut {NoStop}%
\bibitem [{\citenamefont {Eckardt}\ \emph {et~al.}(2010)\citenamefont
  {Eckardt}, \citenamefont {Hauke}, \citenamefont {Soltan-Panahi},
  \citenamefont {Becker}, \citenamefont {Sengstock},\ and\ \citenamefont
  {Lewenstein}}]{eckardt_frustrated_2010}%
  \BibitemOpen
  \bibfield  {author} {\bibinfo {author} {\bibfnamefont {A.}~\bibnamefont
  {Eckardt}}, \bibinfo {author} {\bibfnamefont {P.}~\bibnamefont {Hauke}},
  \bibinfo {author} {\bibfnamefont {P.}~\bibnamefont {Soltan-Panahi}}, \bibinfo
  {author} {\bibfnamefont {C.}~\bibnamefont {Becker}}, \bibinfo {author}
  {\bibfnamefont {K.}~\bibnamefont {Sengstock}},\ and\ \bibinfo {author}
  {\bibfnamefont {M.}~\bibnamefont {Lewenstein}},\ }\href
  {https://doi.org/10.1209/0295-5075/89/10010} {\bibfield  {journal} {\bibinfo
  {journal} {Europhysics Letters}\ }\textbf {\bibinfo {volume} {89}},\ \bibinfo
  {pages} {10010} (\bibinfo {year} {2010})}\BibitemShut {NoStop}%
\bibitem [{\citenamefont {Struck}\ \emph {et~al.}(2013)\citenamefont {Struck},
  \citenamefont {Weinberg}, \citenamefont {Ölschläger}, \citenamefont
  {Windpassinger}, \citenamefont {Simonet}, \citenamefont {Sengstock},
  \citenamefont {Höppner}, \citenamefont {Hauke}, \citenamefont {Eckardt},
  \citenamefont {Lewenstein},\ and\ \citenamefont
  {Mathey}}]{struck_engineering_2013}%
  \BibitemOpen
  \bibfield  {author} {\bibinfo {author} {\bibfnamefont {J.}~\bibnamefont
  {Struck}}, \bibinfo {author} {\bibfnamefont {M.}~\bibnamefont {Weinberg}},
  \bibinfo {author} {\bibfnamefont {C.}~\bibnamefont {Ölschläger}}, \bibinfo
  {author} {\bibfnamefont {P.}~\bibnamefont {Windpassinger}}, \bibinfo {author}
  {\bibfnamefont {J.}~\bibnamefont {Simonet}}, \bibinfo {author} {\bibfnamefont
  {K.}~\bibnamefont {Sengstock}}, \bibinfo {author} {\bibfnamefont
  {R.}~\bibnamefont {Höppner}}, \bibinfo {author} {\bibfnamefont
  {P.}~\bibnamefont {Hauke}}, \bibinfo {author} {\bibfnamefont
  {A.}~\bibnamefont {Eckardt}}, \bibinfo {author} {\bibfnamefont
  {M.}~\bibnamefont {Lewenstein}},\ and\ \bibinfo {author} {\bibfnamefont
  {L.}~\bibnamefont {Mathey}},\ }\href {https://doi.org/10.1038/nphys2750}
  {\bibfield  {journal} {\bibinfo  {journal} {Nature Physics}\ }\textbf
  {\bibinfo {volume} {9}},\ \bibinfo {pages} {738} (\bibinfo {year}
  {2013})}\BibitemShut {NoStop}%
\bibitem [{\citenamefont {Clark}\ \emph {et~al.}(2018)\citenamefont {Clark},
  \citenamefont {Anderson}, \citenamefont {Feng}, \citenamefont {Gaj},
  \citenamefont {Levin},\ and\ \citenamefont {Chin}}]{clark_observation_2018}%
  \BibitemOpen
  \bibfield  {author} {\bibinfo {author} {\bibfnamefont {L.~W.}\ \bibnamefont
  {Clark}}, \bibinfo {author} {\bibfnamefont {B.~M.}\ \bibnamefont {Anderson}},
  \bibinfo {author} {\bibfnamefont {L.}~\bibnamefont {Feng}}, \bibinfo {author}
  {\bibfnamefont {A.}~\bibnamefont {Gaj}}, \bibinfo {author} {\bibfnamefont
  {K.}~\bibnamefont {Levin}},\ and\ \bibinfo {author} {\bibfnamefont
  {C.}~\bibnamefont {Chin}},\ }\href
  {https://doi.org/10.1103/PhysRevLett.121.030402} {\bibfield  {journal}
  {\bibinfo  {journal} {Physical Review Letters}\ }\textbf {\bibinfo {volume}
  {121}},\ \bibinfo {pages} {030402} (\bibinfo {year} {2018})}\BibitemShut
  {NoStop}%
\bibitem [{\citenamefont {Barbiero}\ \emph {et~al.}(2019)\citenamefont
  {Barbiero}, \citenamefont {Schweizer}, \citenamefont {Aidelsburger},
  \citenamefont {Demler}, \citenamefont {Goldman},\ and\ \citenamefont
  {Grusdt}}]{barbiero_coupling_2019}%
  \BibitemOpen
  \bibfield  {author} {\bibinfo {author} {\bibfnamefont {L.}~\bibnamefont
  {Barbiero}}, \bibinfo {author} {\bibfnamefont {C.}~\bibnamefont {Schweizer}},
  \bibinfo {author} {\bibfnamefont {M.}~\bibnamefont {Aidelsburger}}, \bibinfo
  {author} {\bibfnamefont {E.}~\bibnamefont {Demler}}, \bibinfo {author}
  {\bibfnamefont {N.}~\bibnamefont {Goldman}},\ and\ \bibinfo {author}
  {\bibfnamefont {F.}~\bibnamefont {Grusdt}},\ }\href
  {https://doi.org/10.1126/sciadv.aav7444} {\bibfield  {journal} {\bibinfo
  {journal} {Science Advances}\ }\textbf {\bibinfo {volume} {5}},\ \bibinfo
  {pages} {eaav7444} (\bibinfo {year} {2019})}\BibitemShut {NoStop}%
\bibitem [{\citenamefont {Görg}\ \emph {et~al.}(2019)\citenamefont {Görg},
  \citenamefont {Sandholzer}, \citenamefont {Minguzzi}, \citenamefont
  {Desbuquois}, \citenamefont {Messer},\ and\ \citenamefont
  {Esslinger}}]{gorg_realization_2019}%
  \BibitemOpen
  \bibfield  {author} {\bibinfo {author} {\bibfnamefont {F.}~\bibnamefont
  {Görg}}, \bibinfo {author} {\bibfnamefont {K.}~\bibnamefont {Sandholzer}},
  \bibinfo {author} {\bibfnamefont {J.}~\bibnamefont {Minguzzi}}, \bibinfo
  {author} {\bibfnamefont {R.}~\bibnamefont {Desbuquois}}, \bibinfo {author}
  {\bibfnamefont {M.}~\bibnamefont {Messer}},\ and\ \bibinfo {author}
  {\bibfnamefont {T.}~\bibnamefont {Esslinger}},\ }\href
  {https://doi.org/10.1038/s41567-019-0615-4} {\bibfield  {journal} {\bibinfo
  {journal} {Nature Physics}\ }\textbf {\bibinfo {volume} {15}},\ \bibinfo
  {pages} {1161} (\bibinfo {year} {2019})}\BibitemShut {NoStop}%
\bibitem [{\citenamefont {Schweizer}\ \emph {et~al.}(2019)\citenamefont
  {Schweizer}, \citenamefont {Grusdt}, \citenamefont {Berngruber},
  \citenamefont {Barbiero}, \citenamefont {Demler}, \citenamefont {Goldman},
  \citenamefont {Bloch},\ and\ \citenamefont
  {Aidelsburger}}]{schweizer_floquet_2019}%
  \BibitemOpen
  \bibfield  {author} {\bibinfo {author} {\bibfnamefont {C.}~\bibnamefont
  {Schweizer}}, \bibinfo {author} {\bibfnamefont {F.}~\bibnamefont {Grusdt}},
  \bibinfo {author} {\bibfnamefont {M.}~\bibnamefont {Berngruber}}, \bibinfo
  {author} {\bibfnamefont {L.}~\bibnamefont {Barbiero}}, \bibinfo {author}
  {\bibfnamefont {E.}~\bibnamefont {Demler}}, \bibinfo {author} {\bibfnamefont
  {N.}~\bibnamefont {Goldman}}, \bibinfo {author} {\bibfnamefont
  {I.}~\bibnamefont {Bloch}},\ and\ \bibinfo {author} {\bibfnamefont
  {M.}~\bibnamefont {Aidelsburger}},\ }\href
  {https://doi.org/10.1038/s41567-019-0649-7} {\bibfield  {journal} {\bibinfo
  {journal} {Nature Physics}\ }\textbf {\bibinfo {volume} {15}},\ \bibinfo
  {pages} {1168} (\bibinfo {year} {2019})}\BibitemShut {NoStop}%
\bibitem [{\citenamefont {Bakr}\ \emph {et~al.}(2009)\citenamefont {Bakr},
  \citenamefont {Gillen}, \citenamefont {Peng}, \citenamefont {Fölling},\ and\
  \citenamefont {Greiner}}]{bakr_quantum_2009}%
  \BibitemOpen
  \bibfield  {author} {\bibinfo {author} {\bibfnamefont {W.~S.}\ \bibnamefont
  {Bakr}}, \bibinfo {author} {\bibfnamefont {J.~I.}\ \bibnamefont {Gillen}},
  \bibinfo {author} {\bibfnamefont {A.}~\bibnamefont {Peng}}, \bibinfo {author}
  {\bibfnamefont {S.}~\bibnamefont {Fölling}},\ and\ \bibinfo {author}
  {\bibfnamefont {M.}~\bibnamefont {Greiner}},\ }\href
  {https://doi.org/10.1038/nature08482} {\bibfield  {journal} {\bibinfo
  {journal} {Nature}\ }\textbf {\bibinfo {volume} {462}},\ \bibinfo {pages}
  {74} (\bibinfo {year} {2009})}\BibitemShut {NoStop}%
\bibitem [{\citenamefont {Sherson}\ \emph {et~al.}(2010)\citenamefont
  {Sherson}, \citenamefont {Weitenberg}, \citenamefont {Endres}, \citenamefont
  {Cheneau}, \citenamefont {Bloch},\ and\ \citenamefont
  {Kuhr}}]{Sherson-Kuhr-SingleatomresolvedFluorescenceImaging-2010}%
  \BibitemOpen
  \bibfield  {author} {\bibinfo {author} {\bibfnamefont {J.~F.}\ \bibnamefont
  {Sherson}}, \bibinfo {author} {\bibfnamefont {C.}~\bibnamefont {Weitenberg}},
  \bibinfo {author} {\bibfnamefont {M.}~\bibnamefont {Endres}}, \bibinfo
  {author} {\bibfnamefont {M.}~\bibnamefont {Cheneau}}, \bibinfo {author}
  {\bibfnamefont {I.}~\bibnamefont {Bloch}},\ and\ \bibinfo {author}
  {\bibfnamefont {S.}~\bibnamefont {Kuhr}},\ }\href
  {https://doi.org/10.1038/nature09378} {\bibfield  {journal} {\bibinfo
  {journal} {Nature}\ }\textbf {\bibinfo {volume} {467}},\ \bibinfo {pages}
  {68} (\bibinfo {year} {2010})}\BibitemShut {NoStop}%
\bibitem [{\citenamefont {Gross}\ and\ \citenamefont
  {Bakr}(2021)}]{GrossBakr21_NatPhys}%
  \BibitemOpen
  \bibfield  {author} {\bibinfo {author} {\bibfnamefont {C.}~\bibnamefont
  {Gross}}\ and\ \bibinfo {author} {\bibfnamefont {W.}~\bibnamefont {Bakr}},\
  }\href {https://doi.org/10.1038/s41567-021-01370-5} {\bibfield  {journal}
  {\bibinfo  {journal} {Nature Physics}\ }\textbf {\bibinfo {volume} {17}},\
  \bibinfo {pages} {1316} (\bibinfo {year} {2021})}\BibitemShut {NoStop}%
\bibitem [{\citenamefont {Mart\'{\i}nez}\ and\ \citenamefont
  {\"Unal}(2023)}]{Martinez23_arXivWP}%
  \BibitemOpen
  \bibfield  {author} {\bibinfo {author} {\bibfnamefont {M.~F.}\ \bibnamefont
  {Mart\'{\i}nez}}\ and\ \bibinfo {author} {\bibfnamefont {F.~N.}\ \bibnamefont
  {\"Unal}},\ }\href {https://doi.org/10.1103/PhysRevA.108.063314} {\bibfield
  {journal} {\bibinfo  {journal} {Physical Review A}\ }\textbf {\bibinfo
  {volume} {108}},\ \bibinfo {pages} {063314} (\bibinfo {year}
  {2023})}\BibitemShut {NoStop}%
\bibitem [{\citenamefont {Braun}\ \emph {et~al.}(2023)\citenamefont {Braun},
  \citenamefont {Saint-Jalm}, \citenamefont {Hesse}, \citenamefont {Arceri},
  \citenamefont {Bloch},\ and\ \citenamefont {Aidelsburger}}]{Braun23_arXivWP}%
  \BibitemOpen
  \bibfield  {author} {\bibinfo {author} {\bibfnamefont {C.}~\bibnamefont
  {Braun}}, \bibinfo {author} {\bibfnamefont {R.}~\bibnamefont {Saint-Jalm}},
  \bibinfo {author} {\bibfnamefont {A.}~\bibnamefont {Hesse}}, \bibinfo
  {author} {\bibfnamefont {J.}~\bibnamefont {Arceri}}, \bibinfo {author}
  {\bibfnamefont {I.}~\bibnamefont {Bloch}},\ and\ \bibinfo {author}
  {\bibfnamefont {M.}~\bibnamefont {Aidelsburger}},\ }\bibfield  {journal}
  {\bibinfo  {journal} {arXiv preprint}\ }\href
  {https://doi.org/10.48550/arXiv.2304.01980} {10.48550/arXiv.2304.01980}
  (\bibinfo {year} {2023})\BibitemShut {NoStop}%
\bibitem [{\citenamefont {Thompson}\ \emph {et~al.}(2013)\citenamefont
  {Thompson}, \citenamefont {Tiecke}, \citenamefont {Zibrov}, \citenamefont
  {Vuleti{\'c}},\ and\ \citenamefont
  {Lukin}}]{Thompson-Lukin-CoherenceRamanSideband-2013}%
  \BibitemOpen
  \bibfield  {author} {\bibinfo {author} {\bibfnamefont {J.~D.}\ \bibnamefont
  {Thompson}}, \bibinfo {author} {\bibfnamefont {T.~G.}\ \bibnamefont
  {Tiecke}}, \bibinfo {author} {\bibfnamefont {A.~S.}\ \bibnamefont {Zibrov}},
  \bibinfo {author} {\bibfnamefont {V.}~\bibnamefont {Vuleti{\'c}}},\ and\
  \bibinfo {author} {\bibfnamefont {M.~D.}\ \bibnamefont {Lukin}},\ }\href
  {https://doi.org/10.1103/PhysRevLett.110.133001} {\bibfield  {journal}
  {\bibinfo  {journal} {Physical Review Letters}\ }\textbf {\bibinfo {volume}
  {110}},\ \bibinfo {pages} {133001} (\bibinfo {year} {2013})}\BibitemShut
  {NoStop}%
\bibitem [{\citenamefont {Lester}\ \emph {et~al.}(2015)\citenamefont {Lester},
  \citenamefont {Luick}, \citenamefont {Kaufman}, \citenamefont {Reynolds},\
  and\ \citenamefont {Regal}}]{Lester-Regal-RapidProductionUniformly-2015}%
  \BibitemOpen
  \bibfield  {author} {\bibinfo {author} {\bibfnamefont {B.~J.}\ \bibnamefont
  {Lester}}, \bibinfo {author} {\bibfnamefont {N.}~\bibnamefont {Luick}},
  \bibinfo {author} {\bibfnamefont {A.~M.}\ \bibnamefont {Kaufman}}, \bibinfo
  {author} {\bibfnamefont {C.~M.}\ \bibnamefont {Reynolds}},\ and\ \bibinfo
  {author} {\bibfnamefont {C.~A.}\ \bibnamefont {Regal}},\ }\href
  {https://doi.org/10.1103/PhysRevLett.115.073003} {\bibfield  {journal}
  {\bibinfo  {journal} {Physical Review Letters}\ }\textbf {\bibinfo {volume}
  {115}},\ \bibinfo {pages} {073003} (\bibinfo {year} {2015})}\BibitemShut
  {NoStop}%
\bibitem [{\citenamefont {Bernien}\ \emph {et~al.}(2017)\citenamefont
  {Bernien}, \citenamefont {Schwartz}, \citenamefont {Keesling}, \citenamefont
  {Levine}, \citenamefont {Omran}, \citenamefont {Pichler}, \citenamefont
  {Choi}, \citenamefont {Zibrov}, \citenamefont {Endres}, \citenamefont
  {Greiner}, \citenamefont {Vuleti{\'c}},\ and\ \citenamefont
  {Lukin}}]{Bernien-Lukin-ProbingManybodyDynamics-2017}%
  \BibitemOpen
  \bibfield  {author} {\bibinfo {author} {\bibfnamefont {H.}~\bibnamefont
  {Bernien}}, \bibinfo {author} {\bibfnamefont {S.}~\bibnamefont {Schwartz}},
  \bibinfo {author} {\bibfnamefont {A.}~\bibnamefont {Keesling}}, \bibinfo
  {author} {\bibfnamefont {H.}~\bibnamefont {Levine}}, \bibinfo {author}
  {\bibfnamefont {A.}~\bibnamefont {Omran}}, \bibinfo {author} {\bibfnamefont
  {H.}~\bibnamefont {Pichler}}, \bibinfo {author} {\bibfnamefont
  {S.}~\bibnamefont {Choi}}, \bibinfo {author} {\bibfnamefont {A.~S.}\
  \bibnamefont {Zibrov}}, \bibinfo {author} {\bibfnamefont {M.}~\bibnamefont
  {Endres}}, \bibinfo {author} {\bibfnamefont {M.}~\bibnamefont {Greiner}},
  \bibinfo {author} {\bibfnamefont {V.}~\bibnamefont {Vuleti{\'c}}},\ and\
  \bibinfo {author} {\bibfnamefont {M.~D.}\ \bibnamefont {Lukin}},\ }\href
  {https://doi.org/10.1038/nature24622} {\bibfield  {journal} {\bibinfo
  {journal} {Nature}\ }\textbf {\bibinfo {volume} {551}},\ \bibinfo {pages}
  {579} (\bibinfo {year} {2017})}\BibitemShut {NoStop}%
\bibitem [{\citenamefont {Barredo}\ \emph {et~al.}(2018)\citenamefont
  {Barredo}, \citenamefont {Lienhard}, \citenamefont {{de L{\'e}s{\'e}leuc}},
  \citenamefont {Lahaye},\ and\ \citenamefont
  {Browaeys}}]{Barredo-Browaeys-SyntheticThreedimensionalAtomic-2018}%
  \BibitemOpen
  \bibfield  {author} {\bibinfo {author} {\bibfnamefont {D.}~\bibnamefont
  {Barredo}}, \bibinfo {author} {\bibfnamefont {V.}~\bibnamefont {Lienhard}},
  \bibinfo {author} {\bibfnamefont {S.}~\bibnamefont {{de L{\'e}s{\'e}leuc}}},
  \bibinfo {author} {\bibfnamefont {T.}~\bibnamefont {Lahaye}},\ and\ \bibinfo
  {author} {\bibfnamefont {A.}~\bibnamefont {Browaeys}},\ }\href
  {https://doi.org/10.1038/s41586-018-0450-2} {\bibfield  {journal} {\bibinfo
  {journal} {Nature}\ }\textbf {\bibinfo {volume} {561}},\ \bibinfo {pages}
  {79} (\bibinfo {year} {2018})}\BibitemShut {NoStop}%
\bibitem [{\citenamefont
  {Andersen}(2022)}]{Andersen-OpticalTweezersBottomup-2022}%
  \BibitemOpen
  \bibfield  {author} {\bibinfo {author} {\bibfnamefont {M.~F.}\ \bibnamefont
  {Andersen}},\ }\href {https://doi.org/10.1080/23746149.2022.2064231}
  {\bibfield  {journal} {\bibinfo  {journal} {Advances in Physics: X}\ }\textbf
  {\bibinfo {volume} {7}},\ \bibinfo {pages} {2064231} (\bibinfo {year}
  {2022})}\BibitemShut {NoStop}%
\bibitem [{\citenamefont {Trisnadi}\ \emph {et~al.}(2022)\citenamefont
  {Trisnadi}, \citenamefont {Zhang}, \citenamefont {Weiss},\ and\ \citenamefont
  {Chin}}]{Trisnadi-Chin-DesignConstructionQuantum-2022}%
  \BibitemOpen
  \bibfield  {author} {\bibinfo {author} {\bibfnamefont {J.}~\bibnamefont
  {Trisnadi}}, \bibinfo {author} {\bibfnamefont {M.}~\bibnamefont {Zhang}},
  \bibinfo {author} {\bibfnamefont {L.}~\bibnamefont {Weiss}},\ and\ \bibinfo
  {author} {\bibfnamefont {C.}~\bibnamefont {Chin}},\ }\href
  {https://doi.org/10.1063/5.0100088} {\bibfield  {journal} {\bibinfo
  {journal} {Review of Scientific Instruments}\ }\textbf {\bibinfo {volume}
  {93}},\ \bibinfo {pages} {083203} (\bibinfo {year} {2022})}\BibitemShut
  {NoStop}%
\bibitem [{\citenamefont {H{\"u}bner}\ \emph {et~al.}(2023)\citenamefont
  {H{\"u}bner}, \citenamefont {Dauer}, \citenamefont {Eggert}, \citenamefont
  {Kollath},\ and\ \citenamefont
  {Sheikhan}}]{Hubner-Sheikhan-MomentumresolvedFloquetengineeredPair-2023}%
  \BibitemOpen
  \bibfield  {author} {\bibinfo {author} {\bibfnamefont {F.}~\bibnamefont
  {H{\"u}bner}}, \bibinfo {author} {\bibfnamefont {C.}~\bibnamefont {Dauer}},
  \bibinfo {author} {\bibfnamefont {S.}~\bibnamefont {Eggert}}, \bibinfo
  {author} {\bibfnamefont {C.}~\bibnamefont {Kollath}},\ and\ \bibinfo {author}
  {\bibfnamefont {A.}~\bibnamefont {Sheikhan}},\ }\href
  {https://doi.org/10.1103/PhysRevA.108.023307} {\bibfield  {journal} {\bibinfo
   {journal} {Physical Review A}\ }\textbf {\bibinfo {volume} {108}},\ \bibinfo
  {pages} {023307} (\bibinfo {year} {2023})}\BibitemShut {NoStop}%
\bibitem [{\citenamefont {Benhemou}\ \emph {et~al.}(2023)\citenamefont
  {Benhemou}, \citenamefont {Nixon}, \citenamefont {Deger}, \citenamefont
  {Schneider},\ and\ \citenamefont {Pachos}}]{benhemou23_arxiv_blackhole}%
  \BibitemOpen
  \bibfield  {author} {\bibinfo {author} {\bibfnamefont {A.}~\bibnamefont
  {Benhemou}}, \bibinfo {author} {\bibfnamefont {G.}~\bibnamefont {Nixon}},
  \bibinfo {author} {\bibfnamefont {A.}~\bibnamefont {Deger}}, \bibinfo
  {author} {\bibfnamefont {U.}~\bibnamefont {Schneider}},\ and\ \bibinfo
  {author} {\bibfnamefont {J.~K.}\ \bibnamefont {Pachos}},\ }\bibfield
  {journal} {\bibinfo  {journal} {arXiv preprint}\ }\href
  {https://doi.org/10.48550/arXiv.2312.14058} {10.48550/arXiv.2312.14058}
  (\bibinfo {year} {2023})\BibitemShut {NoStop}%
\bibitem [{\citenamefont {Goldman}\ and\ \citenamefont
  {Dalibard}(2014)}]{goldman_periodically_2014}%
  \BibitemOpen
  \bibfield  {author} {\bibinfo {author} {\bibfnamefont {N.}~\bibnamefont
  {Goldman}}\ and\ \bibinfo {author} {\bibfnamefont {J.}~\bibnamefont
  {Dalibard}},\ }\href {https://doi.org/10.1103/PhysRevX.4.031027} {\bibfield
  {journal} {\bibinfo  {journal} {Physical Review X}\ }\textbf {\bibinfo
  {volume} {4}},\ \bibinfo {pages} {031027} (\bibinfo {year}
  {2014})}\BibitemShut {NoStop}%
\bibitem [{\citenamefont {Bukov}\ \emph {et~al.}(2015)\citenamefont {Bukov},
  \citenamefont {D'Alessio},\ and\ \citenamefont
  {Polkovnikov}}]{bukov_universal_2015}%
  \BibitemOpen
  \bibfield  {author} {\bibinfo {author} {\bibfnamefont {M.}~\bibnamefont
  {Bukov}}, \bibinfo {author} {\bibfnamefont {L.}~\bibnamefont {D'Alessio}},\
  and\ \bibinfo {author} {\bibfnamefont {A.}~\bibnamefont {Polkovnikov}},\
  }\bibfield  {journal} {\bibinfo  {journal} {Advances in Physics}\ }\textbf
  {\bibinfo {volume} {64}},\ \href
  {https://doi.org/10.1080/00018732.2015.1055918}
  {10.1080/00018732.2015.1055918} (\bibinfo {year} {2015})\BibitemShut
  {NoStop}%
\bibitem [{\citenamefont {Eckardt}\ and\ \citenamefont
  {Anisimovas}(2015)}]{eckardt_high-frequency_2015}%
  \BibitemOpen
  \bibfield  {author} {\bibinfo {author} {\bibfnamefont {A.}~\bibnamefont
  {Eckardt}}\ and\ \bibinfo {author} {\bibfnamefont {E.}~\bibnamefont
  {Anisimovas}},\ }\href {https://doi.org/10.1088/1367-2630/17/9/093039}
  {\bibfield  {journal} {\bibinfo  {journal} {New Journal of Physics}\ }\textbf
  {\bibinfo {volume} {17}},\ \bibinfo {pages} {093039} (\bibinfo {year}
  {2015})}\BibitemShut {NoStop}%
\bibitem [{\citenamefont {Holthaus}(2016)}]{holthaus_floquet_2016}%
  \BibitemOpen
  \bibfield  {author} {\bibinfo {author} {\bibfnamefont {M.}~\bibnamefont
  {Holthaus}},\ }\href {https://doi.org/10.1088/0953-4075/49/1/013001}
  {\bibfield  {journal} {\bibinfo  {journal} {Journal of Physics B: Atomic,
  Molecular and Optical Physics}\ }\textbf {\bibinfo {volume} {49}},\ \bibinfo
  {pages} {013001} (\bibinfo {year} {2016})}\BibitemShut {NoStop}%
\bibitem [{\citenamefont {Souza}\ \emph {et~al.}(2001)\citenamefont {Souza},
  \citenamefont {Marzari},\ and\ \citenamefont
  {Vanderbilt}}]{Vanderbilt01_PRB_wannier}%
  \BibitemOpen
  \bibfield  {author} {\bibinfo {author} {\bibfnamefont {I.}~\bibnamefont
  {Souza}}, \bibinfo {author} {\bibfnamefont {N.}~\bibnamefont {Marzari}},\
  and\ \bibinfo {author} {\bibfnamefont {D.}~\bibnamefont {Vanderbilt}},\
  }\href {https://doi.org/10.1103/PhysRevB.65.035109} {\bibfield  {journal}
  {\bibinfo  {journal} {Physical Review B}\ }\textbf {\bibinfo {volume} {65}},\
  \bibinfo {pages} {035109} (\bibinfo {year} {2001})}\BibitemShut {NoStop}%
\bibitem [{\citenamefont {Struck}\ \emph {et~al.}(2012)\citenamefont {Struck},
  \citenamefont {Ölschläger}, \citenamefont {Weinberg}, \citenamefont
  {Hauke}, \citenamefont {Simonet}, \citenamefont {Eckardt}, \citenamefont
  {Lewenstein}, \citenamefont {Sengstock},\ and\ \citenamefont
  {Windpassinger}}]{struck_tunable_2012}%
  \BibitemOpen
  \bibfield  {author} {\bibinfo {author} {\bibfnamefont {J.}~\bibnamefont
  {Struck}}, \bibinfo {author} {\bibfnamefont {C.}~\bibnamefont
  {Ölschläger}}, \bibinfo {author} {\bibfnamefont {M.}~\bibnamefont
  {Weinberg}}, \bibinfo {author} {\bibfnamefont {P.}~\bibnamefont {Hauke}},
  \bibinfo {author} {\bibfnamefont {J.}~\bibnamefont {Simonet}}, \bibinfo
  {author} {\bibfnamefont {A.}~\bibnamefont {Eckardt}}, \bibinfo {author}
  {\bibfnamefont {M.}~\bibnamefont {Lewenstein}}, \bibinfo {author}
  {\bibfnamefont {K.}~\bibnamefont {Sengstock}},\ and\ \bibinfo {author}
  {\bibfnamefont {P.}~\bibnamefont {Windpassinger}},\ }\href
  {https://doi.org/10.1103/PhysRevLett.108.225304} {\bibfield  {journal}
  {\bibinfo  {journal} {Physical Review Letters}\ }\textbf {\bibinfo {volume}
  {108}},\ \bibinfo {pages} {225304} (\bibinfo {year} {2012})}\BibitemShut
  {NoStop}%
\bibitem [{\citenamefont {Wang}\ \emph {et~al.}(2021)\citenamefont {Wang},
  \citenamefont {Dong}, \citenamefont {{\"U}nal},\ and\ \citenamefont
  {Eckardt}}]{Wang21_NJP}%
  \BibitemOpen
  \bibfield  {author} {\bibinfo {author} {\bibfnamefont {B.}~\bibnamefont
  {Wang}}, \bibinfo {author} {\bibfnamefont {X.-Y.}\ \bibnamefont {Dong}},
  \bibinfo {author} {\bibfnamefont {F.~N.}\ \bibnamefont {{\"U}nal}},\ and\
  \bibinfo {author} {\bibfnamefont {A.}~\bibnamefont {Eckardt}},\ }\href
  {https://doi.org/10.1088/1367-2630/abf9b2} {\bibfield  {journal} {\bibinfo
  {journal} {New Journal of Physics}\ }\textbf {\bibinfo {volume} {23}},\
  \bibinfo {pages} {063017} (\bibinfo {year} {2021})}\BibitemShut {NoStop}%
\bibitem [{\citenamefont {Young}\ \emph {et~al.}(2022)\citenamefont {Young},
  \citenamefont {Eckner}, \citenamefont {Schine}, \citenamefont {Childs},\ and\
  \citenamefont {Kaufman}}]{Young-Kaufman-Tweezerprogrammable2DQuantum-2022}%
  \BibitemOpen
  \bibfield  {author} {\bibinfo {author} {\bibfnamefont {A.~W.}\ \bibnamefont
  {Young}}, \bibinfo {author} {\bibfnamefont {W.~J.}\ \bibnamefont {Eckner}},
  \bibinfo {author} {\bibfnamefont {N.}~\bibnamefont {Schine}}, \bibinfo
  {author} {\bibfnamefont {A.~M.}\ \bibnamefont {Childs}},\ and\ \bibinfo
  {author} {\bibfnamefont {A.~M.}\ \bibnamefont {Kaufman}},\ }\href
  {https://doi.org/10.1126/science.abo0608} {\bibfield  {journal} {\bibinfo
  {journal} {Science}\ }\textbf {\bibinfo {volume} {377}},\ \bibinfo {pages}
  {885} (\bibinfo {year} {2022})}\BibitemShut {NoStop}%
\bibitem [{\citenamefont {Wang}\ \emph {et~al.}(2018)\citenamefont {Wang},
  \citenamefont {Ünal},\ and\ \citenamefont {Eckardt}}]{wang_floquet_2018}%
  \BibitemOpen
  \bibfield  {author} {\bibinfo {author} {\bibfnamefont {B.}~\bibnamefont
  {Wang}}, \bibinfo {author} {\bibfnamefont {F.~N.}\ \bibnamefont {Ünal}},\
  and\ \bibinfo {author} {\bibfnamefont {A.}~\bibnamefont {Eckardt}},\ }\href
  {https://doi.org/10.1103/PhysRevLett.120.243602} {\bibfield  {journal}
  {\bibinfo  {journal} {Physical Review Letters}\ }\textbf {\bibinfo {volume}
  {120}},\ \bibinfo {pages} {243602} (\bibinfo {year} {2018})}\BibitemShut
  {NoStop}%
\bibitem [{\citenamefont {Su}\ \emph {et~al.}(1979)\citenamefont {Su},
  \citenamefont {Schrieffer},\ and\ \citenamefont {Heeger}}]{su_solitons_1979}%
  \BibitemOpen
  \bibfield  {author} {\bibinfo {author} {\bibfnamefont {W.~P.}\ \bibnamefont
  {Su}}, \bibinfo {author} {\bibfnamefont {J.~R.}\ \bibnamefont {Schrieffer}},\
  and\ \bibinfo {author} {\bibfnamefont {A.~J.}\ \bibnamefont {Heeger}},\
  }\href {https://doi.org/10.1103/PhysRevLett.42.1698} {\bibfield  {journal}
  {\bibinfo  {journal} {Physical Review Letters}\ }\textbf {\bibinfo {volume}
  {42}},\ \bibinfo {pages} {1698} (\bibinfo {year} {1979})}\BibitemShut
  {NoStop}%
\bibitem [{\citenamefont {Ryu}\ \emph {et~al.}(2010)\citenamefont {Ryu},
  \citenamefont {Schnyder}, \citenamefont {Furusaki},\ and\ \citenamefont
  {Ludwig}}]{ryu_topological_2010}%
  \BibitemOpen
  \bibfield  {author} {\bibinfo {author} {\bibfnamefont {S.}~\bibnamefont
  {Ryu}}, \bibinfo {author} {\bibfnamefont {A.~P.}\ \bibnamefont {Schnyder}},
  \bibinfo {author} {\bibfnamefont {A.}~\bibnamefont {Furusaki}},\ and\
  \bibinfo {author} {\bibfnamefont {A.~W.~W.}\ \bibnamefont {Ludwig}},\ }\href
  {https://doi.org/10.1088/1367-2630/12/6/065010} {\bibfield  {journal}
  {\bibinfo  {journal} {New Journal of Physics}\ }\textbf {\bibinfo {volume}
  {12}},\ \bibinfo {pages} {065010} (\bibinfo {year} {2010})}\BibitemShut
  {NoStop}%
\bibitem [{\citenamefont {Atala}\ \emph {et~al.}(2013)\citenamefont {Atala},
  \citenamefont {Aidelsburger}, \citenamefont {Barreiro}, \citenamefont
  {Abanin}, \citenamefont {Kitagawa}, \citenamefont {Demler},\ and\
  \citenamefont {Bloch}}]{atala13_NatPhys}%
  \BibitemOpen
  \bibfield  {author} {\bibinfo {author} {\bibfnamefont {M.}~\bibnamefont
  {Atala}}, \bibinfo {author} {\bibfnamefont {M.}~\bibnamefont {Aidelsburger}},
  \bibinfo {author} {\bibfnamefont {J.}~\bibnamefont {Barreiro}}, \bibinfo
  {author} {\bibfnamefont {D.}~\bibnamefont {Abanin}}, \bibinfo {author}
  {\bibfnamefont {T.}~\bibnamefont {Kitagawa}}, \bibinfo {author}
  {\bibfnamefont {E.}~\bibnamefont {Demler}},\ and\ \bibinfo {author}
  {\bibfnamefont {I.}~\bibnamefont {Bloch}},\ }\href
  {https://doi.org/10.1038/nphys2790} {\bibfield  {journal} {\bibinfo
  {journal} {Nature Physics}\ }\textbf {\bibinfo {volume} {9}},\ \bibinfo
  {pages} {795–800} (\bibinfo {year} {2013})}\BibitemShut {NoStop}%
\bibitem [{\citenamefont {Nakajima}\ \emph {et~al.}(2016)\citenamefont
  {Nakajima}, \citenamefont {Tomita}, \citenamefont {Taie}, \citenamefont
  {Ichinose}, \citenamefont {Ozawa}, \citenamefont {Wang}, \citenamefont
  {Troyer},\ and\ \citenamefont {Takahashi}}]{nakajima_topological_2016}%
  \BibitemOpen
  \bibfield  {author} {\bibinfo {author} {\bibfnamefont {S.}~\bibnamefont
  {Nakajima}}, \bibinfo {author} {\bibfnamefont {T.}~\bibnamefont {Tomita}},
  \bibinfo {author} {\bibfnamefont {S.}~\bibnamefont {Taie}}, \bibinfo {author}
  {\bibfnamefont {T.}~\bibnamefont {Ichinose}}, \bibinfo {author}
  {\bibfnamefont {H.}~\bibnamefont {Ozawa}}, \bibinfo {author} {\bibfnamefont
  {L.}~\bibnamefont {Wang}}, \bibinfo {author} {\bibfnamefont {M.}~\bibnamefont
  {Troyer}},\ and\ \bibinfo {author} {\bibfnamefont {Y.}~\bibnamefont
  {Takahashi}},\ }\href {https://doi.org/10.1038/nphys3622} {\bibfield
  {journal} {\bibinfo  {journal} {Nature Physics}\ }\textbf {\bibinfo {volume}
  {12}},\ \bibinfo {pages} {296} (\bibinfo {year} {2016})}\BibitemShut
  {NoStop}%
\bibitem [{\citenamefont {Lohse}\ \emph {et~al.}(2016)\citenamefont {Lohse},
  \citenamefont {Schweizer}, \citenamefont {Zilberberg}, \citenamefont
  {Aidelsburger},\ and\ \citenamefont {Bloch}}]{lohse_thouless_2016}%
  \BibitemOpen
  \bibfield  {author} {\bibinfo {author} {\bibfnamefont {M.}~\bibnamefont
  {Lohse}}, \bibinfo {author} {\bibfnamefont {C.}~\bibnamefont {Schweizer}},
  \bibinfo {author} {\bibfnamefont {O.}~\bibnamefont {Zilberberg}}, \bibinfo
  {author} {\bibfnamefont {M.}~\bibnamefont {Aidelsburger}},\ and\ \bibinfo
  {author} {\bibfnamefont {I.}~\bibnamefont {Bloch}},\ }\href
  {https://doi.org/10.1038/nphys3584} {\bibfield  {journal} {\bibinfo
  {journal} {Nature Physics}\ }\textbf {\bibinfo {volume} {12}},\ \bibinfo
  {pages} {350} (\bibinfo {year} {2016})}\BibitemShut {NoStop}%
\bibitem [{\citenamefont {{de L{\'e}s{\'e}leuc}}\ \emph
  {et~al.}(2019)\citenamefont {{de L{\'e}s{\'e}leuc}}, \citenamefont
  {Lienhard}, \citenamefont {Scholl}, \citenamefont {Barredo}, \citenamefont
  {Weber}, \citenamefont {Lang}, \citenamefont {B{\"u}chler}, \citenamefont
  {Lahaye},\ and\ \citenamefont
  {Browaeys}}]{deLeseleuc-Browaeys-ObservationSymmetryprotectedTopological-2019}%
  \BibitemOpen
  \bibfield  {author} {\bibinfo {author} {\bibfnamefont {S.}~\bibnamefont {{de
  L{\'e}s{\'e}leuc}}}, \bibinfo {author} {\bibfnamefont {V.}~\bibnamefont
  {Lienhard}}, \bibinfo {author} {\bibfnamefont {P.}~\bibnamefont {Scholl}},
  \bibinfo {author} {\bibfnamefont {D.}~\bibnamefont {Barredo}}, \bibinfo
  {author} {\bibfnamefont {S.}~\bibnamefont {Weber}}, \bibinfo {author}
  {\bibfnamefont {N.}~\bibnamefont {Lang}}, \bibinfo {author} {\bibfnamefont
  {H.~P.}\ \bibnamefont {B{\"u}chler}}, \bibinfo {author} {\bibfnamefont
  {T.}~\bibnamefont {Lahaye}},\ and\ \bibinfo {author} {\bibfnamefont
  {A.}~\bibnamefont {Browaeys}},\ }\href
  {https://doi.org/10.1126/science.aav9105} {\bibfield  {journal} {\bibinfo
  {journal} {Science}\ }\textbf {\bibinfo {volume} {365}},\ \bibinfo {pages}
  {775} (\bibinfo {year} {2019})}\BibitemShut {NoStop}%
\bibitem [{\citenamefont {Maffei}\ \emph {et~al.}(2018)\citenamefont {Maffei},
  \citenamefont {Dauphin}, \citenamefont {Cardano}, \citenamefont
  {Lewenstein},\ and\ \citenamefont
  {Massignan}}]{Maffei-Massignan-TopologicalCharacterizationChiral-2018}%
  \BibitemOpen
  \bibfield  {author} {\bibinfo {author} {\bibfnamefont {M.}~\bibnamefont
  {Maffei}}, \bibinfo {author} {\bibfnamefont {A.}~\bibnamefont {Dauphin}},
  \bibinfo {author} {\bibfnamefont {F.}~\bibnamefont {Cardano}}, \bibinfo
  {author} {\bibfnamefont {M.}~\bibnamefont {Lewenstein}},\ and\ \bibinfo
  {author} {\bibfnamefont {P.}~\bibnamefont {Massignan}},\ }\href
  {https://doi.org/10.1088/1367-2630/aa9d4c} {\bibfield  {journal} {\bibinfo
  {journal} {New Journal of Physics}\ }\textbf {\bibinfo {volume} {20}},\
  \bibinfo {pages} {013023} (\bibinfo {year} {2018})}\BibitemShut {NoStop}%
\bibitem [{\citenamefont {Xie}\ \emph {et~al.}(2019)\citenamefont {Xie},
  \citenamefont {Gou}, \citenamefont {Xiao}, \citenamefont {Gadway},\ and\
  \citenamefont {Yan}}]{Xie-Yan-TopologicalCharacterizationsExtended-2019}%
  \BibitemOpen
  \bibfield  {author} {\bibinfo {author} {\bibfnamefont {D.}~\bibnamefont
  {Xie}}, \bibinfo {author} {\bibfnamefont {W.}~\bibnamefont {Gou}}, \bibinfo
  {author} {\bibfnamefont {T.}~\bibnamefont {Xiao}}, \bibinfo {author}
  {\bibfnamefont {B.}~\bibnamefont {Gadway}},\ and\ \bibinfo {author}
  {\bibfnamefont {B.}~\bibnamefont {Yan}},\ }\href
  {https://doi.org/10.1038/s41534-019-0159-6} {\bibfield  {journal} {\bibinfo
  {journal} {npj Quantum Information}\ }\textbf {\bibinfo {volume} {5}},\
  \bibinfo {pages} {1} (\bibinfo {year} {2019})}\BibitemShut {NoStop}%
\bibitem [{\citenamefont {{Peierls}}(1933)}]{Peierls33}%
  \BibitemOpen
  \bibfield  {author} {\bibinfo {author} {\bibfnamefont {R.}~\bibnamefont
  {{Peierls}}},\ }\href {https://doi.org/10.1007/BF01342591} {\bibfield
  {journal} {\bibinfo  {journal} {Zeitschrift fur Physik}\ }\textbf {\bibinfo
  {volume} {80}},\ \bibinfo {pages} {763} (\bibinfo {year} {1933})}\BibitemShut
  {NoStop}%
\bibitem [{\citenamefont {Oka}\ and\ \citenamefont
  {Aoki}(2009)}]{OkaAoki09_PRB}%
  \BibitemOpen
  \bibfield  {author} {\bibinfo {author} {\bibfnamefont {T.}~\bibnamefont
  {Oka}}\ and\ \bibinfo {author} {\bibfnamefont {H.}~\bibnamefont {Aoki}},\
  }\href {https://doi.org/10.1103/PhysRevB.79.081406} {\bibfield  {journal}
  {\bibinfo  {journal} {Physical Review B}\ }\textbf {\bibinfo {volume} {79}},\
  \bibinfo {pages} {081406} (\bibinfo {year} {2009})}\BibitemShut {NoStop}%
\bibitem [{\citenamefont {Korsch}\ \emph {et~al.}(2006)\citenamefont {Korsch},
  \citenamefont {Klumpp},\ and\ \citenamefont
  {Witthaut}}]{Korsch-Witthaut-TwodimensionalBesselFunctions-2006}%
  \BibitemOpen
  \bibfield  {author} {\bibinfo {author} {\bibfnamefont {H.~J.}\ \bibnamefont
  {Korsch}}, \bibinfo {author} {\bibfnamefont {A.}~\bibnamefont {Klumpp}},\
  and\ \bibinfo {author} {\bibfnamefont {D.}~\bibnamefont {Witthaut}},\ }\href
  {https://doi.org/10.1088/0305-4470/39/48/008} {\bibfield  {journal} {\bibinfo
   {journal} {Journal of Physics A: Mathematical and General}\ }\textbf
  {\bibinfo {volume} {39}},\ \bibinfo {pages} {14947} (\bibinfo {year}
  {2006})}\BibitemShut {NoStop}%
\bibitem [{\citenamefont {Atala}\ \emph {et~al.}(2014)\citenamefont {Atala},
  \citenamefont {Aidelsburger}, \citenamefont {Lohse}, \citenamefont
  {Barreiro}, \citenamefont {Paredes},\ and\ \citenamefont
  {Bloch}}]{Atala-Bloch-ObservationChiralCurrents-2014}%
  \BibitemOpen
  \bibfield  {author} {\bibinfo {author} {\bibfnamefont {M.}~\bibnamefont
  {Atala}}, \bibinfo {author} {\bibfnamefont {M.}~\bibnamefont {Aidelsburger}},
  \bibinfo {author} {\bibfnamefont {M.}~\bibnamefont {Lohse}}, \bibinfo
  {author} {\bibfnamefont {J.~T.}\ \bibnamefont {Barreiro}}, \bibinfo {author}
  {\bibfnamefont {B.}~\bibnamefont {Paredes}},\ and\ \bibinfo {author}
  {\bibfnamefont {I.}~\bibnamefont {Bloch}},\ }\href
  {https://doi.org/10.1038/nphys2998} {\bibfield  {journal} {\bibinfo
  {journal} {Nature Physics}\ }\textbf {\bibinfo {volume} {10}},\ \bibinfo
  {pages} {588} (\bibinfo {year} {2014})}\BibitemShut {NoStop}%
\bibitem [{\citenamefont {Kennedy}\ \emph {et~al.}(2015)\citenamefont
  {Kennedy}, \citenamefont {Burton}, \citenamefont {Chung},\ and\ \citenamefont
  {Ketterle}}]{Kennedy-Ketterle-ObservationBoseEinstein-2015}%
  \BibitemOpen
  \bibfield  {author} {\bibinfo {author} {\bibfnamefont {C.~J.}\ \bibnamefont
  {Kennedy}}, \bibinfo {author} {\bibfnamefont {W.~C.}\ \bibnamefont {Burton}},
  \bibinfo {author} {\bibfnamefont {W.~C.}\ \bibnamefont {Chung}},\ and\
  \bibinfo {author} {\bibfnamefont {W.}~\bibnamefont {Ketterle}},\ }\href
  {https://doi.org/10.1038/nphys3421} {\bibfield  {journal} {\bibinfo
  {journal} {Nature Physics}\ }\textbf {\bibinfo {volume} {11}},\ \bibinfo
  {pages} {859} (\bibinfo {year} {2015})}\BibitemShut {NoStop}%
\bibitem [{\citenamefont {Tai}\ \emph {et~al.}(2017)\citenamefont {Tai},
  \citenamefont {Lukin}, \citenamefont {Rispoli}, \citenamefont {Schittko},
  \citenamefont {Menke}, \citenamefont {{Dan Borgnia}}, \citenamefont {Preiss},
  \citenamefont {Grusdt}, \citenamefont {Kaufman},\ and\ \citenamefont
  {Greiner}}]{Tai-Greiner-MicroscopyInteractingHarper-2017}%
  \BibitemOpen
  \bibfield  {author} {\bibinfo {author} {\bibfnamefont {M.~E.}\ \bibnamefont
  {Tai}}, \bibinfo {author} {\bibfnamefont {A.}~\bibnamefont {Lukin}}, \bibinfo
  {author} {\bibfnamefont {M.}~\bibnamefont {Rispoli}}, \bibinfo {author}
  {\bibfnamefont {R.}~\bibnamefont {Schittko}}, \bibinfo {author}
  {\bibfnamefont {T.}~\bibnamefont {Menke}}, \bibinfo {author} {\bibnamefont
  {{Dan Borgnia}}}, \bibinfo {author} {\bibfnamefont {P.~M.}\ \bibnamefont
  {Preiss}}, \bibinfo {author} {\bibfnamefont {F.}~\bibnamefont {Grusdt}},
  \bibinfo {author} {\bibfnamefont {A.~M.}\ \bibnamefont {Kaufman}},\ and\
  \bibinfo {author} {\bibfnamefont {M.}~\bibnamefont {Greiner}},\ }\href
  {https://doi.org/10.1038/nature22811} {\bibfield  {journal} {\bibinfo
  {journal} {Nature}\ }\textbf {\bibinfo {volume} {546}},\ \bibinfo {pages}
  {519} (\bibinfo {year} {2017})}\BibitemShut {NoStop}%
\bibitem [{\citenamefont {Endres}\ \emph {et~al.}(2016)\citenamefont {Endres},
  \citenamefont {Bernien}, \citenamefont {Keesling}, \citenamefont {Levine},
  \citenamefont {Anschuetz}, \citenamefont {Krajenbrink}, \citenamefont
  {Senko}, \citenamefont {Vuletic}, \citenamefont {Greiner},\ and\
  \citenamefont {Lukin}}]{Endres-Lukin-AtombyatomAssemblyDefectfree-2016}%
  \BibitemOpen
  \bibfield  {author} {\bibinfo {author} {\bibfnamefont {M.}~\bibnamefont
  {Endres}}, \bibinfo {author} {\bibfnamefont {H.}~\bibnamefont {Bernien}},
  \bibinfo {author} {\bibfnamefont {A.}~\bibnamefont {Keesling}}, \bibinfo
  {author} {\bibfnamefont {H.}~\bibnamefont {Levine}}, \bibinfo {author}
  {\bibfnamefont {E.~R.}\ \bibnamefont {Anschuetz}}, \bibinfo {author}
  {\bibfnamefont {A.}~\bibnamefont {Krajenbrink}}, \bibinfo {author}
  {\bibfnamefont {C.}~\bibnamefont {Senko}}, \bibinfo {author} {\bibfnamefont
  {V.}~\bibnamefont {Vuletic}}, \bibinfo {author} {\bibfnamefont
  {M.}~\bibnamefont {Greiner}},\ and\ \bibinfo {author} {\bibfnamefont {M.~D.}\
  \bibnamefont {Lukin}},\ }\href {https://doi.org/10.1126/science.aah3752}
  {\bibfield  {journal} {\bibinfo  {journal} {Science}\ }\textbf {\bibinfo
  {volume} {354}},\ \bibinfo {pages} {1024} (\bibinfo {year}
  {2016})}\BibitemShut {NoStop}%
\bibitem [{\citenamefont {Barredo}\ \emph {et~al.}(2016)\citenamefont
  {Barredo}, \citenamefont {{de L{\'e}s{\'e}leuc}}, \citenamefont {Lienhard},
  \citenamefont {Lahaye},\ and\ \citenamefont
  {Browaeys}}]{Barredo-Browaeys-AtombyatomAssemblerDefectfree-2016}%
  \BibitemOpen
  \bibfield  {author} {\bibinfo {author} {\bibfnamefont {D.}~\bibnamefont
  {Barredo}}, \bibinfo {author} {\bibfnamefont {S.}~\bibnamefont {{de
  L{\'e}s{\'e}leuc}}}, \bibinfo {author} {\bibfnamefont {V.}~\bibnamefont
  {Lienhard}}, \bibinfo {author} {\bibfnamefont {T.}~\bibnamefont {Lahaye}},\
  and\ \bibinfo {author} {\bibfnamefont {A.}~\bibnamefont {Browaeys}},\ }\href
  {https://doi.org/10.1126/science.aah3778} {\bibfield  {journal} {\bibinfo
  {journal} {Science}\ }\textbf {\bibinfo {volume} {354}},\ \bibinfo {pages}
  {1021} (\bibinfo {year} {2016})}\BibitemShut {NoStop}%
\bibitem [{\citenamefont {Chisholm}\ \emph {et~al.}(2018)\citenamefont
  {Chisholm}, \citenamefont {Thomas}, \citenamefont {Deb},\ and\ \citenamefont
  {Kj{\ae}rgaard}}]{Chisholm-Kjaergaard-ThreedimensionalSteerableOptical-2018}%
  \BibitemOpen
  \bibfield  {author} {\bibinfo {author} {\bibfnamefont {C.~S.}\ \bibnamefont
  {Chisholm}}, \bibinfo {author} {\bibfnamefont {R.}~\bibnamefont {Thomas}},
  \bibinfo {author} {\bibfnamefont {A.~B.}\ \bibnamefont {Deb}},\ and\ \bibinfo
  {author} {\bibfnamefont {N.}~\bibnamefont {Kj{\ae}rgaard}},\ }\href
  {https://doi.org/10.1063/1.5041481} {\bibfield  {journal} {\bibinfo
  {journal} {Review of Scientific Instruments}\ }\textbf {\bibinfo {volume}
  {89}},\ \bibinfo {pages} {103105} (\bibinfo {year} {2018})}\BibitemShut
  {NoStop}%
\bibitem [{\citenamefont {Omran}\ \emph {et~al.}(2019)\citenamefont {Omran},
  \citenamefont {Levine}, \citenamefont {Keesling}, \citenamefont {Semeghini},
  \citenamefont {Wang}, \citenamefont {Ebadi}, \citenamefont {Bernien},
  \citenamefont {Zibrov}, \citenamefont {Pichler}, \citenamefont {Choi},
  \citenamefont {Cui}, \citenamefont {Rossignolo}, \citenamefont {Rembold},
  \citenamefont {Montangero}, \citenamefont {Calarco}, \citenamefont {Endres},
  \citenamefont {Greiner}, \citenamefont {Vuleti{\'c}},\ and\ \citenamefont
  {Lukin}}]{Omran-Lukin-GenerationManipulationSchrodinger-2019}%
  \BibitemOpen
  \bibfield  {author} {\bibinfo {author} {\bibfnamefont {A.}~\bibnamefont
  {Omran}}, \bibinfo {author} {\bibfnamefont {H.}~\bibnamefont {Levine}},
  \bibinfo {author} {\bibfnamefont {A.}~\bibnamefont {Keesling}}, \bibinfo
  {author} {\bibfnamefont {G.}~\bibnamefont {Semeghini}}, \bibinfo {author}
  {\bibfnamefont {T.~T.}\ \bibnamefont {Wang}}, \bibinfo {author}
  {\bibfnamefont {S.}~\bibnamefont {Ebadi}}, \bibinfo {author} {\bibfnamefont
  {H.}~\bibnamefont {Bernien}}, \bibinfo {author} {\bibfnamefont {A.~S.}\
  \bibnamefont {Zibrov}}, \bibinfo {author} {\bibfnamefont {H.}~\bibnamefont
  {Pichler}}, \bibinfo {author} {\bibfnamefont {S.}~\bibnamefont {Choi}},
  \bibinfo {author} {\bibfnamefont {J.}~\bibnamefont {Cui}}, \bibinfo {author}
  {\bibfnamefont {M.}~\bibnamefont {Rossignolo}}, \bibinfo {author}
  {\bibfnamefont {P.}~\bibnamefont {Rembold}}, \bibinfo {author} {\bibfnamefont
  {S.}~\bibnamefont {Montangero}}, \bibinfo {author} {\bibfnamefont
  {T.}~\bibnamefont {Calarco}}, \bibinfo {author} {\bibfnamefont
  {M.}~\bibnamefont {Endres}}, \bibinfo {author} {\bibfnamefont
  {M.}~\bibnamefont {Greiner}}, \bibinfo {author} {\bibfnamefont
  {V.}~\bibnamefont {Vuleti{\'c}}},\ and\ \bibinfo {author} {\bibfnamefont
  {M.~D.}\ \bibnamefont {Lukin}},\ }\href
  {https://doi.org/10.1126/science.aax9743} {\bibfield  {journal} {\bibinfo
  {journal} {Science}\ }\textbf {\bibinfo {volume} {365}},\ \bibinfo {pages}
  {570} (\bibinfo {year} {2019})}\BibitemShut {NoStop}%
\bibitem [{\citenamefont {Bluvstein}\ \emph {et~al.}(2022)\citenamefont
  {Bluvstein}, \citenamefont {Levine}, \citenamefont {Semeghini}, \citenamefont
  {Wang}, \citenamefont {Ebadi}, \citenamefont {Kalinowski}, \citenamefont
  {Keesling}, \citenamefont {Maskara}, \citenamefont {Pichler}, \citenamefont
  {Greiner}, \citenamefont {Vuleti{\'c}},\ and\ \citenamefont
  {Lukin}}]{Bluvstein-Lukin-QuantumProcessorBased-2022}%
  \BibitemOpen
  \bibfield  {author} {\bibinfo {author} {\bibfnamefont {D.}~\bibnamefont
  {Bluvstein}}, \bibinfo {author} {\bibfnamefont {H.}~\bibnamefont {Levine}},
  \bibinfo {author} {\bibfnamefont {G.}~\bibnamefont {Semeghini}}, \bibinfo
  {author} {\bibfnamefont {T.~T.}\ \bibnamefont {Wang}}, \bibinfo {author}
  {\bibfnamefont {S.}~\bibnamefont {Ebadi}}, \bibinfo {author} {\bibfnamefont
  {M.}~\bibnamefont {Kalinowski}}, \bibinfo {author} {\bibfnamefont
  {A.}~\bibnamefont {Keesling}}, \bibinfo {author} {\bibfnamefont
  {N.}~\bibnamefont {Maskara}}, \bibinfo {author} {\bibfnamefont
  {H.}~\bibnamefont {Pichler}}, \bibinfo {author} {\bibfnamefont
  {M.}~\bibnamefont {Greiner}}, \bibinfo {author} {\bibfnamefont
  {V.}~\bibnamefont {Vuleti{\'c}}},\ and\ \bibinfo {author} {\bibfnamefont
  {M.~D.}\ \bibnamefont {Lukin}},\ }\href
  {https://doi.org/10.1038/s41586-022-04592-6} {\bibfield  {journal} {\bibinfo
  {journal} {Nature}\ }\textbf {\bibinfo {volume} {604}},\ \bibinfo {pages}
  {451} (\bibinfo {year} {2022})}\BibitemShut {NoStop}%
\bibitem [{\citenamefont {Spar}\ \emph {et~al.}(2022)\citenamefont {Spar},
  \citenamefont {{Guardado-Sanchez}}, \citenamefont {Chi}, \citenamefont
  {Yan},\ and\ \citenamefont
  {Bakr}}]{Spar-Bakr-RealizationFermiHubbardOptical-2022}%
  \BibitemOpen
  \bibfield  {author} {\bibinfo {author} {\bibfnamefont {B.~M.}\ \bibnamefont
  {Spar}}, \bibinfo {author} {\bibfnamefont {E.}~\bibnamefont
  {{Guardado-Sanchez}}}, \bibinfo {author} {\bibfnamefont {S.}~\bibnamefont
  {Chi}}, \bibinfo {author} {\bibfnamefont {Z.~Z.}\ \bibnamefont {Yan}},\ and\
  \bibinfo {author} {\bibfnamefont {W.~S.}\ \bibnamefont {Bakr}},\ }\href
  {https://doi.org/10.1103/PhysRevLett.128.223202} {\bibfield  {journal}
  {\bibinfo  {journal} {Physical Review Letters}\ }\textbf {\bibinfo {volume}
  {128}},\ \bibinfo {pages} {223202} (\bibinfo {year} {2022})}\BibitemShut
  {NoStop}%
\bibitem [{\citenamefont {Reitter}\ \emph {et~al.}(2017)\citenamefont
  {Reitter}, \citenamefont {N\"ager}, \citenamefont {Wintersperger},
  \citenamefont {Str\"ater}, \citenamefont {Bloch}, \citenamefont {Eckardt},\
  and\ \citenamefont {Schneider}}]{Reitter2017}%
  \BibitemOpen
  \bibfield  {author} {\bibinfo {author} {\bibfnamefont {M.}~\bibnamefont
  {Reitter}}, \bibinfo {author} {\bibfnamefont {J.}~\bibnamefont {N\"ager}},
  \bibinfo {author} {\bibfnamefont {K.}~\bibnamefont {Wintersperger}}, \bibinfo
  {author} {\bibfnamefont {C.}~\bibnamefont {Str\"ater}}, \bibinfo {author}
  {\bibfnamefont {I.}~\bibnamefont {Bloch}}, \bibinfo {author} {\bibfnamefont
  {A.}~\bibnamefont {Eckardt}},\ and\ \bibinfo {author} {\bibfnamefont
  {U.}~\bibnamefont {Schneider}},\ }\href
  {https://doi.org/10.1103/PhysRevLett.119.200402} {\bibfield  {journal}
  {\bibinfo  {journal} {Physical Review Letters}\ }\textbf {\bibinfo {volume}
  {119}},\ \bibinfo {pages} {200402} (\bibinfo {year} {2017})}\BibitemShut
  {NoStop}%
\bibitem [{\citenamefont {Weinberg}\ \emph {et~al.}(2015)\citenamefont
  {Weinberg}, \citenamefont {\"Olschl\"ager}, \citenamefont {Str\"ater},
  \citenamefont {Prelle}, \citenamefont {Eckardt}, \citenamefont {Sengstock},\
  and\ \citenamefont {Simonet}}]{Weinberg2015}%
  \BibitemOpen
  \bibfield  {author} {\bibinfo {author} {\bibfnamefont {M.}~\bibnamefont
  {Weinberg}}, \bibinfo {author} {\bibfnamefont {C.}~\bibnamefont
  {\"Olschl\"ager}}, \bibinfo {author} {\bibfnamefont {C.}~\bibnamefont
  {Str\"ater}}, \bibinfo {author} {\bibfnamefont {S.}~\bibnamefont {Prelle}},
  \bibinfo {author} {\bibfnamefont {A.}~\bibnamefont {Eckardt}}, \bibinfo
  {author} {\bibfnamefont {K.}~\bibnamefont {Sengstock}},\ and\ \bibinfo
  {author} {\bibfnamefont {J.}~\bibnamefont {Simonet}},\ }\href
  {https://doi.org/10.1103/PhysRevA.92.043621} {\bibfield  {journal} {\bibinfo
  {journal} {Physical Review A}\ }\textbf {\bibinfo {volume} {92}},\ \bibinfo
  {pages} {043621} (\bibinfo {year} {2015})}\BibitemShut {NoStop}%
\bibitem [{\citenamefont {Mandel}\ \emph {et~al.}(2003)\citenamefont {Mandel},
  \citenamefont {Greiner}, \citenamefont {Widera}, \citenamefont {Rom},
  \citenamefont {H\"ansch},\ and\ \citenamefont {Bloch}}]{Mandel_SpinDep}%
  \BibitemOpen
  \bibfield  {author} {\bibinfo {author} {\bibfnamefont {O.}~\bibnamefont
  {Mandel}}, \bibinfo {author} {\bibfnamefont {M.}~\bibnamefont {Greiner}},
  \bibinfo {author} {\bibfnamefont {A.}~\bibnamefont {Widera}}, \bibinfo
  {author} {\bibfnamefont {T.}~\bibnamefont {Rom}}, \bibinfo {author}
  {\bibfnamefont {T.~W.}\ \bibnamefont {H\"ansch}},\ and\ \bibinfo {author}
  {\bibfnamefont {I.}~\bibnamefont {Bloch}},\ }\href
  {https://doi.org/10.1103/PhysRevLett.91.010407} {\bibfield  {journal}
  {\bibinfo  {journal} {Physical Review Letters}\ }\textbf {\bibinfo {volume}
  {91}},\ \bibinfo {pages} {010407} (\bibinfo {year} {2003})}\BibitemShut
  {NoStop}%
\bibitem [{\citenamefont {Duan}\ \emph {et~al.}(2003)\citenamefont {Duan},
  \citenamefont {Demler},\ and\ \citenamefont
  {Lukin}}]{Duan-Lukin-ControllingSpinExchange-2003}%
  \BibitemOpen
  \bibfield  {author} {\bibinfo {author} {\bibfnamefont {L.-M.}\ \bibnamefont
  {Duan}}, \bibinfo {author} {\bibfnamefont {E.}~\bibnamefont {Demler}},\ and\
  \bibinfo {author} {\bibfnamefont {M.~D.}\ \bibnamefont {Lukin}},\ }\href
  {https://doi.org/10.1103/PhysRevLett.91.090402} {\bibfield  {journal}
  {\bibinfo  {journal} {Physical Review Letters}\ }\textbf {\bibinfo {volume}
  {91}},\ \bibinfo {pages} {090402} (\bibinfo {year} {2003})}\BibitemShut
  {NoStop}%
\bibitem [{\citenamefont {{Garc{\'i}a-Ripoll}}\ and\ \citenamefont
  {Cirac}(2003)}]{Garcia-Ripoll-Cirac-SpinDynamicsBosons-2003}%
  \BibitemOpen
  \bibfield  {author} {\bibinfo {author} {\bibfnamefont {J.~J.}\ \bibnamefont
  {{Garc{\'i}a-Ripoll}}}\ and\ \bibinfo {author} {\bibfnamefont {J.~I.}\
  \bibnamefont {Cirac}},\ }\href {https://doi.org/10.1088/1367-2630/5/1/376}
  {\bibfield  {journal} {\bibinfo  {journal} {New Journal of Physics}\ }\textbf
  {\bibinfo {volume} {5}},\ \bibinfo {pages} {76} (\bibinfo {year}
  {2003})}\BibitemShut {NoStop}%
\bibitem [{\citenamefont {Slager}\ \emph {et~al.}(2024)\citenamefont {Slager},
  \citenamefont {Bouhon},\ and\ \citenamefont
  {\"Unal}}]{Slager22_arXivAnomEuFloq}%
  \BibitemOpen
  \bibfield  {author} {\bibinfo {author} {\bibfnamefont {R.-J.}\ \bibnamefont
  {Slager}}, \bibinfo {author} {\bibfnamefont {A.}~\bibnamefont {Bouhon}},\
  and\ \bibinfo {author} {\bibfnamefont {F.~N.}\ \bibnamefont {\"Unal}},\
  }\href {https://doi.org/10.1038/s41467-024-45302-2} {\bibfield  {journal}
  {\bibinfo  {journal} {Nature Communications}\ }\textbf {\bibinfo {volume}
  {15}},\ \bibinfo {pages} {1144} (\bibinfo {year} {2024})}\BibitemShut
  {NoStop}%
\bibitem [{\citenamefont {Cubitt}\ \emph {et~al.}(2018)\citenamefont {Cubitt},
  \citenamefont {Montanaro},\ and\ \citenamefont
  {Piddock}}]{Cubitt-Piddock-UniversalQuantumHamiltonians-2018}%
  \BibitemOpen
  \bibfield  {author} {\bibinfo {author} {\bibfnamefont {T.~S.}\ \bibnamefont
  {Cubitt}}, \bibinfo {author} {\bibfnamefont {A.}~\bibnamefont {Montanaro}},\
  and\ \bibinfo {author} {\bibfnamefont {S.}~\bibnamefont {Piddock}},\ }\href
  {https://doi.org/10.1073/pnas.1804949115} {\bibfield  {journal} {\bibinfo
  {journal} {PNAS}\ }\textbf {\bibinfo {volume} {115}},\ \bibinfo {pages}
  {9497} (\bibinfo {year} {2018})}\BibitemShut {NoStop}%
\bibitem [{\citenamefont {Lucas}(2014)}]{Lucas-IsingFormulationsMany-2014}%
  \BibitemOpen
  \bibfield  {author} {\bibinfo {author} {\bibfnamefont {A.}~\bibnamefont
  {Lucas}},\ }\href {https://doi.org/10.3389/fphy.2014.00005} {\bibfield
  {journal} {\bibinfo  {journal} {Frontiers in Physics}\ }\textbf {\bibinfo
  {volume} {2}},\ \bibinfo {pages} {5} (\bibinfo {year} {2014})}\BibitemShut
  {NoStop}%
\end{thebibliography}%

\end{document}